\documentclass[preprint,dvitex]{ptephy_v1}

\def\simge{\mathrel{%
       \rlap{\raise 0.511ex \hbox{$>$}}{\lower 0.511ex \hbox{$\sim$}}}}
\def\simle{\mathrel{
       \rlap{\raise 0.511ex \hbox{$<$}}{\lower 0.511ex \hbox{$\sim$}}}}

\preprintnumber{UTHEP-753, UTCCS-P-135, KYUSHU-HET-218, J-PARC-TH-0232} 

\begin{document}

\title{Latent heat and pressure gap at the first-order deconfining phase transition of SU(3) Yang-Mills theory using the small flow-time expansion method}

\author[1]{Mizuki Shirogane}
\affil[1]{Graduate School of Science and Technology, Niigata University, Niigata 950-2181, Japan}

\author[2]{Shinji Ejiri}
\affil[2]{Department of Physics, Niigata University, Niigata 950-2181, Japan}

\author[3]{Ryo Iwami}
\affil[3]{Track Maintenance of Shinkansen, Rail Maintenance 1st Department, East Japan Railway Company Niigata Branch, Niigata 950-0086, Japan}

\author[4]{Kazuyuki Kanaya}
\affil[4]{Tomonaga Center for the History of the Universe, University of Tsukuba, Tsukuba, Ibaraki 305-8571, Japan}

\author[5,6]{Masakiyo Kitazawa}
\affil[5]{Department of Physics, Osaka University, Toyonaka, Osaka 560-0043, Japan}
\affil[6]{J-PARC Branch, KEK Theory Center, Institute of Particle and Nuclear Studies, KEK, 203-1, Shirakata, Tokai, Ibaraki, 319-1106, Japan}

\author[7]{Hiroshi Suzuki}
\affil[7]{Department of Physics, Kyushu University, 744 Motooka, Nishi-ku, Fukuoka 819-0395, Japan}

\author[8]{Yusuke Taniguchi}
\affil[8]{Center for Computational Sciences, University of Tsukuba, Tsukuba, Ibaraki 305-8577, Japan}

\author[9]{Takashi Umeda}
\affil[9]{Graduate School of Humanities and Social Sciences, Hiroshima University, Higashihiroshima, Hiroshima 739-8524, Japan}

\date{\today}

\begin{abstract}
We study latent heat and pressure gap between the hot and cold phases at the first-order deconfining phase transition temperature of the SU(3) Yang-Mills theory. 
Performing simulations on lattices with various spatial volumes and lattice spacings, we calculate the gaps of the energy density and pressure
using the small flow time expansion (SF$t$X) method.
We find that the latent heat $\Delta \epsilon$ in the continuum limit is 
$\Delta \epsilon /T^4 = 1.117 \pm 0.040$ for the aspect ratio $N_s/N_t=8$
and $1.349 \pm 0.038$ for $N_s/N_t=6$
at the transition temperature $T=T_c$.
We also confirm that the pressure gap is consistent with zero, as expected from the dynamical balance of two phases at~$T_c$. 
From hysteresis curves of the energy density near $T_c$, we show that the energy density in the (metastable) deconfined phase is sensitive to the spatial volume, 
while that in the confined phase is insensitive.
Furthermore, we examine the effect of alternative procedures in the SF$t$X method --- the order of the continuum and the vanishing flow time extrapolations, and also the renormalization scale and higher order corrections in the matching coefficients.
We confirm that the final results are all well consistent with each other for these alternatives.
\end{abstract}

\subjectindex{B01, B38, B64}

\maketitle

\section{Introduction}
\label{intro}

First-order phase transitions appear in various important thermodynamic systems including the high density QCD and many-flavor QCD,
and are associated with various phenomena such as phase coexistence and latent heat.
It is worth developing numerical techniques to investigate first-order phase transitions.
The SU(3) Yang-Mills theory, i.e., the quenched approximation of QCD, at finite temperature has a first-order deconfining phase transition, 
and provides us with a good testing ground for that.

At a first-order phase transition point, metastable states corresponding to two phases coexist at the same time.
To keep a dynamical balance between them, their pressures must be the same.
Therefore, one can check the validity of computational methods by measuring the pressure in each metastable state.
In fact, the problem of the nonvanishing pressure gap has played an important role
in early development of lattice methods to study thermodynamic quantities~\cite{karsch,satz,burgers,ejiri98,engels00,ejiri03}.
On the other hand, the energy density takes different values in each phase.
The difference between them is the latent heat, which is one of the fundamental observables characterizing the first-order phase transition.

In Refs.~\cite{ejiri98, shirogane16}, we studied the first-order transition of the SU(3) Yang-Mills theory using the derivative method \cite{karsch}.
We have numerically confirmed that the pressure gap vanishes when we adopt the nonperturbatively evaluated values of the Karsch coefficients (anisotropy coefficients), and then computed the latent heat using the nonperturbative Karsch coefficients.

In this paper, we study the first-order phase transition of the SU(3) Yang-Mills theory 
adopting a new technique to calculate thermodynamic observables:
the small flow time expansion (SF$t$X) method based on the gradient flow~\cite{Suzuki:2013gza,Makino:2014taa}.
The gradient flow is an evolution of the fields in terms of a fictitious ``time'' $t$ \cite{Narayanan:2006rf,Luscher:2009eq,Luscher:2010iy,Luscher:2011bx,Luscher:2013cpa}.
Fields at positive flow-time $t>0$ can be viewed as smeared fields averaged over a mean-square physical radius of $\sqrt{8t}$ in four dimensions, and the operators constructed by flowed fields at $t>0$ are shown to be free from the ultraviolet divergences and short-distance singularities. 
Making use of this strict finiteness, the SF$t$X method provides us with a general method to correctly calculate any renormalized observables on the lattice. 
The SF$t$X method has been successfully applied to the evaluation of 
thermodynamic quantities in the Yang-Mills gauge theory~\cite{Asakawa:2013laa,Kitazawa:2016dsl,Hirakida:2018uoy,Iritani2019} and in QCD with $2+1$ flavors of dynamical quarks~\cite{WHOT2017b, WHOT2017,Taniguchi:2020mgg,Lat2019-kanaya}.
In the case of SU(3) Yang-Mills theory we study, agreement with other methods has been established within a $1\%$ level~\cite{Iritani2019}.
The method has also been applied recently to study various observables associated with the energy-momentum tensor~\cite{Kitazawa:2017qab,Yanagihara:2018qqg,Kitazawa:2019otp,Yanagihara:2020tvs,Taniguchi:2017ibr,Taniguchi:2018}.

We apply the SF$t$X method to calculate the energy density and pressure separately in hot and cold phases at the first-order deconfining phase transition temperature of the SU(3) Yang-Mills theory. 
We perform simulations on lattices with several lattice spacings and spatial volumes to carry out the continuum extrapolation and also to study the finite-volume effect. 
On confirming that the pressure gap is consistent with zero, we evaluate the latent heat. 

We also test several alternative procedures in the SF$t$X method:
To obtain a physical result by the SF$t$X method, a double extrapolation to the continuum and vanishing flow-time limits, $a\to0$ and $t\to0$, is required. 
When a proper fitting range avoiding small $t$ singularities at $a>0$ is chosen, the final results should be insensitive to the order of these limits. 
We confirm this for the latent heat and pressure gap in the SU(3) Yang-Mills theory.
The second point to be addressed is the quality of the matching coefficients that relate the physical observables and flowed operators in the SF$t$X method. 
We test the next-to-leading order (NLO) and next-to-next-to-leading order (NNLO) expressions as well as the dependence on the choice of the renormalization scale for the matching coefficients.
Though the final results should be insensitive to the details of the matching coefficients, the width of the available fitting range for the extrapolations and thus the numerical quality of the final results are affected by the choice of them.
We show that the final results for the latent heat and pressure gap are insensitive to the choice of these alternatives.
We also show that the NNLO matching coefficients help to reduce lattice artifacts in the calculation of the latent heat.

In the next section, we introduce the SF$t$X method and several details of the method to be discussed in this paper.
Then, our simulations at the first-order transition of SU(3) Yang-Mills theory are explained in Sec.~\ref{simulation}. 
In~Sec.~\ref{results}, we show the results of the latent heat and pressure gap, mainly using the NNLO matching coefficients, and test several alternative procedures in the SF$t$X method.
We then study the hysteresis of the energy density near the phase transition temperature and its spatial volume dependence in Sec.~\ref{hysteresis}.
Our conclusions are given in Sec.~\ref{conclusion}.
Appendix~\ref{sec:leading} is devoted to showing the results of the latent heat and pressure gap with the NLO matching coefficients, with which we study the effect due to the truncation of the perturbative series for the matching coefficients.
In Appendix~\ref{plaquette}, we discuss the choice of lattice operator for the field strength.

\section{Small flow-time expansion method}
\label{method}

Our flow equation for the gradient flow is the simplest one proposed in Ref.~\cite{Luscher:2010iy}.
The ``flowed'' gauge field $B_{\mu}^a(t,x)$ at flow time $t$ is obtained by solving the flow equation 
\begin{equation}
   \partial_tB_\mu^a(t,x)=D_\nu G_{\nu\mu}^a(t,x) \equiv
\partial_\nu G_{\nu\mu}^a(t,x)+f^{abc}B_\nu^b(t,x)G_{\nu\mu}^c(t,x) 
\label{eq:(1.5)}
\end{equation}
with the initial condition $B_{\mu}^a(0,x)=A_{\mu}^a(x)$, where 
\begin{equation}
G_{\mu\nu}^a(t,x) \equiv \partial_\mu B_\nu^a(t,x)-\partial_\nu B_\mu^a(t,x)+f^{abc}B_\mu^b(t,x)B_\nu^c(t,x)
\end{equation}
is the flowed field strength constructed from $B_\mu^a(t,x)$.
Because Eq.~(\ref{eq:(1.5)}) is a kind of diffusion equation, we can regard $B_{\mu}^a(t,x)$ as a smeared field of the original gauge field $A_{\mu}^a(x)$ over a physical range of $\sqrt{8t}$ in four dimensions.
It was shown that operators constructed from $B_{\mu}^a(t,x)$ (``flowed operators'') have no ultraviolet divergences nor short-distance singularities at finite and positive $t$.
Therefore, the gradient flow defines a kind of renormalization scheme, which is formulated nonperturbatively and thus can be calculated directly on the lattice.

The small flow-time expansion (SF$t$X) method provides us with a general method to evaluate any renormalized observables in terms of flowed operators at small and positive flow time $t$, which are strictly finite and thus can be calculated on the lattice without further renormalization~\cite{Suzuki:2013gza}.
In asymptotic free theories such as QCD, we can apply the perturbation theory to calculate the coefficients (``matching coefficients'') relating the renormalized observables to the flowed operators at small flow time $t$ \cite{Luscher:2011bx}. 
Contamination of $O(t)$ terms can be removed by a vanishing flow-time extrapolation $t\to0$.

\subsection{Energy-momentum tensor}

To calculate the energy-momentum tensor by the SF$t$X method, we consider the following flowed operators at flow time $t$ which are gauge-invariant and local,
\begin{eqnarray}
U_{\mu\nu}(t,x) &\equiv& G^a_{\mu\rho}(t,x)G^a_{\nu\rho}(t,x)
-\frac{1}{4}\delta_{\mu\nu}G^a_{\rho\sigma}(t,x)G^a_{\rho\sigma}(t,x),
\label{eq:UE1}
\\ 
E(t,x) &\equiv& \frac{1}{4}G^a_{\mu\nu}(t,x)G^a_{\mu\nu}(t,x) .
\label{eq:UE}
\end{eqnarray}
With these dimension-four operators, the correctly renormalized energy-momentum tensor $T_{\mu \nu}^R$ is given by~\cite{Suzuki:2013gza}
\begin{eqnarray}
T_{\mu\nu}^R(x)
=\lim_{t\to0}\left\{ c_1(t) \,U_{\mu\nu}(t,x)
   +4c_2(t)\, \delta_{\mu\nu}
   \left[E(t,x)-\left\langle E(t,x)\right\rangle_0 \right]\right\},
\label{eq:EMT2}
\end{eqnarray}
where $\left\langle E(t,x)\right\rangle_0$ is the zero-temperature subtraction.
Here, the matching coefficients $c_1(t)$ and $c_2(t)$ are calculated by the 
perturbation theory~\cite{Luscher:2011bx} as%
\footnote{Note that our convention for~$c_2(t)$ differs from that
of~Ref.~\cite{Harlander:2018zpi}. Our $c_2(t)$ corresponds
to~$c_2(t)+(1/4)c_1(t)$ in~Ref.~\cite{Harlander:2018zpi}.}
\begin{equation}
   c_1(t)=\frac{1}{g^2}\sum_{\ell=0}^\infty k_1^{(\ell)}(\mu,t)
   \left[\frac{g^2}{(4\pi)^2}\right]^\ell,
   \qquad
   c_2(t)=\frac{1}{g^2}\sum_{\ell=1}^\infty k_2^{(\ell)}(\mu,t)
   \left[\frac{g^2}{(4\pi)^2}\right]^\ell,
\label{eq:(1.6)}
\end{equation}
where $g=g(\mu)$ is the renormalized gauge coupling in the $\overline{\text{MS}}$ scheme at the renormalization scale $\mu$.
Because 
$F_{\mu\rho}^a(x)F_{\nu\rho}^a(x)=G_{\mu\rho}^a(t,x)G_{\nu\rho}^a(t,x)+O(t)$ 
in the tree-level approximation, we have $k_1^{(0)}=1$.
On the other hand, there is no ``$k_2^{(0)}$'' because the tree-level energy-momentum tensor is traceless---the trace anomaly emerges from the one-loop order.
Therefore, 
the next-to-leading order (\textbf{NLO}) expression for $c_1(t)$ contains $k_1^{(0)}$ and $k_1^{(1)}$, and that for $c_2(t)$ contains $k_2^{(1)}$ and $k_2^{(2)}$,
while the next-to-next-to-leading order (\textbf{NNLO}) expression for $c_1(t)$ contains terms up to $k_1^{(2)}$, and that for $c_2(t)$ contains terms up to $k_2^{(3)}$.
We mainly adopt NNLO matching coefficients in this study. 
To study the effect due to the truncation of perturbative series, we also perform calculation using NLO matching coefficients in Appendix~\ref{sec:leading}.

The coefficients $k_i^{(1)}$ in the one-loop level are calculated in Refs.~\cite{Suzuki:2013gza,Makino:2014taa,Suzuki:2015bqa}, 
and $k_i^{(2)}$ in the two-loop level in Refs.~\cite{Harlander:2018zpi}. (See also Ref.~\cite{AHLNP}.)
As pointed out in Ref.~\cite{Suzuki:2013gza}, in pure gauge Yang-Mills theories, $k_2^{(\ell+1)}$ can be deduced by $\ell$-loop coefficients using the trace anomaly. 
A concrete form of $k_2^{(3)}$ is given in Ref.~\cite{Iritani2019}. 
Collecting these results for the case of pure gauge Yang-Mills theories, we have
\begin{eqnarray}
   k_1^{(1)}(\mu,t) &=& -\beta_0L(\mu,t)-\frac{7}{3}C_A
   =C_A\left(-\frac{11}{3}L(\mu,t)-\frac{7}{3}\right),
\label{eq:(2.11)}
\\
   k_2^{(1)}(\mu,t) &=& \frac{1}{8}\beta_0
   =\frac{11}{24}C_A, 
\label{eq:(2.12)}
\\
   k_1^{(2)}(\mu,t) &=&
   -\beta_1L(\mu,t) +C_A^2
   \left(
   -\frac{14\,482}{405}
   -\frac{16\,546}{135}\ln2
   +\frac{1187}{10}\ln3
   \right) \nonumber \\
   &=& C_A^2\left(
   -\frac{34}{3}L(\mu,t)
   -\frac{14\,482}{405}
   -\frac{16\,546}{135}\ln2
   +\frac{1187}{10}\ln3
   \right),
\label{eq:(2.13)}
\\
   k_2^{(2)}(\mu,t)
   &=&\frac{1}{8}\beta_1-\frac{7}{16}\beta_0C_A
   =C_A^2
   \left(
   -\frac{3}{16}
   \right),
\label{eq:(2.14)}
\\
   k_2^{(3)}(\mu,t) &=&
   \frac{1}{8}\beta_2
   -\frac{7}{16}\beta_1C_A
   +\beta_0C_A^2
   \left(
   -\frac{3}{16}L(\mu,t)
   -\frac{1427}{1440}+\frac{87}{40}\ln2-\frac{27}{20}\ln3
   \right) \nonumber \\
   &=& C_A^3\left(
   -\frac{11}{16}L(\mu,t)
   -\frac{2849}{1440}
   +\frac{319}{40}\ln2
   -\frac{99}{20}\ln3
   \right),
\label{eq:(2.18)}
\end{eqnarray}
where
\begin{equation}
\beta_0 =\frac{11}{3}C_A, \ \ 
\beta_1 =\frac{34}{3}C_A^2, \ \ {\rm and} \ \ 
\beta_2 =\frac{2857}{54}C_A^3
\end{equation}
are the first three coefficients of the beta function in Yang-Mills theories, and 
\begin{equation}
   L(\mu,t)\equiv\ln(2\mu^2t)+\gamma_{\text{E}}
\end{equation} 
was introduced in Ref.~\cite{Harlander:2018zpi}
with $\gamma_{\text{E}}$ the Euler-Mascheroni constant. 
The factor $C_A$ is the quadratic Casimir for the adjoint representation defined by
\begin{equation}
   f^{acd}f^{bcd}=C_A\delta^{ab}, 
\label{eq:(2.6)}
\end{equation}
and $C_A=N_c$ for the gauge group SU($N_c$).

\subsection{Latent heat and pressure gap}

The energy density and the pressure are obtained from the diagonal elements of the energy-momentum tensor, 
\begin{eqnarray}
\epsilon = -\left\langle T_{00}^R(x)\right\rangle, 
\hspace{5mm}
p = \frac{1}{3} \sum_{i=1,2,3}\langle T_{ii}^R(x)\rangle .
\end{eqnarray}
Separating configurations in the Monte Carlo time history into the hot and cold phases (see Sec.~\ref{sec:separation}), 
and adopting the multipoint reweighting method~\cite{FS89,iwami15} to fine-tune the coupling parameter to the critical point, 
we calculate the energy density and the pressure in each phase just at the transition temperature, 
to estimate the latent heat and the pressure gap defined by
\begin{equation}
\Delta \epsilon = \epsilon^{\rm (hot)} - \epsilon^{\rm (cold)}
, \qquad
\Delta p = p^{\rm (hot)} - p^{\rm (cold)},
\end{equation}
where
$\epsilon^{\rm (hot/cold)}$ and $p^{\rm (hot/cold)}$ indicate $\epsilon$ and $p$ in the hot and cold phases, respectively.

In this paper, we calculate these quantities in the following conventional combinations of the trace anomaly and the entropy density:
\begin{eqnarray}
\frac{ \Delta ( \epsilon -3p )}{T^4} 
\ \ {\rm and} \ \ 
\frac{ \Delta ( \epsilon +p )}{T^4}.
\label{eq:Delta}
\end{eqnarray}
When $\Delta p=0$ as expected, 
$\Delta (\epsilon -3p)/T^4$ and $\Delta (\epsilon +p)/T^4$ should coincide with each other. 
We note that $\Delta (\epsilon -3p)/T^4$ is computed by the operator $E(t,x)$ 
with the matching coefficient $c_2(t)$, while 
$\Delta (\epsilon +p)/T^4$ is computed by the operator $U_{\mu \nu}(t,x)$ with $c_1(t)$.

\subsection{Renormalization scale}

The matching coefficients $c_i(t)$ given in the previous subsection are written in terms of the renormalization scale~$\mu$.
However, because the matching coefficients relate the energy-momentum tensor and flowed operators, both physical and thus both finite and independent of the renormalization scale~$\mu$, the matching coefficients themselves should also be finite and independent of $\mu$---the explicit $\mu$ dependence from $L(\mu,t)$ should be canceled by that of the running coupling $g$, in principle.
In practice, however, because the perturbative series for the matching coefficients are truncated at a finite order, the choice of $\mu$ may affect the magnitude of systematic errors due to the neglected higher order terms. 

In early studies with the SF$t$X method~\cite{Asakawa:2013laa,WHOT2017b,Kitazawa:2016dsl,Kitazawa:2017qab}, the $\mu_d$ scale 
\begin{equation}
\mu= \mu_d(t) \equiv \frac{1}{\sqrt{8t}}, \hspace{5mm}
L(\mu,t)=-2\ln2+\gamma_{\text{E}}
\end{equation}
has been adopted because $\mu_d$ is a natural scale of local flowed operators that have a physical smearing extent of $\sqrt{8t}$.
On the other hand, in~Ref.~\cite{Harlander:2018zpi}, it is argued that
\begin{equation}
  \mu= \mu_0(t) \equiv \frac{1}{\sqrt{2t \, e^{\gamma_{\text{E}}}}}  \simeq 1.499\,\mu_d(t), 
\hspace{5mm} L(\mu,t)=0,
\end{equation}
would be a good choice of $\mu$ because it keeps the magnitude of two-loop contributions small.
The $\mu_0$ scale was tested in quenched QCD~\cite{Iritani2019} and in $2+1$-flavor QCD~\cite{Taniguchi:2020mgg,Lat2019-kanaya}, and it was reported that the $\mu_0$ scale may improve the signal when the lattice is not quite fine~\cite{Taniguchi:2020mgg,Lat2019-kanaya}.

In this paper, we examine both choices, $\mu=\mu_0(t)$ and $\mu_d(t)$. 
From the difference in the results between these two choices, we learn the magnitude of the effect due to truncation of higher-order terms in the perturbation theory.

\subsection{Continuum and vanishing flow-time extrapolations}
\label{sec:method12}

To obtain a physical result with the SF$t$X method, the continuum extrapolation $a\to0$ and the vanishing flow-time extrapolation $t\to0$ are both mandatory to remove contamination of unwanted higher-dimension operators in Eq.~(\ref{eq:EMT2}) as well as errors from the lattice regularization.
In Refs.~\cite{Kitazawa:2016dsl,Hirakida:2018uoy,Iritani2019,Kitazawa:2017qab,Yanagihara:2018qqg,Yanagihara:2020tvs}, 
this double extrapolation is carried out by first taking $a \to 0$ at each flow time $t$, and then take $t \to 0$ using the continuum-extrapolated results.
In Refs.~\cite{WHOT2017b,WHOT2017,Taniguchi:2020mgg,Kitazawa:2019otp}, $t\to0$ is taken first at fixed finite $a$, reserving the continuum extrapolation for a later stage. 
In the following, we call the way to first take $t\to0$ at fixed finite $a$ and then take $a\to0$ ``\textbf{method~1},'' and the way to first take $a\to0$ at fixed finite $t$ and then take $t\to0$ ``\textbf{method~2}.''

On finite lattices, additional unwanted operators may contaminate the right-hand side of Eq.~(\ref{eq:EMT2}) due to the lattice artifacts. 
To the leading order of the lattice spacing $a$, we expect that the lattice operator in the curly bracket of Eq.~(\ref{eq:EMT2}) is expanded as
\begin{eqnarray}
   T_{\mu\nu}(t,x,a)
   &=&T_{\mu\nu}^R(x) + t S_{\mu\nu}(x) 
   +A_{\mu\nu}(x)\frac{a^2}{t}+C_{\mu\nu}(x)(aT)^2 \nonumber\\  &&
   +D_{\mu\nu}(x)(a\Lambda_{\mathrm{QCD}})^2
   +a^2S'_{\mu\nu}(x)+O(a^4,t^2),
\label{eq:a2overt}
\end{eqnarray}
where $T_{\mu\nu}^R$ is the physical energy-momentum tensor, $S_{\mu\nu}$ and $S'_{\mu\nu}$ are contaminations of dimension-six operators with the same quantum number, and $A_{\mu\nu}$, $C_{\mu\nu}$, and $D_{\mu\nu}$ are those from dimension-four operators. 
In higher orders of $a$, we also have terms proportional to $(a^2/t)^2$ etc.~\cite{Fodor14}.
In Eq.~(\ref{eq:a2overt}), the term proportional to $a^2/t$ is singular at $t=0$ and thus is problematic when we want to take $t\to0$ at finite $a$ in method~1.
When we take the $a\to0$ limit first, the term proportional to $a^2/t$ is removed in principle. 
However, in practice, the $a^2/t$ term is also problematic in method~2 when we want to take $a\to0$ at small $t$ where the $a^2/t$ term dominates the data---we cannot perform the $a\to0$ extrapolation reliably at small $t$. 
Therefore, in both methods, we need to find a range of $t$ at each $a$ where the $a^2/t$ and more singular terms are negligible.
Let us call the range of $t$ in which the $a^2/t$ and more singular terms as well as the $O(t^2)$ terms look negligible the ``linear window''~\cite{WHOT2017b,Taniguchi:2020mgg}.

On the other hand, when we can find such a range of $t$ and perform $t\to0$ and $a\to0$ extrapolations using data in this range only, because the flowed operator on the lattice will be well approximated as $T_{\mu\nu}(t,a) \approx T_{\mu\nu}^R + tS_{\mu\nu} + a^2 C' _{\mu\nu}+ O(a^4, t^2)$ in this range, the final results of method~1 and method~2 for $T_{\mu\nu}^R$ should be identical.
The consistency of both methods is thus a good test of the SF$t$X method.
In the following, we adopt both method~1 and method~2 using data in linear windows and perform extrapolations linear in $t$ and $a^2$. 
When both methods are consistent with each other, we get strong support in our identification of the linear windows.

Another issue in the study of thermodynamic properties is the possible dependence on the physical volume of the system.
We have to keep the physical volume fixed in the $a\to0$ and $t\to0$ extrapolations to identify finite-volume effects.
In our study, $T$ is adjusted to the transition temperature $T_c$ whose value is a physical constant.
On a lattice with the size $N_s^3 \times N_t$, the lattice spacing defined as $a = 1/(N_t T_c)$ is thus changed by changing $N_t$, 
and the spatial volume in physical units, $V = (N_s a)^3 = N_s^3/(N_t T_c)^3$, is fixed when the aspect ratio $N_s/N_t$ is fixed.
In this paper, we take the double extrapolation for the cases of $N_s/N_t=6$ and 8 to study the finite-volume effect.

\begin{figure}[t]
\centering
\includegraphics[width=7.5cm]{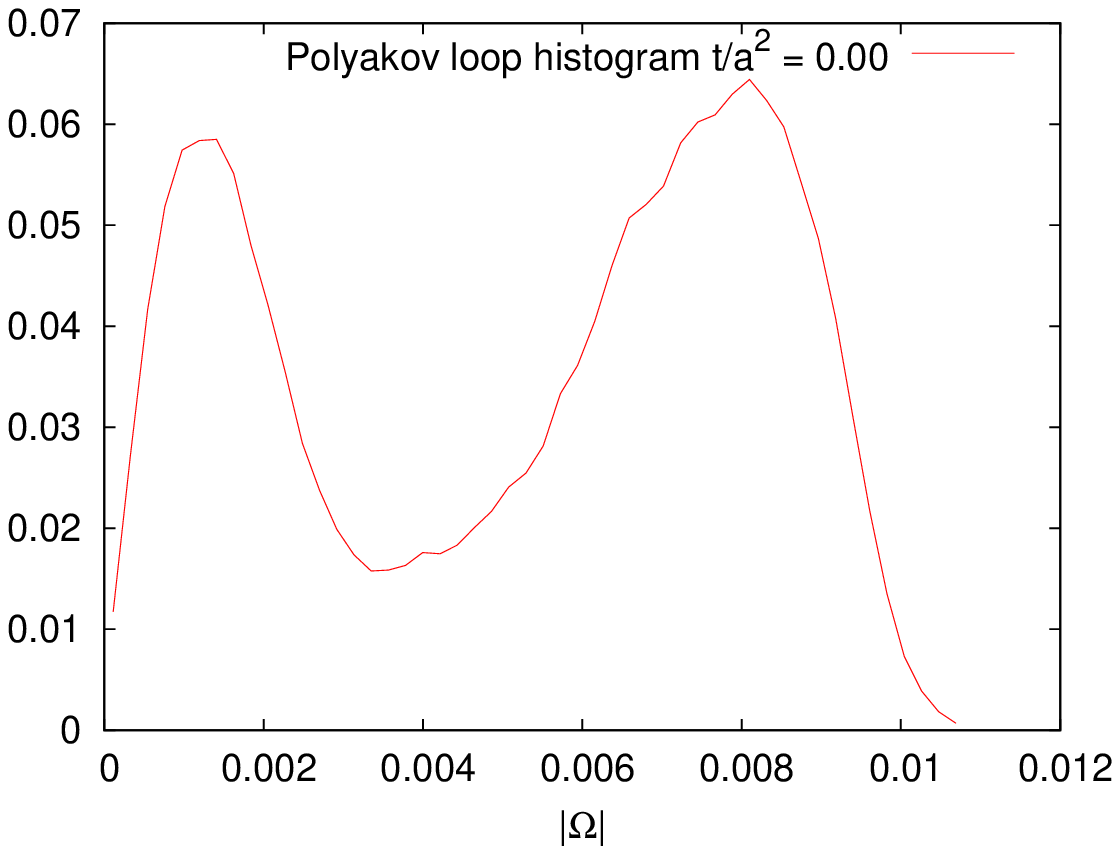}
\includegraphics[width=7.5cm]{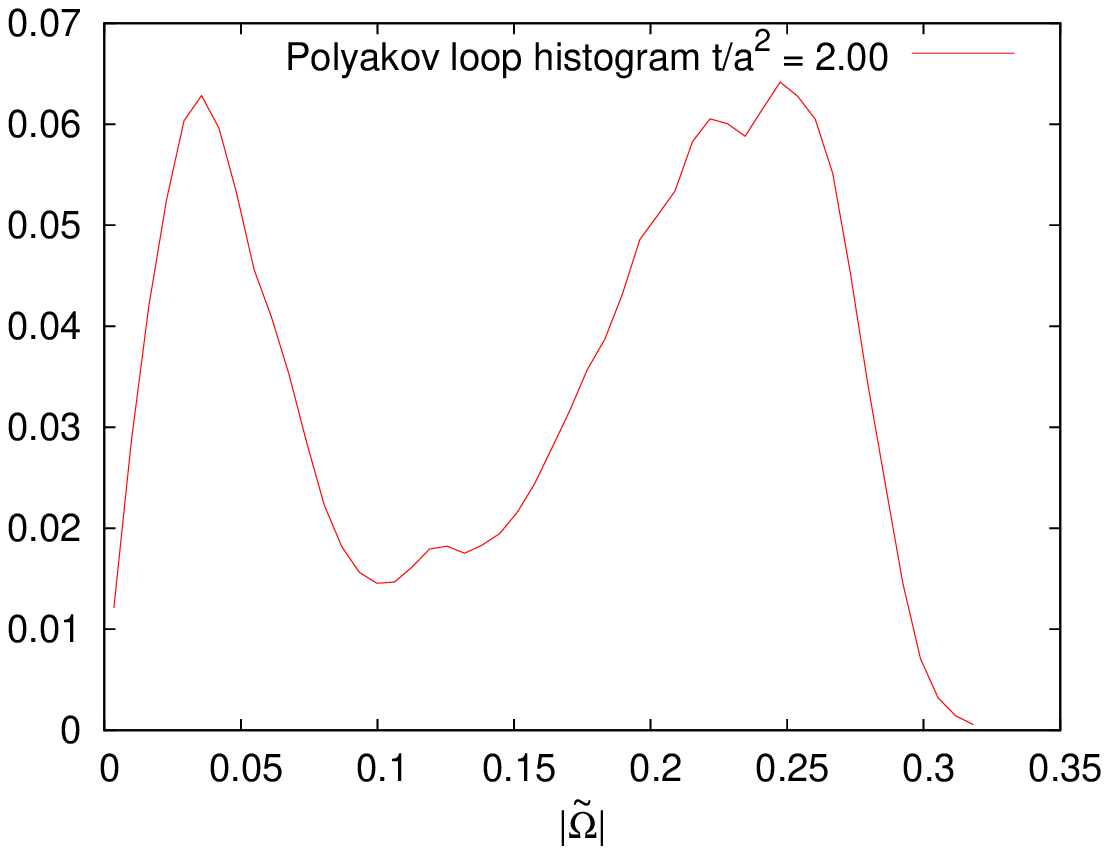}
\caption{
Histograms of the original Polyakov loop (left) and the flowed Polyakov loop after gradient flow to $t/a^2 =2.0$ (right) at the transition point $T=T_c$ on the $96^3 \times 12$ lattice.
}
\label{fig:plhist9612}
\end{figure}

\begin{figure}[t]
\centering
\includegraphics[width=7.5cm]{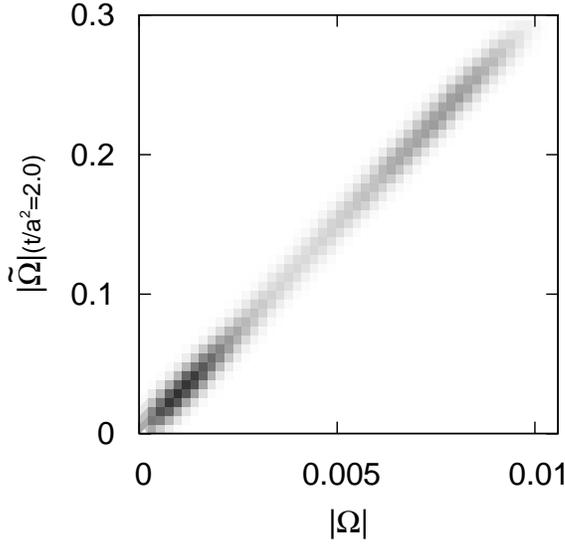}
\caption{
Double distribution of the original Polyakov loop $|\Omega|$ and the flowed Polyakov loop $|\tilde\Omega|_{t/a^2=2.0}$ measured on the $96^3 \times 12$ lattice. The probability observed at each simulation point was translated to that at the transition point by the multipoint reweighting method.
}
\label{fig:polcor9612}
\end{figure}

\begin{table}[thb]
  \small
\centering
\caption{Simulation parameters: 
the spatial lattice size $N_s$, the temporal lattice size $N_t$,
coupling parameter $\beta=6/g_0^2$ and the critical coupling $\beta_c$.
The measurements have been performed $N_{\mathrm{conf}}$ times at 
intervals of $N_{\rm sep}$ sweeps.
The data for $48^3\times8$, $64^3\times8$, and $64^3\times12$ are taken from our previous study~\cite{shirogane16}.
}
\label{tab1}
\begin{tabular}{cccccr}
\hline
$N_s$ & $N_t$ & $\beta$ & $\beta_c$& $N_{\rm sep}$ & $N_{\rm conf}$ \\
\hline
48  &  8  &  6.056 &  6.06160(18)  &   20  &    10000  \\
    &     &  6.058 &               &       &    10000  \\
    &     &  6.060 &               &       &    10000  \\
    &     &  6.062 &               &       &    10000  \\
    &     &  6.065 &               &       &    11000  \\
    &     &  6.067 &               &       &    10000  \\
\hline
64  &  8  &  6.0585&  6.06247(14)  &   20  &    4750   \\
    &     &  6.061 &               &       &    103000 \\
    &     &  6.063 &               &       &    12500  \\
    &     &  6.065 &               &       &    25500  \\
    &     &  6.068 &               &       &    81000  \\
\hline
48  &  12 &  6.333 &  6.33472(13)  &   50  &    52500  \\
    &     &  6.335 &               &       &    40000  \\
    &     &  6.337 &               &       &    52500  \\
\hline
64  &  12 &  6.332 &  6.33493(17)  &   50  &    6000  \\
    &     &  6.3335&               &       &    17500 \\
    &     &  6.335 &               &       &    5800  \\
    &     &  6.3375&               &       &    9600  \\
    &     &  6.339 &               &       &    8800  \\
\hline
72  &  12 &  6.334 &  6.33531(10)  &   50  &    8000  \\
    &     &  6.335 &               &       &    8000  \\
    &     &  6.337 &               &       &    8000  \\
\hline
96  &  12 &  6.334 &  6.33532(11)  &   50  &    20550 \\
    &     &  6.335 &               &       &    12000 \\
    &     &  6.336 &               &       &    14400 \\
\hline
96  &  16 &  6.543 &  6.54667(20)  &   50  &    4350  \\
    &     &  6.545 &               &       &    4800  \\
    &     &  6.547 &               &       &    4250  \\
\hline
128 &  16 &  6.544 &  6.54616(23)  &   200 &    2750  \\
    &     &  6.545 &               &       &    3000  \\
    &     &  6.546 &               &       &    3000  \\
    &     &  6.547 &               &       &    3000  \\
    &     &  6.548 &               &       &    3000  \\
\hline
\end{tabular}
\end{table}

\section{Numerical simulations}
\label{simulation}

\subsection{Simulation parameters}
\label{parameter}

We perform simulations of the SU(3) Yang-Mills theory with the standard Wilson action at several inverse gauge couplings $\beta=6/g^2$
around the deconfining transition point $\beta_c$. 
We have studied lattices with temporal lattice sizes of $N_t=8$, $12$ and $16$ with several different spatial lattice sizes $N_s$.
Some of the configurations are taken from~Ref.~\cite{shirogane16}.
Although we have studied $N_t=6$ lattices too, we do not use them in this study because the linear window turned out not to be clear enough for $N_t=6$.
Our simulation parameters are summarized in~Table~\ref{tab1}.
The configurations are generated by a pseudo-heat-bath algorithm followed by over-relaxation updates.
Each measurement is separated by~$N_{\rm sep}$ heat-bath sweeps.
Data are taken at 3 to 6 $\beta$ values for each~$(N_s, N_t)$,
and are combined using the multipoint reweighting method~\cite{FS89,iwami15}.
The statistical errors are estimated by the jackknife method with the bin size chosen such that the errors are saturated.

We define the transition point as the peak position of the Polyakov loop susceptibility 
$\chi_{\Omega}=N_s^3 (\langle \Omega^2 \rangle -\langle \Omega \rangle ^2)$.
The results of the critical point~$\beta_c$ determined in this way are also listed in~Table~\ref{tab1}.

We employ the discretized representation of the 
flow equation~(\ref{eq:(1.5)}) defined from the Wilson gauge action~\cite{Luscher:2010iy}.
The numerical solution of the flow equation is obtained by the third-order Runge-Kutta method.
For the field strength~$G_{\mu\nu}(t,x)$ in the observables, several candidates are available for the lattice operator.
From a test of the plaquette and the clover-type representations for~$G_{\mu\nu}(t,x)$ given in~Appendix~\ref{plaquette}, we adopt the clover-type representation.

\subsection{Phase separation around the first-order transition point}
\label{sec:separation}

To evaluate the latent heat and the pressure gap, we need to separate the configurations around the first-order transition point into the hot and cold phases. 
In the left panel of~Fig.~\ref{fig:plhist9612}, we show the histogram 
of the absolute value of the Polyakov loop~$|\Omega|$ obtained on the~$96^3 \times 12$ lattice, where~$\Omega$ is averaged over the spatial coordinates. 
The coupling parameter $\beta$ is adjusted to the transition point~$\beta_c$ using the multipoint reweighting method. 
The two peaks in~Fig.~\ref{fig:plhist9612} correspond to the hot and cold phases. 
As discussed in~Ref.~\cite{shirogane16}, though the two peaks of the Polyakov loop histogram are well separated, the peaks of the plaquette histogram are overlapping. 
It is thus meaningful to use the Polyakov loop to classify configurations into the hot and cold phases. 

In the right panel of~Fig.~\ref{fig:plhist9612}, we show the histogram of the flowed Polyakov loop $|\tilde\Omega|_{t/a^2=2.0}$, 
which is constructed with the flowed gauge field at~$t/a^2 =2.0$, as a typical example.
We find that, though the scale of the horizontal axis is different, its shape is quite similar to that shown in the left panel. 
The same similarity is also observed at other values of $t/a^2$. 
This suggests a strong correlation between the original Polyakov loop~$\Omega$ and the flowed Polyakov loop $\tilde\Omega$. 
In~Fig.~\ref{fig:polcor9612}, we show the 2D distribution of~$(|\Omega|,|\tilde\Omega|_{t/a^2=2.0})$ 
measured on the~$96^3 \times 12$ lattice.
A darker tone means a higher probability. 
We find that $\Omega$ and~$\tilde\Omega$ have a strong linear correlation with each other.
These results show that the gradient flow does not affect the separation, and thus there is no merit in using the flowed Polyakov loop for the phase separation.

We thus follow the method of~Ref.~\cite{shirogane16} to classify configurations into the hot and cold phases: 
First, we remove short-time-range fluctuations by averaging $|\Omega|$ over several configurations (101 configurations, except for the~$128^3\times16$ lattice, for which we use 45 configurations) around the current configuration.  
We then classify the configurations into hot, cold, and mixed phases by the value of this time-smeared Polyakov loop.
With this classification, we observe many flip-flops between the hot and cold phases during our Monte Carlo steps, while mixed configurations in which the two phases coexist are found to be rare on our lattices.
We discard the configurations in the mixed phase in the following. 

After the phase separation, we carry out the gradient flow on each of the configurations to measure flowed operators given in~Eqs.~(\ref{eq:UE1}) and~(\ref{eq:UE}). 
We then combine their expectation values in each phase by the multipoint reweighting method to obtain their values just at the transition point~$\beta_c$.


\begin{figure}[t]
\centering
\includegraphics[width=7.5cm]{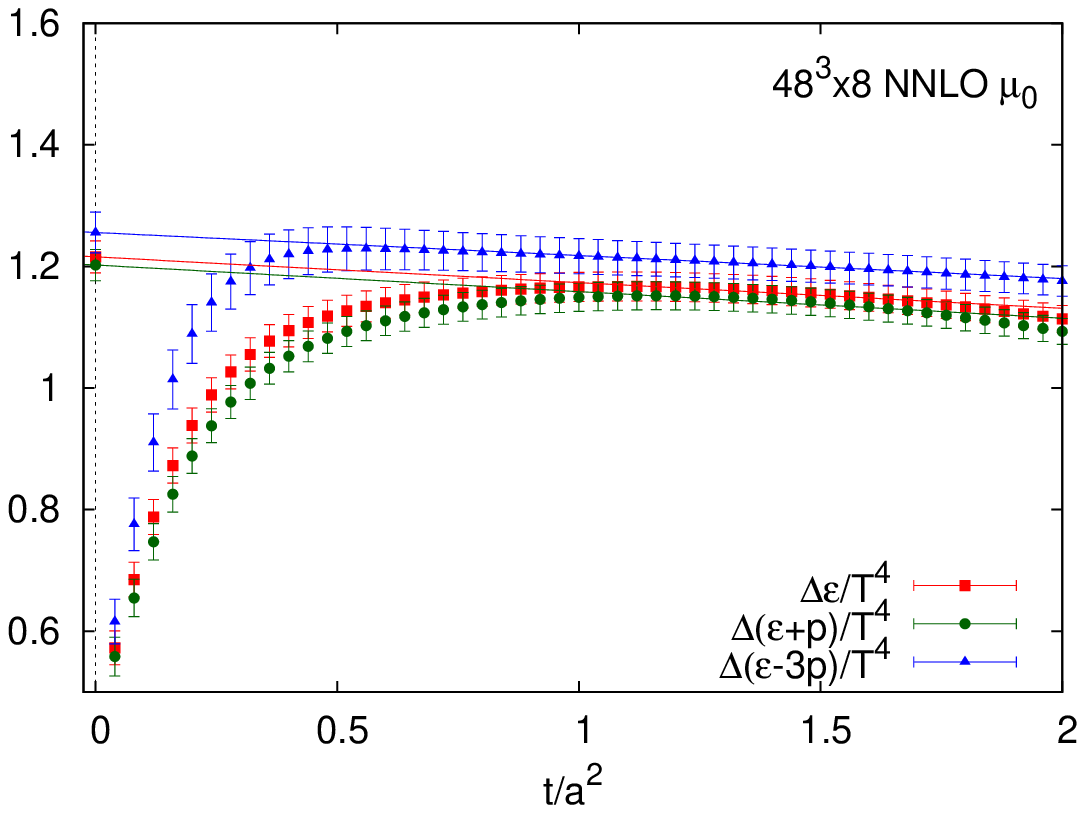}
\hspace{-3mm}
\includegraphics[width=7.5cm]{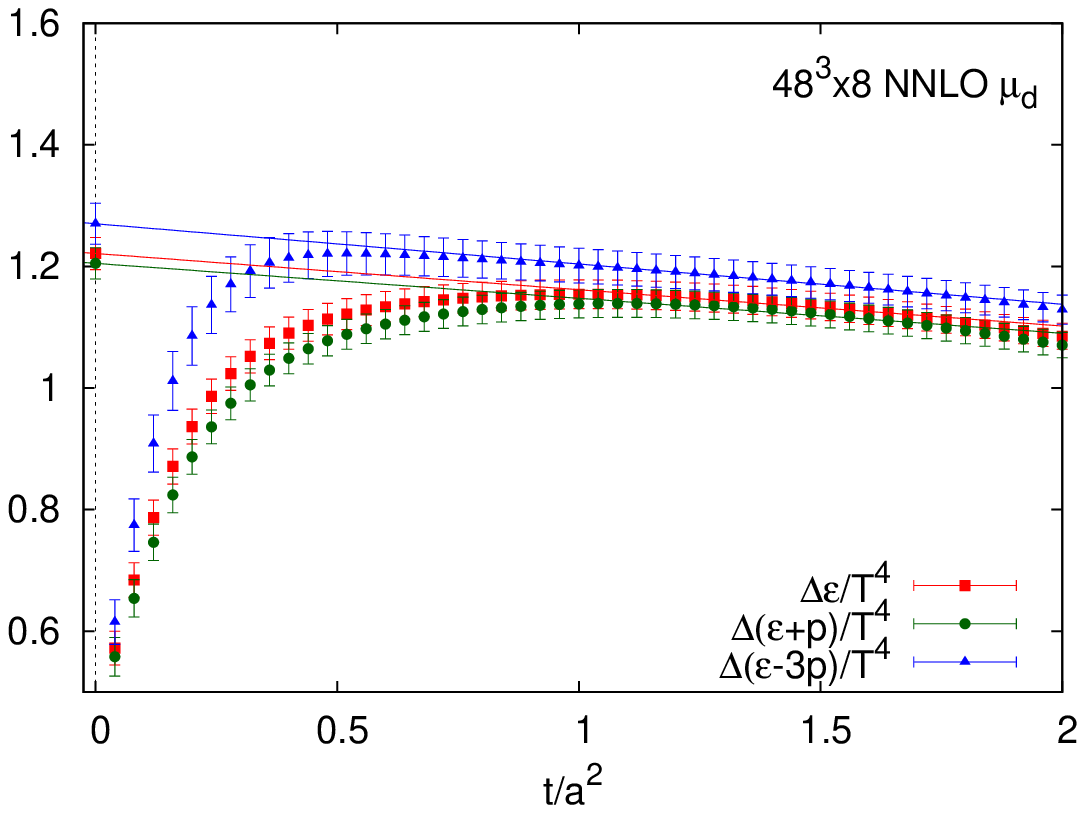}
\\
\includegraphics[width=7.5cm]{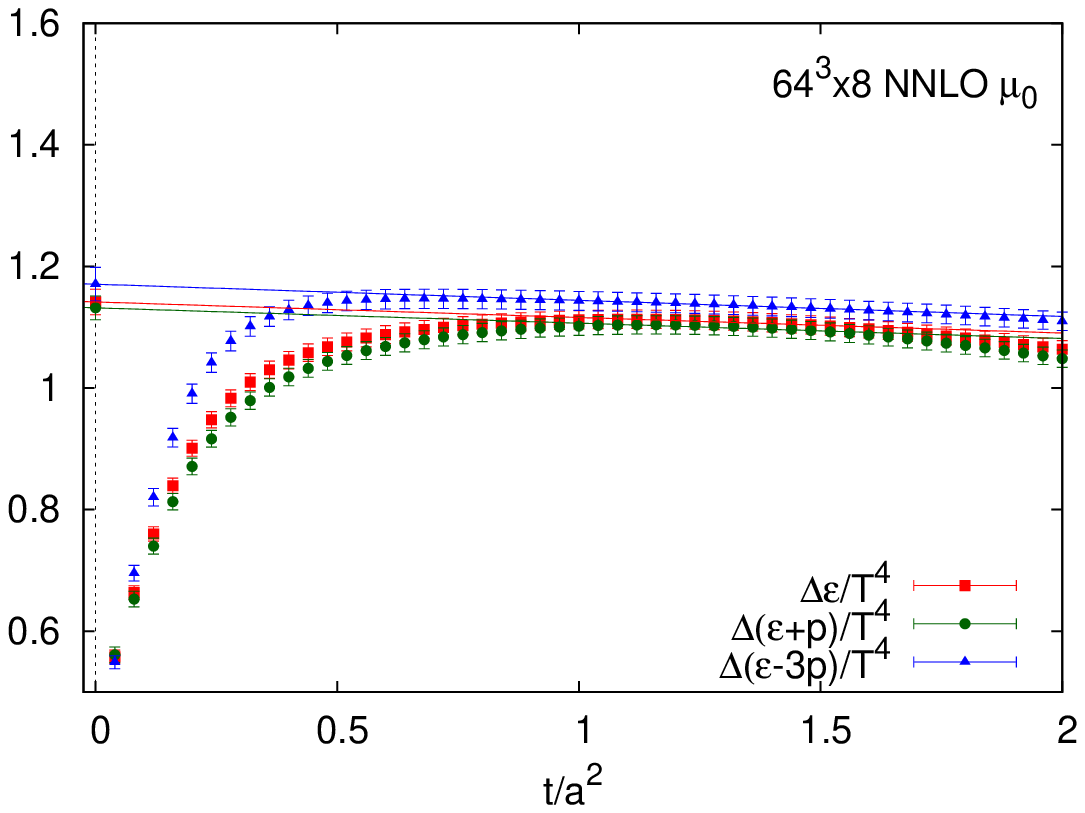}
\hspace{-3mm} 
\includegraphics[width=7.5cm]{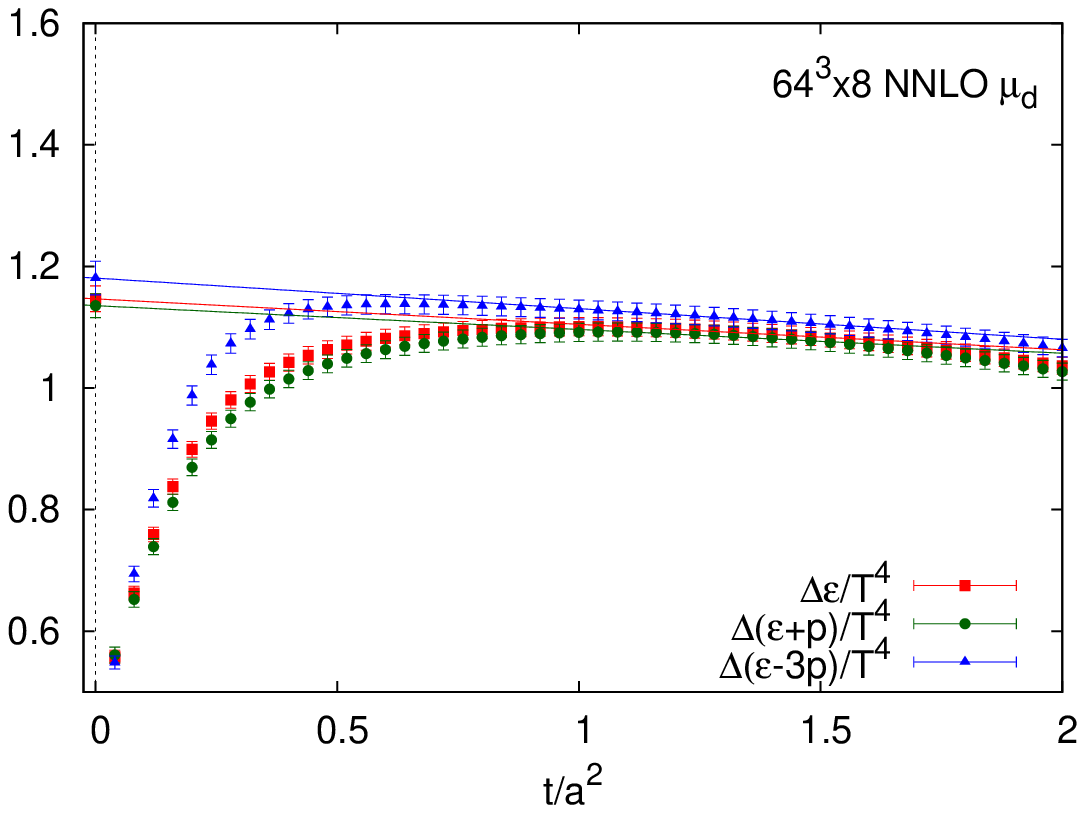}
\caption{
  $\Delta (\epsilon -3p)/T^4$ (blue), $\Delta (\epsilon +p)/T^4$ (green), and $\Delta \epsilon /T^4$ (red) calculated on the~$48^3 \times 8$ (top) and $64^3 \times 8$ (bottom) lattices.
  The left and right panels show the results obtained with $\mu=\mu_0$ and $\mu=\mu_d$, respectively. 
The horizontal axis is the flow time in lattice units~$t/a^2$.
The symbols at $t/a^2=0$ and the straight lines are the results of $t \to 0$ linear extrapolation.
}
\label{fig:flow8}
\end{figure}

\begin{figure}[t]
\centering
\includegraphics[width=7.5cm]{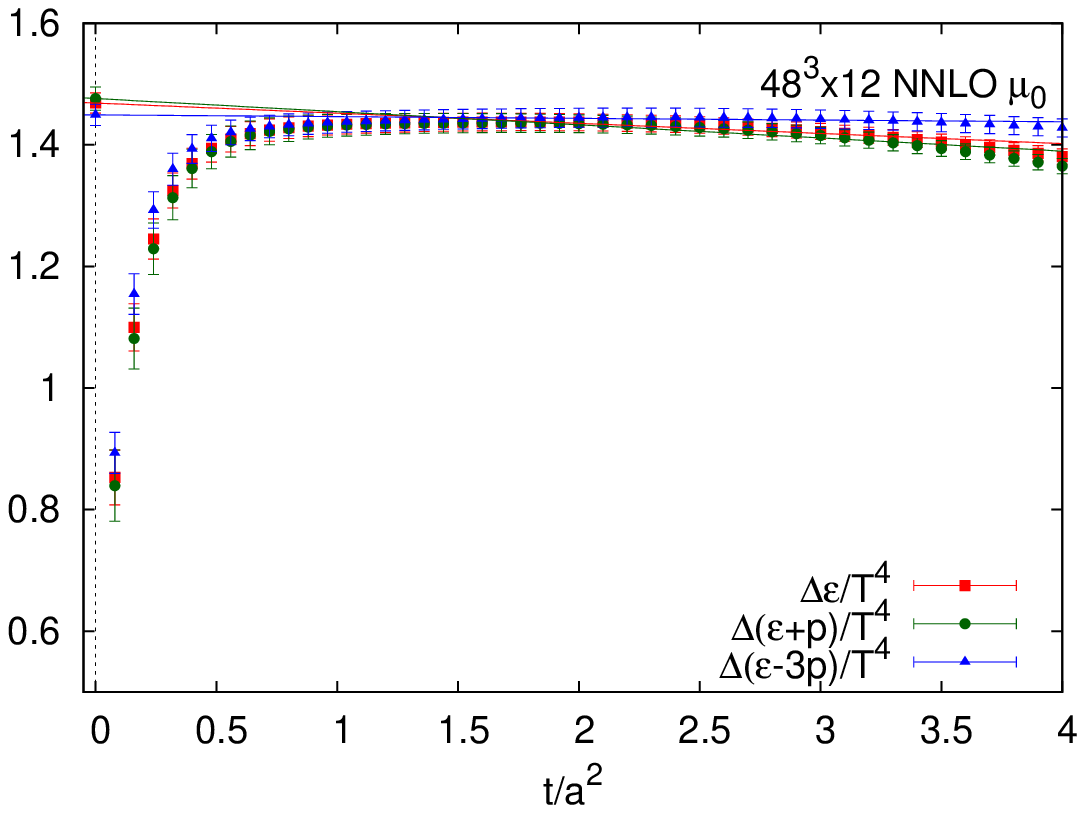}
\hspace{-3mm}
\includegraphics[width=7.5cm]{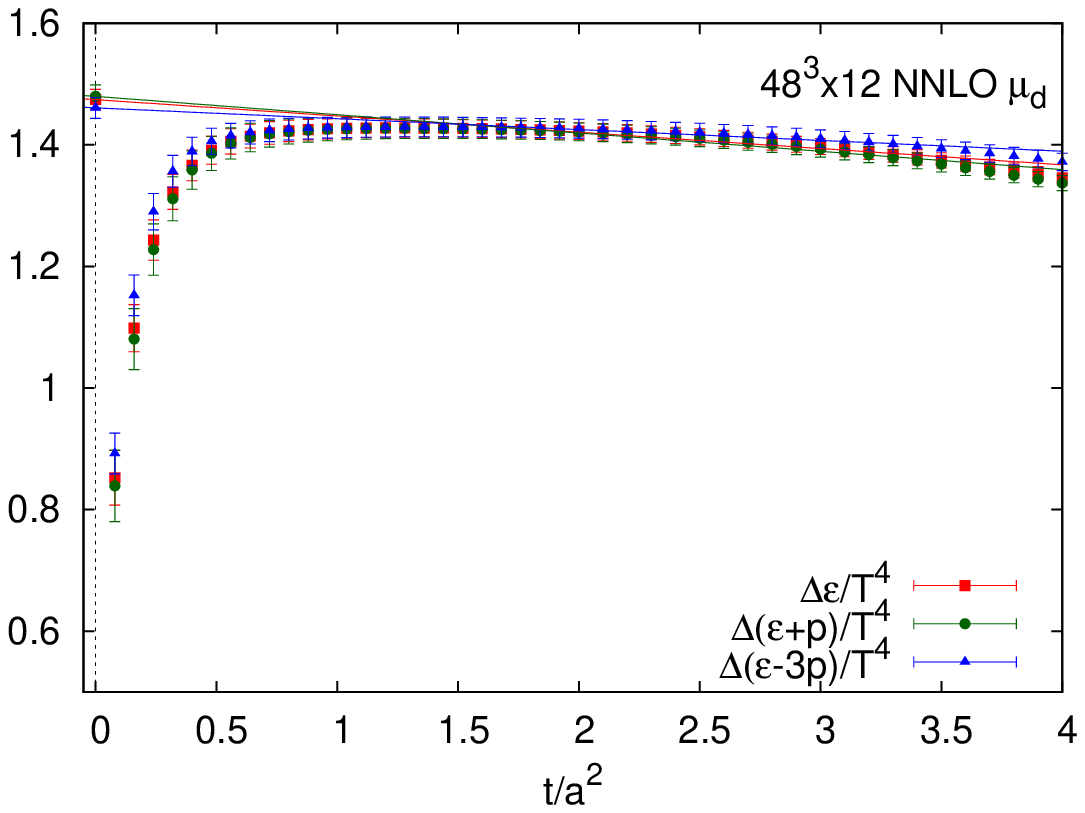}
\\
\includegraphics[width=7.5cm]{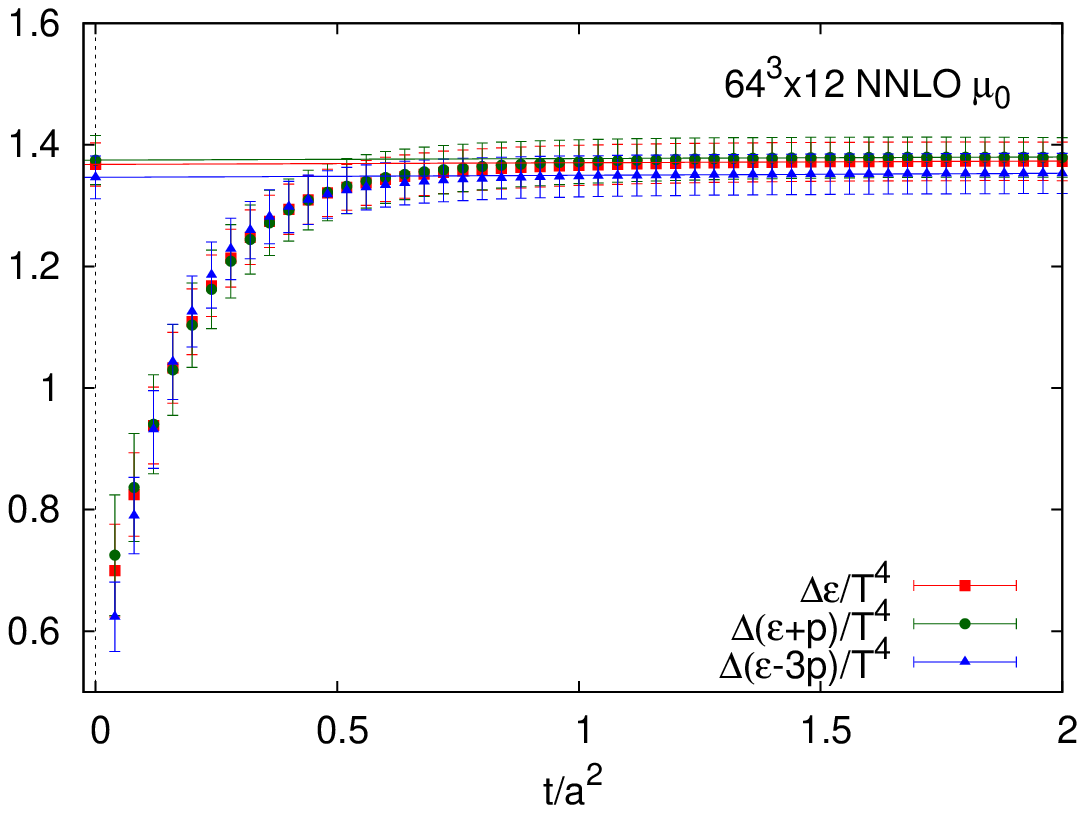}
\hspace{-3mm}
\includegraphics[width=7.5cm]{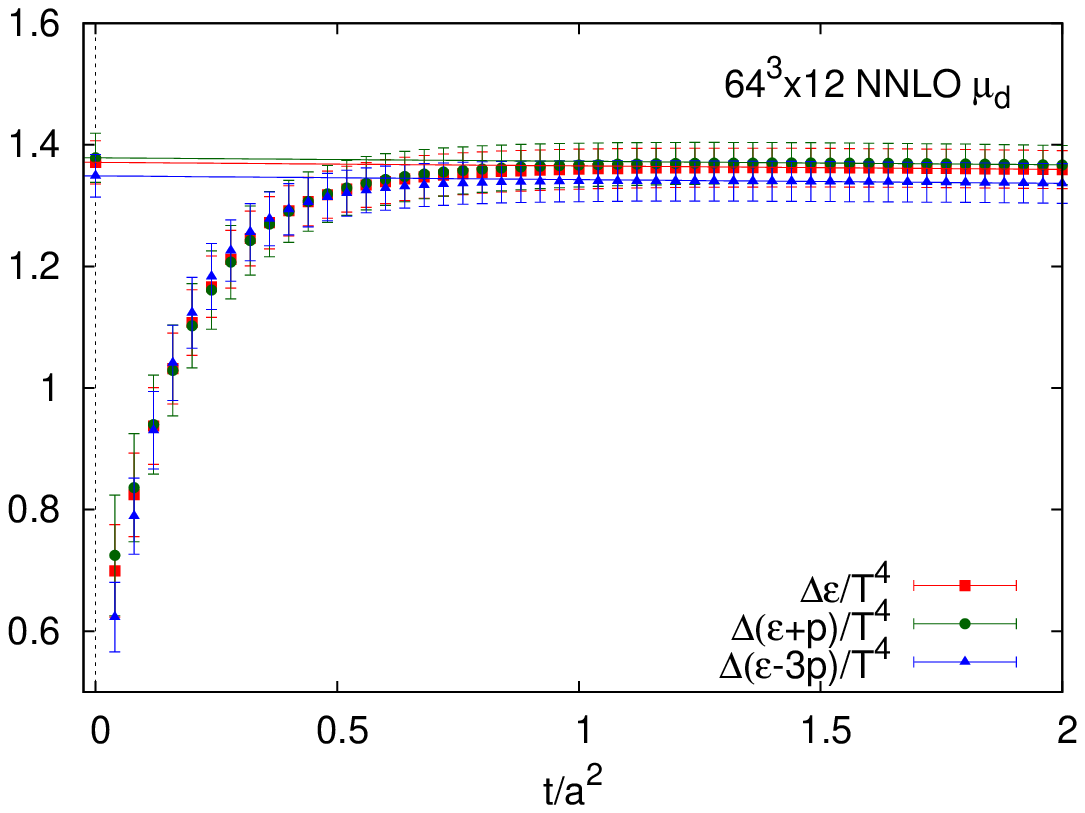}
\\
\includegraphics[width=7.5cm]{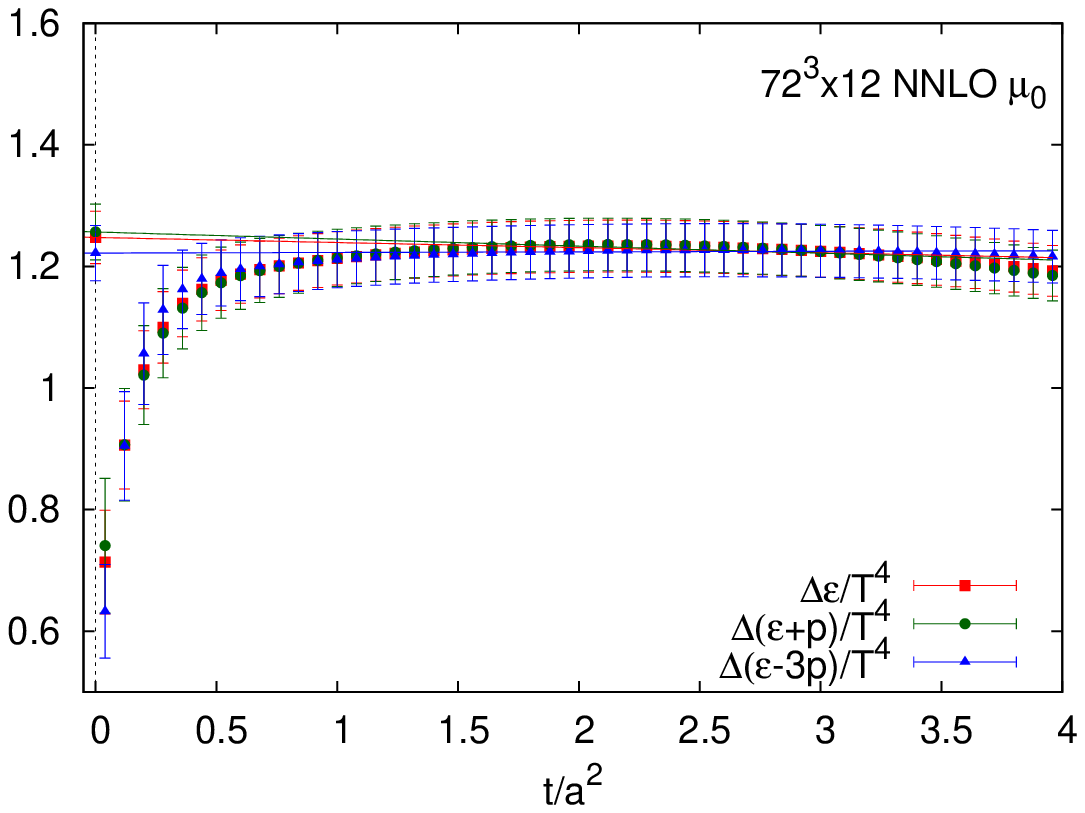}
\hspace{-3mm}
\includegraphics[width=7.5cm]{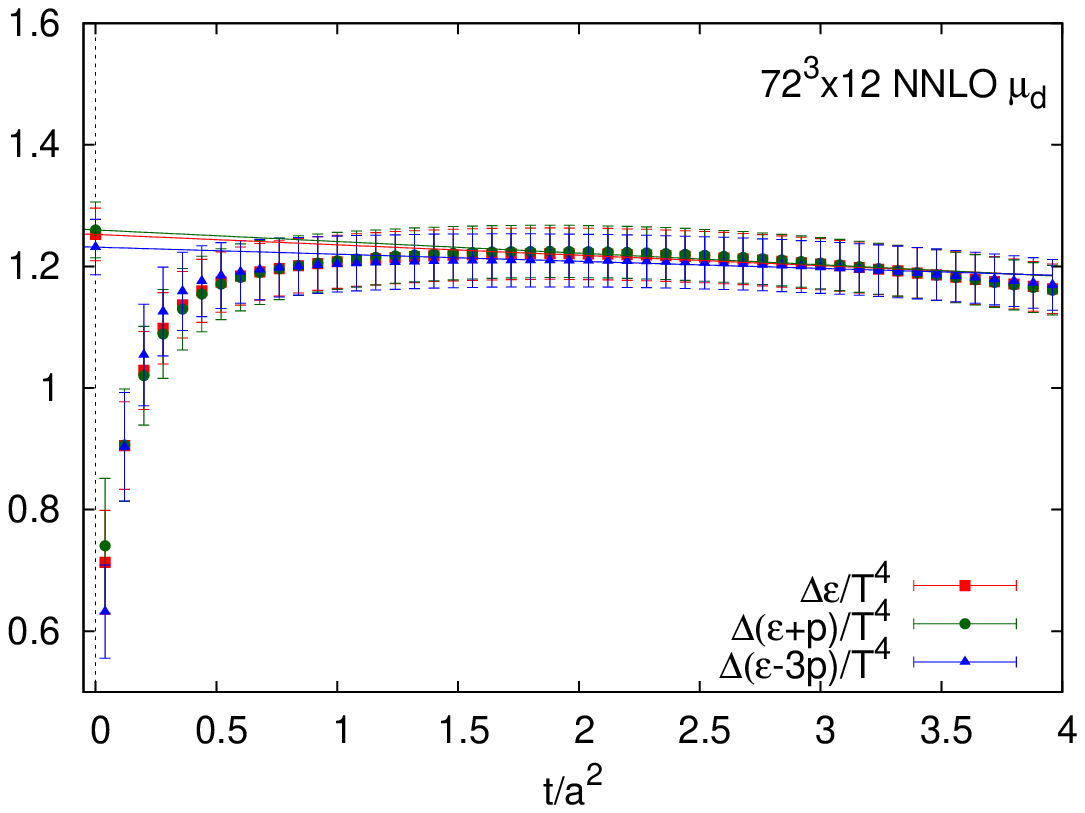}
\\
\includegraphics[width=7.5cm]{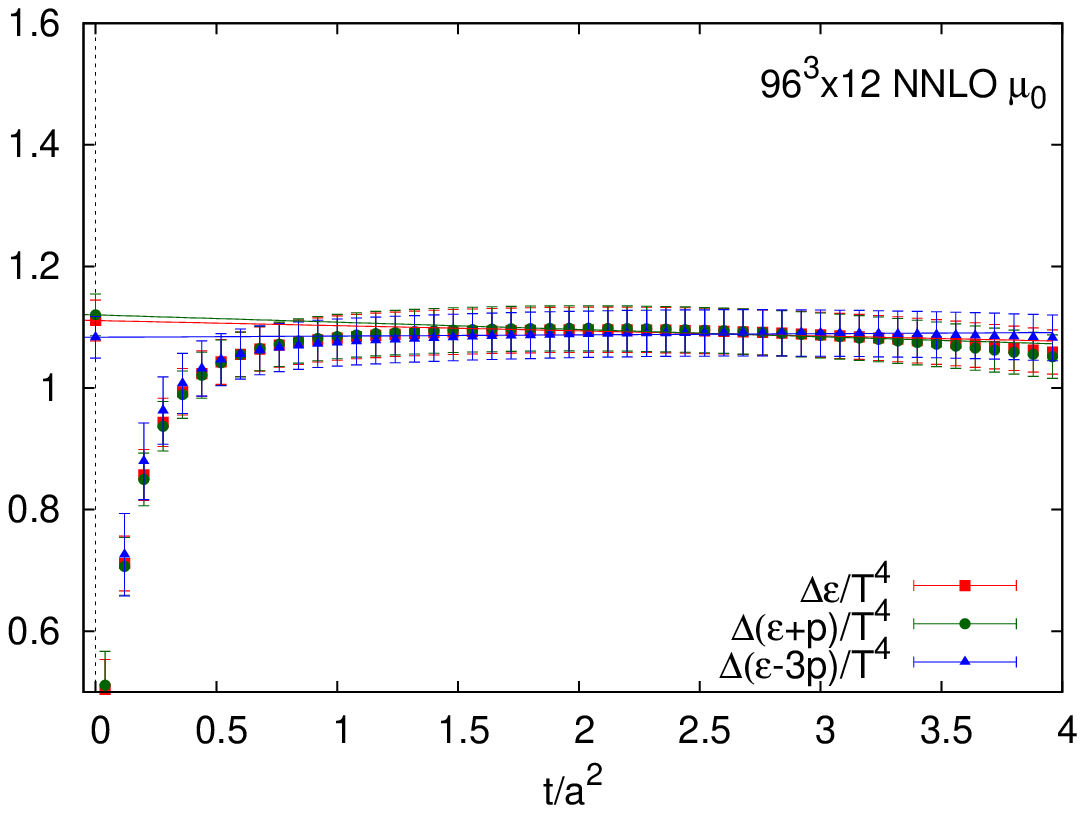}
\hspace{-3mm}
\includegraphics[width=7.5cm]{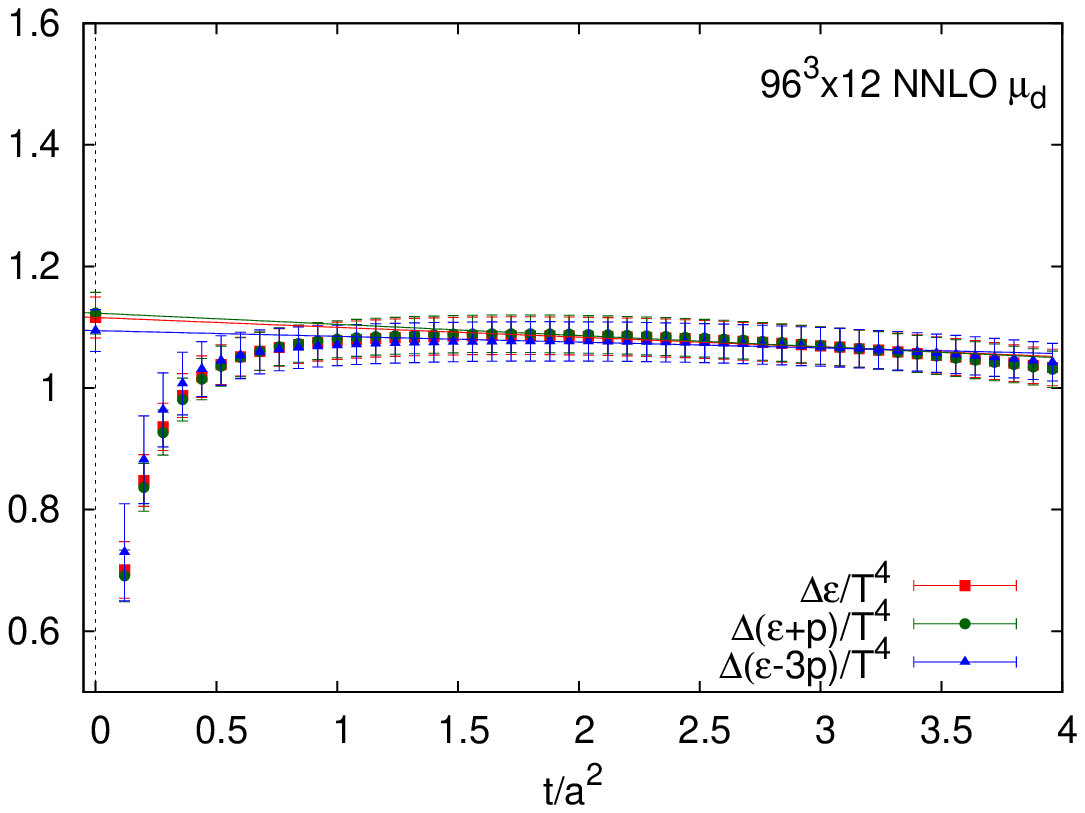}
\caption{
The same as Fig.~\ref{fig:flow8}, but on the $48^3 \times 12$, $64^3 \times 12$, $72^3 \times 12$, $96^3 \times 12$ lattices (from top to bottom). 
}
\label{fig:flow12}
\end{figure}

\begin{figure}[t]
\centering
\includegraphics[width=7.5cm]{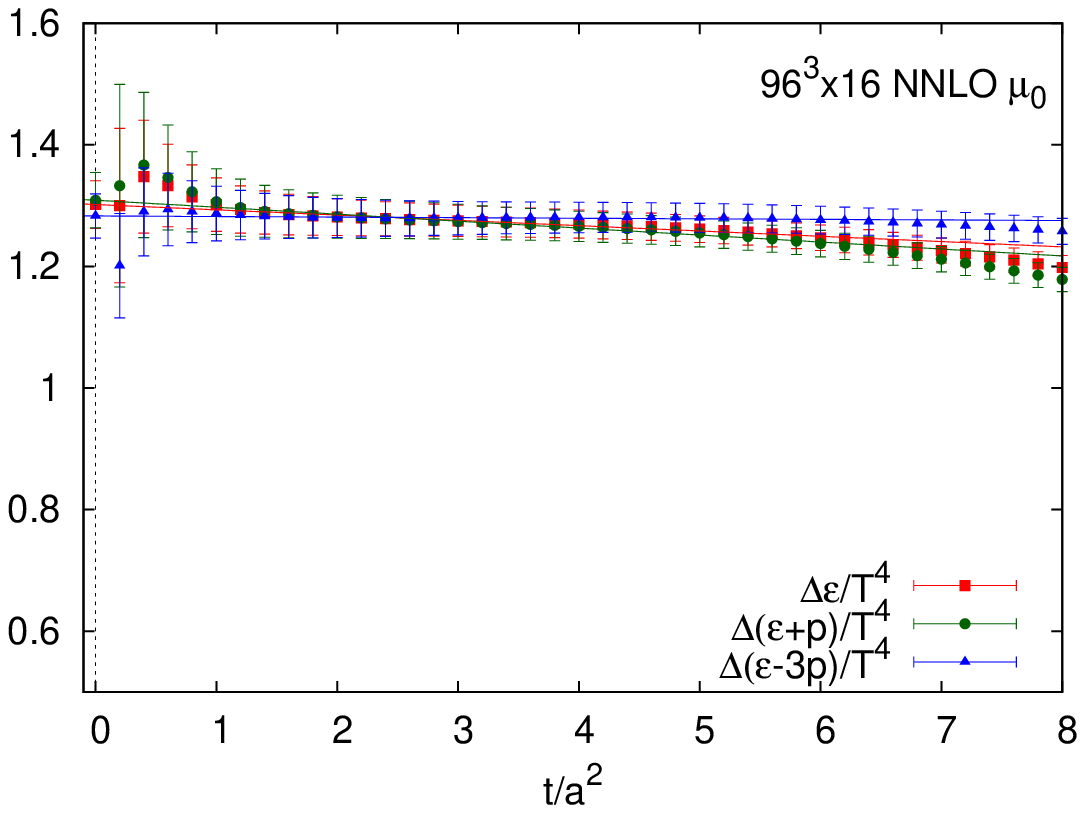}
\hspace{-3mm}
\includegraphics[width=7.5cm]{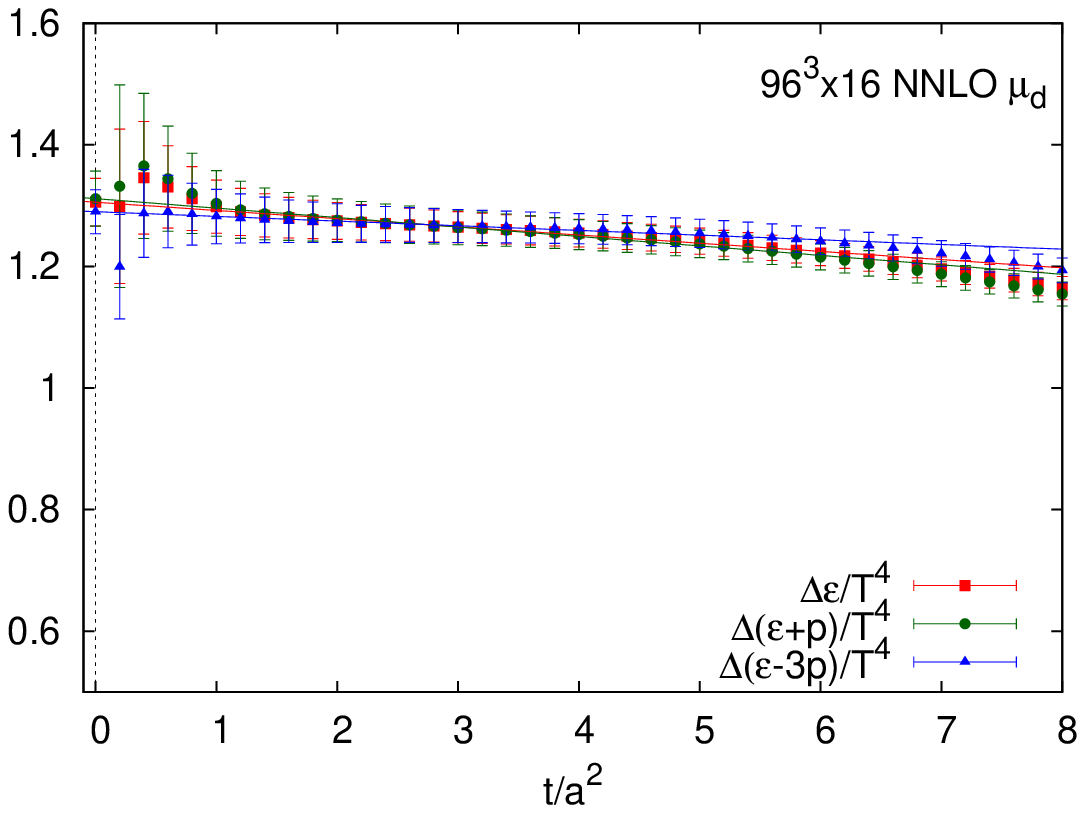}
\\
\includegraphics[width=7.5cm]{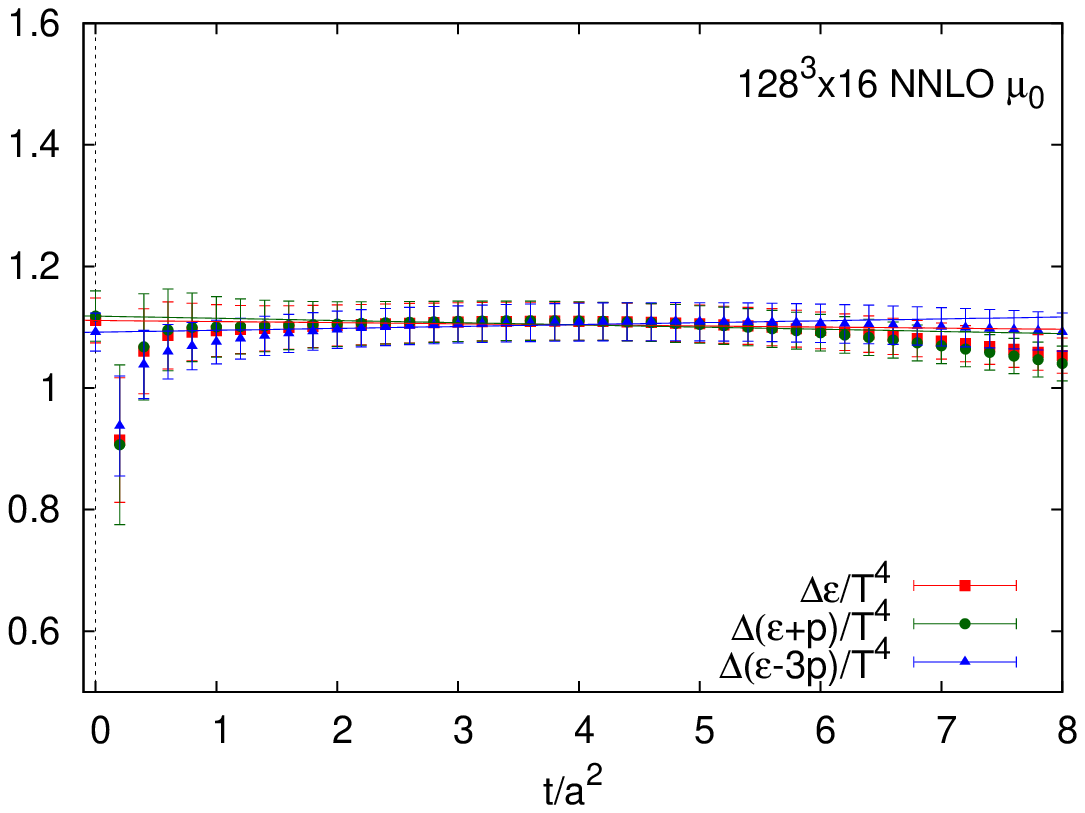}
\hspace{-3mm}
\includegraphics[width=7.5cm]{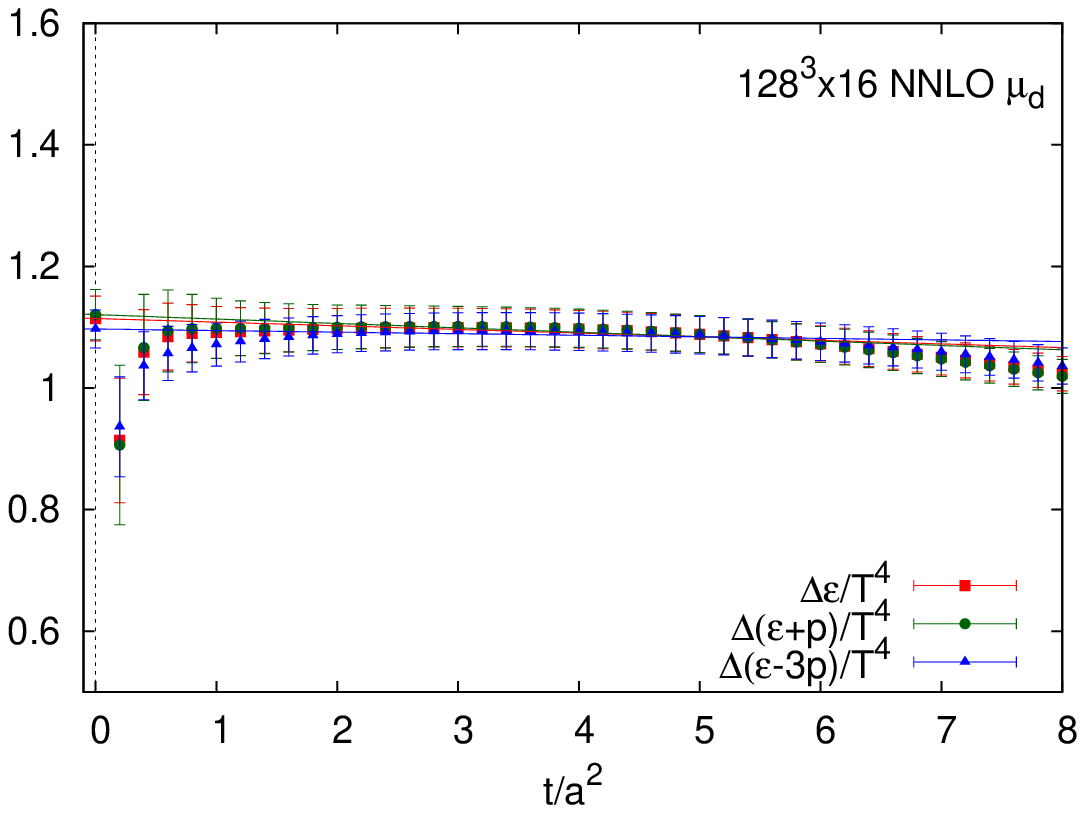}
\caption{
The same as Fig.~\ref{fig:flow8}, but  on the $96^3 \times 16$ (top) and $128^3\times16$ (bottom) lattices.
}
\label{fig:flow16}
\end{figure}

\section{Results of latent heat and pressure gap}
\label{results}

\subsection{Latent heat and pressure gap on finite lattices}

In Figs.~\ref{fig:flow8}--\ref{fig:flow16},
we show the results of $\Delta(\epsilon -3 p)/T^4$, $\Delta(\epsilon + p)/T^4$ and $\Delta \epsilon/T^4$ 
as functions of $t/a^2$, using the NNLO matching coefficients.
The blue, green and red symbols are the results of $\Delta(\epsilon -3 p)/T^4$, $\Delta(\epsilon + p)/T^4$ and $\Delta \epsilon/T^4$, respectively.
The plots in the left panels are the results adopting the renormalization scale $\mu=\mu_0$ and those in the right panels adopting $\mu=\mu_d$.
The rapid decrease of these observables at small $t$ signals the appearance of the $O(a^2/t)$ lattice discretization effect discussed in Sec.~\ref{sec:method12}. 
Similar behavior at similar values of $t$ has been reported for $(\epsilon -3 p)/T^4$ and $(\epsilon + p)/T^4$ themselves in previous studies of SU(3) Yang-Mills theory~\cite{Asakawa:2013laa,Kitazawa:2016dsl,Iritani2019}.
We thus disregard the data in this region in the $t\to0$ and $a\to0$ extrapolations.
We note that the results behave almost linearly for $t/a^2\gtrsim1.0$, while the $O(t)$ error sometimes starts to contaminate at large $t/a^2$.
We also note that, though the dependence on the renormalization scale $\mu$ is small, there is a general tendency that the $\mu_0$ scale (left panels) leads to a slightly smaller slope towards the $t\to0$ limit than the $\mu_d$ scale (right panels).

It should be noted that, within the statistical errors, the results of $\Delta(\epsilon -3 p)/T^4$ and $\Delta(\epsilon + p)/T^4$ are consistent with each other in a wide range of $t$.
This means that the pressure gap between two phases $\Delta p$ vanishes, which must be satisfied in the final results after the double extrapolation of $t\to0$ and $a\to0$.
This suggests that, for these observables, the errors from the lattice discretization and the small flow-time expansion are well under control in these ranges of $t$.

To study the influence of the truncation of perturbative series for the matching coefficients, we repeat the calculation with the NLO matching coefficients following the original procedure of~Ref.~\cite{Suzuki:2013gza}. 
The results are summarized in Appendix~\ref{sec:leading}.
We find that the NNLO matching coefficients drastically improve the signal over the NLO matching coefficients in the sense that the $\Delta p$ at finite $t$ and $a$ becomes much smaller and also the observables show smaller slope in $t$. 
As discussed in the next section, we also confirm that the results of $\Delta(\epsilon -3 p)/T^4$, $\Delta(\epsilon + p)/T^4$, and $\Delta \epsilon/T^4$ extrapolated to the $t\to0$ limit are consistent between the NNLO and NLO analyses.


\begin{figure}[t]
\centering
\includegraphics[width=7.5cm]{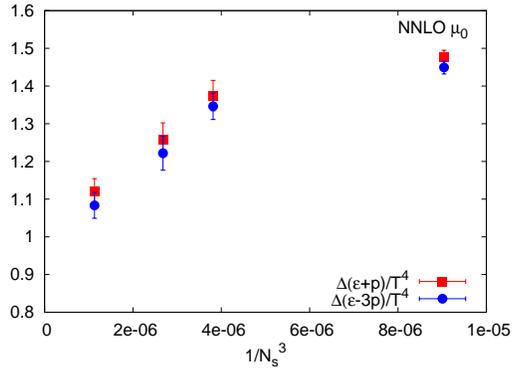}
\caption{Spatial volume dependence of $\Delta (\epsilon +p)/T^4$ (red) and $\Delta (\epsilon -3p)/T^4$ (blue) with the NNLO matching coefficients adopting $\mu=\mu_0$ on $N_s^3 \times 12$ lattices with $N_s=48,$ 64, 72, and 96.
}
\label{fig:vdepnt12}
\end{figure}

\clearpage

\subsection{Method 1: $t\to0$ followed by $a\to0$}
\label{tto0}

Let us first analyze the data adopting method~1, i.e., we first take the small flow-time limit $t\to0$ on each lattice, and then take the continuum limit $a\to0$.
We note that the data shown in Figs.~\ref{fig:flow8}--\ref{fig:flow16} are very linear in a wide range of $t$ within the statistical error. 
In this study, we choose the fit range as follows: 
First, we require the fit range to satisfy $t T_c^2 \le 0.025$. 
This value of the upper bound is determined by consulting the linearity of the data and also from the consistency with the fit range adopted in method~2 to be discussed in~Sec.~\ref{continuum}. 
We confirm that this upper bound satisfies the requirement that the smearing by the gradient flow is not overlapping around the periodic lattice ($t/a^2 < t_{1/2}$ defined in~Ref.~\cite{WHOT2017b}).
We also require $t/a^2 \ge \sqrt{2}$ so that the smearing radius by the gradient flow well covers the nearest-neighbor lattice sites. 
For the case of~$N_t=8$, however, because this makes the available range too narrow, we relaxed it to the minimum requirement of $t/a^2 \ge 1$. 
The fit ranges that we adopt as well as the results of linear $t\to0$ extrapolations are given in~Table~\ref{tab2}.
Corresponding linear fit lines are shown in Figs.~\ref{fig:flow8}--\ref{fig:flow16}.
In~Table~\ref{tab2}, we also show the results with the NLO matching coefficients discussed in~Appendix~\ref{sec:leading}. 
The errors in~Table~\ref{tab2} are statistical only.
We have also tested different fit ranges, but the variation of the results due to the fit range turned out to be within the statistical errors given in~Table~\ref{tab2}.

From~Table~\ref{tab2}, we note that results of different renormalization scales as well as with NNLO and NLO matching coefficients are all very consistent with each other.
We also find that the differences between $\Delta (\epsilon -3p) /T^4$ and $\Delta (\epsilon +p) /T^4$ are less than about one sigma for all lattices, in accordance with the expectation that $\Delta p$ should vanish.


\begin{figure}[t]
\centering
\hspace*{-7mm}
\includegraphics[width=5.5cm]{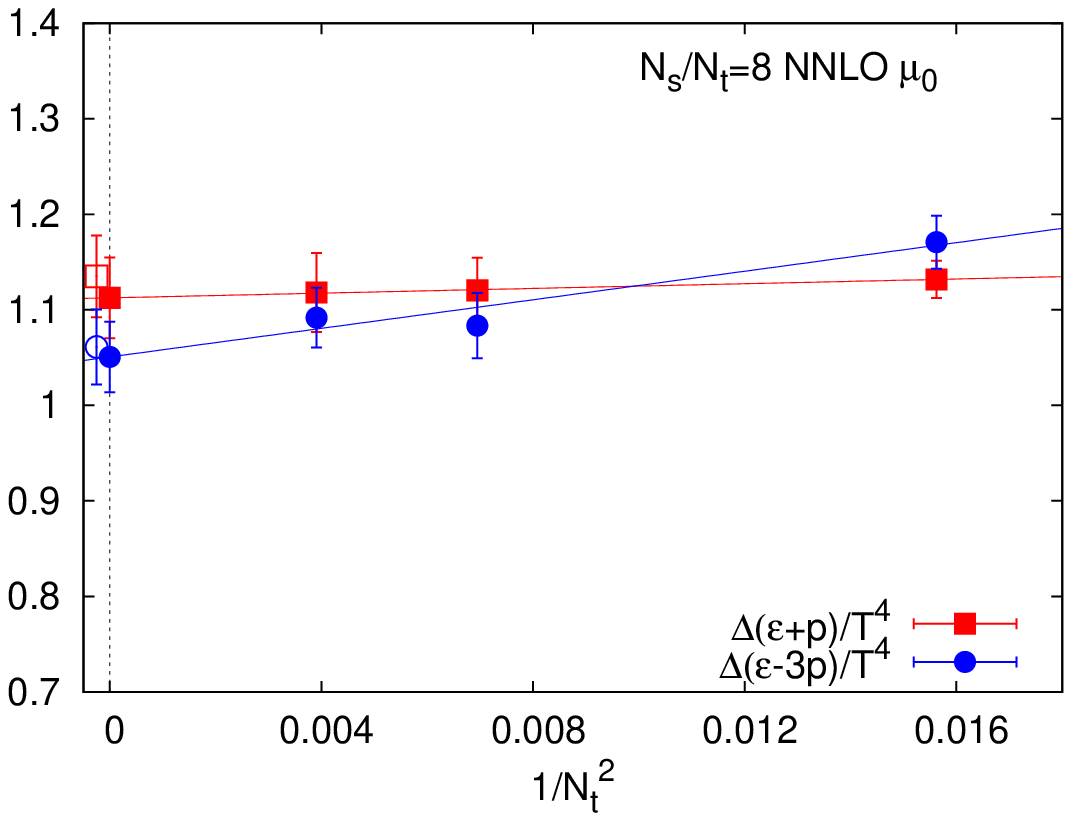}
\hspace{-7mm}
\includegraphics[width=5.5cm]{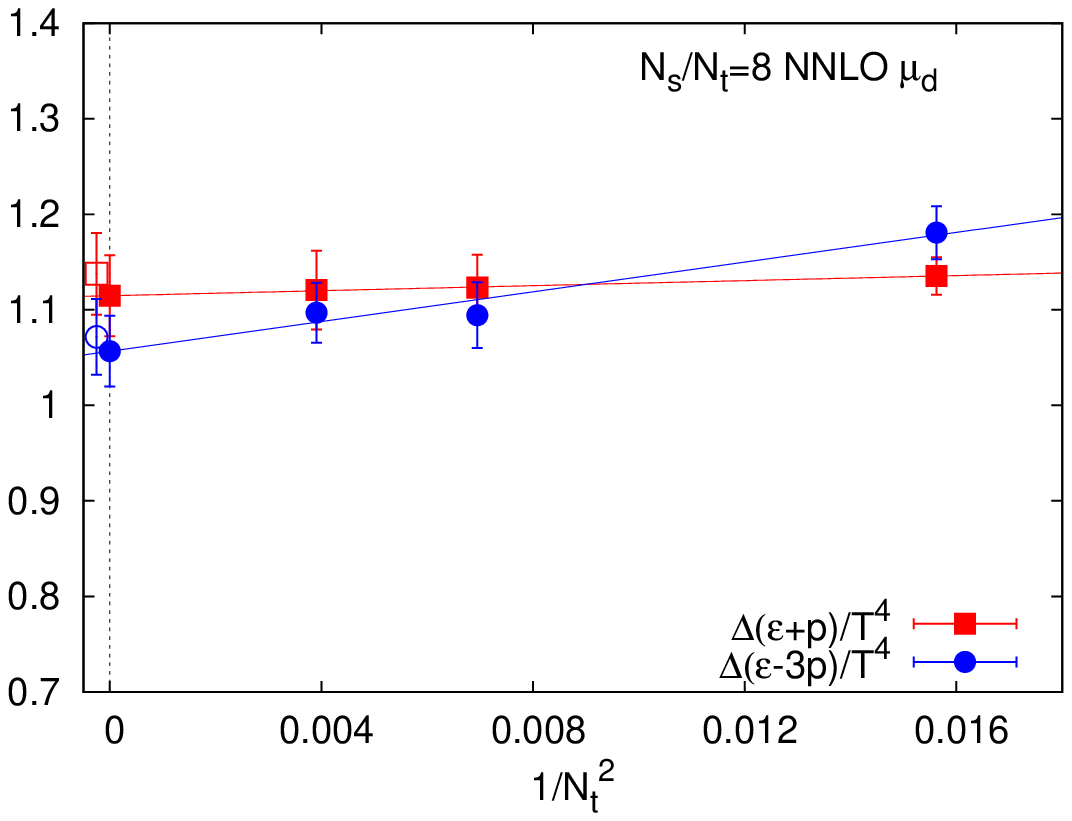}
\hspace{-7mm}
\includegraphics[width=5.5cm]{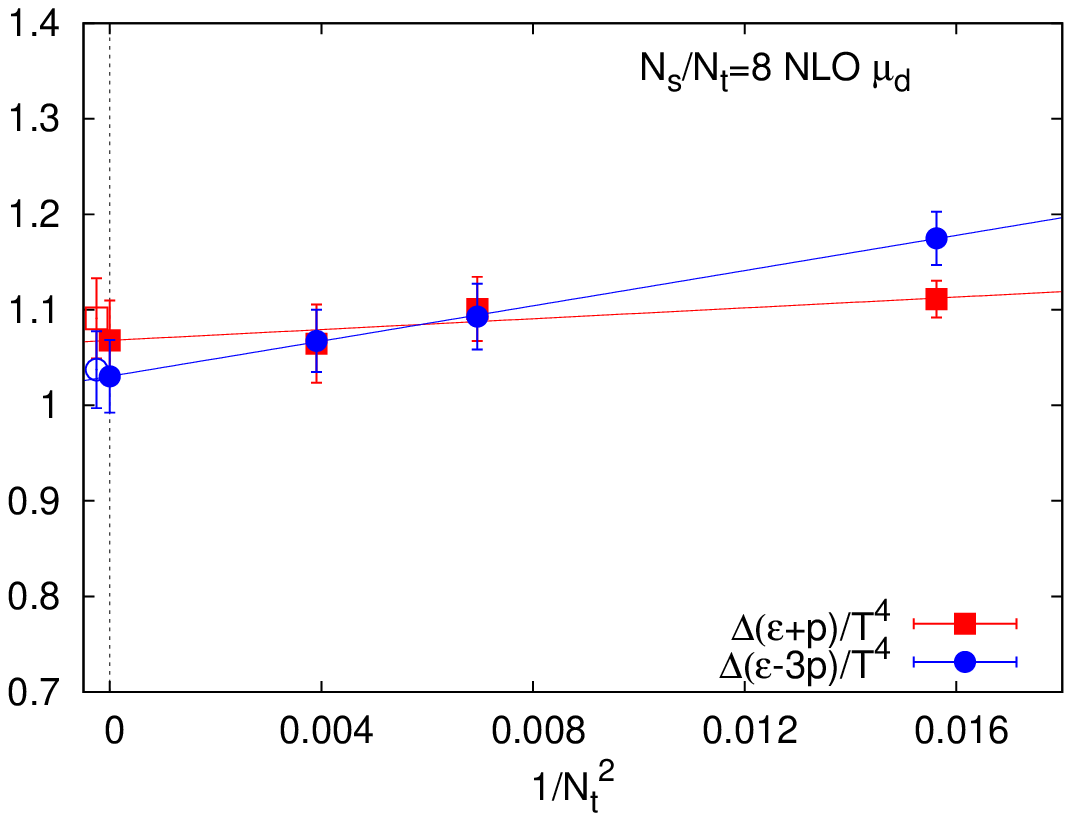}
\caption{$\Delta (\epsilon+p)/T^4$ (square) and $\Delta (\epsilon-3p)/T^4$ (circle) at $t=0$ for the aspect ratio $N_s/N_t=8$. 
The horizontal axis is $1/N_t^2 = (T_c a)^2$.
The open square and open circle at $1/N_t^2 = 0$ are the results of method 2, first taking $a\to0$ and then $t\to0$.
The left and middle panels are the results with NNLO matching coefficients adopting $\mu=\mu_0$ and $\mu_d$, respectively.
The right panel is the results with NLO matching coefficients adopting $\mu=\mu_d$. 
}
\label{fig:conlim8}
\end{figure}

\begin{figure}[t]
\centering
\hspace*{-7mm}
\includegraphics[width=5.5cm]{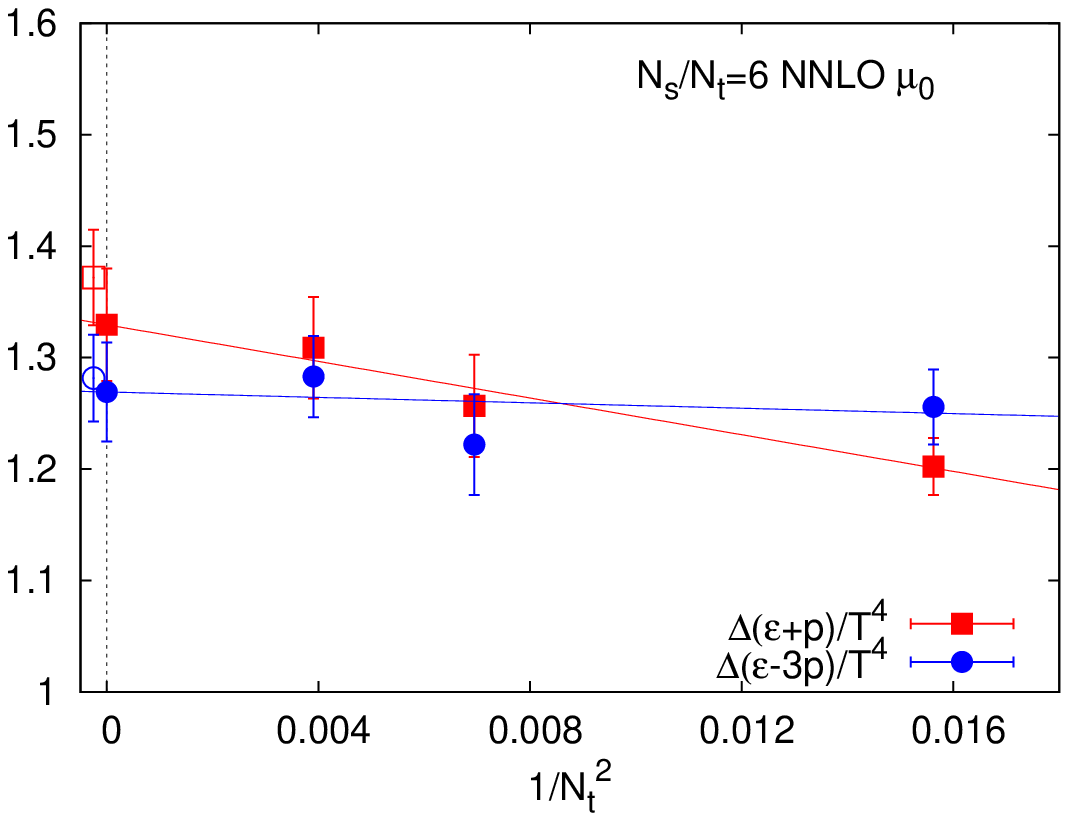}
\hspace{-7mm}
\includegraphics[width=5.5cm]{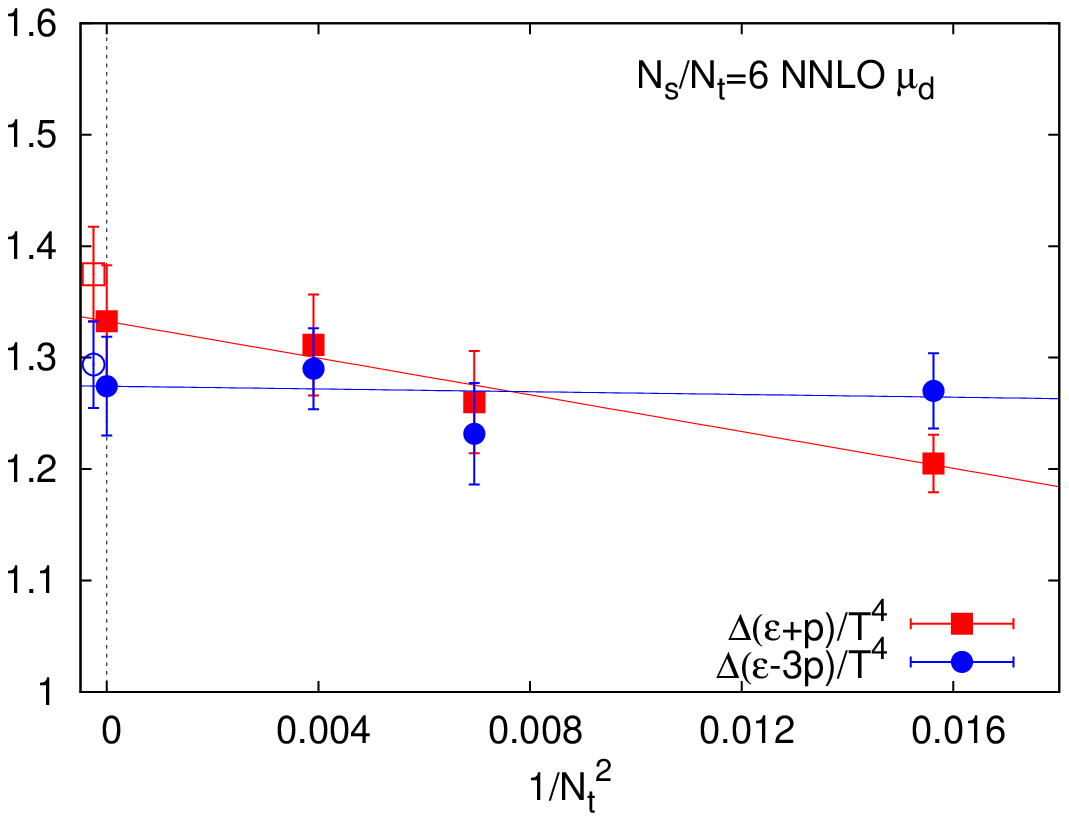}
\hspace{-7mm}
\includegraphics[width=5.5cm]{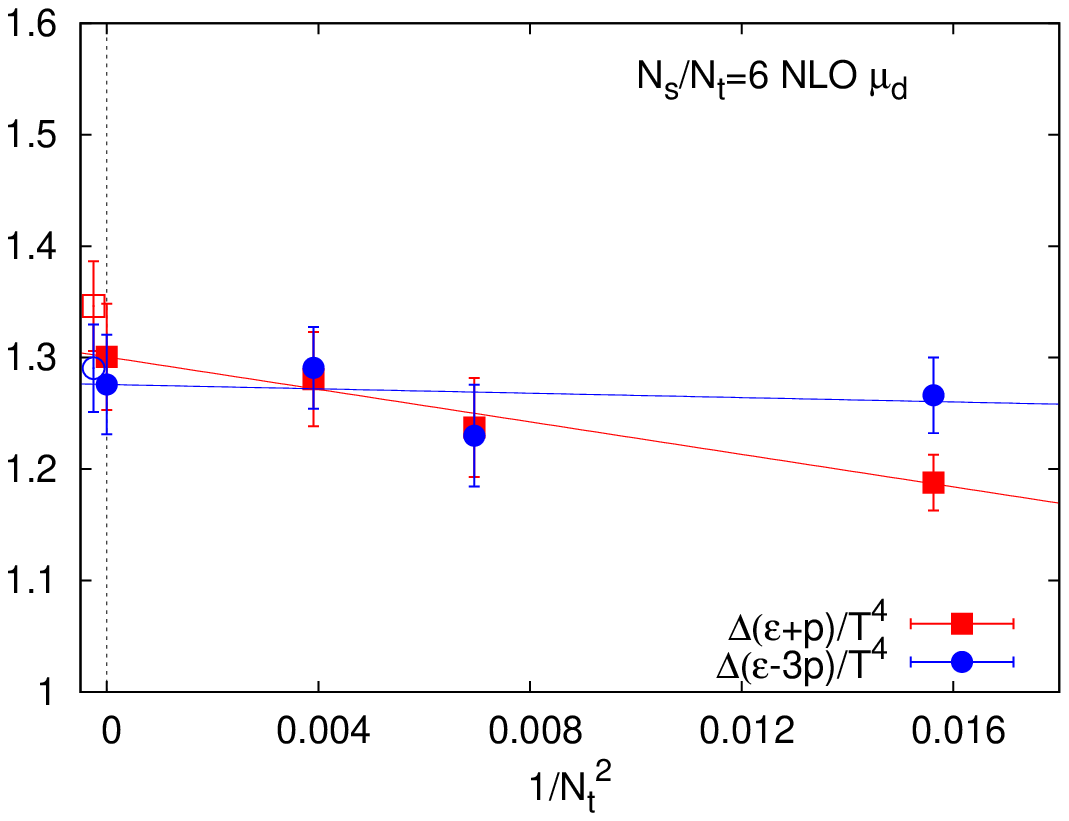}
\caption{The same as Fig.~\ref{fig:conlim8}, but for the aspect ratio $N_s/N_t=6$. 
}
\label{fig:conlim6}
\end{figure}

\begin{table}[t]
  \small
\caption{
The results of $t\to0$ extrapolation of $\Delta(\epsilon -3p)/T^4$ and $\Delta(\epsilon +p)/T^4$ on each lattice. 
Errors are statistical only. Systematic errors due to the choice of the fit range are smaller than the statistical errors.
}
\label{tab2}
\begin{center}
\begin{tabular}{cccccccccccc}
\hline
$N_s$ & $N_t$ & Fit range && 
\multicolumn{2}{c}{NNLO with $\mu_0$} &&
\multicolumn{2}{c}{NNLO with $\mu_d$} &&
\multicolumn{2}{c}{NLO with $\mu_d$}
\\
&& $t/a^2$ && $\frac{\Delta(\epsilon -3p)}{T^4}$ & $\frac{\Delta(\epsilon +p)}{T^4}$ && $\frac{\Delta(\epsilon -3p)}{T^4}$ & $\frac{\Delta(\epsilon +p)}{T^4}$ && $\frac{\Delta(\epsilon -3p)}{T^4}$ & $\frac{\Delta(\epsilon +p)}{T^4}$  
\\
\hline
48 & 8  & 1.0--1.6 &&  1.256(34) & 1.202(26) && 1.270(34) & 1.205(26) && 1.266(34) & 1.188(25)\\ 
64 & 8  & 1.0--1.6 &&  1.171(28) & 1.132(20) && 1.181(28) & 1.135(20) && 1.175(28) & 1.111(19)\\
48 & 12 & 1.4--3.6 &&  1.449(17) & 1.476(19) && 1.461(17) & 1.480(19) && 1.451(18) & 1.441(18)\\
64 & 12 & 1.4--2.0 &&  1.347(35) & 1.375(40) && 1.349(35) & 1.379(40) && 1.325(35) & 1.307(40)\\
72 & 12 & 1.4--3.6 &&  1.222(45) & 1.257(46) && 1.232(46) & 1.260(46) && 1.230(46) & 1.237(44)\\
96 & 12 & 1.4--3.6 &&  1.083(34) & 1.120(34) && 1.094(34) & 1.123(34) && 1.093(34) & 1.101(34)\\
96 & 16 & 1.4--6.4 &&  1.283(36) & 1.309(45) && 1.290(36) & 1.311(45) && 1.291(37) & 1.281(42)\\
128& 16 & 1.4--6.4 &&  1.092(31) & 1.118(41) && 1.097(31) & 1.121(41) && 1.067(33) & 1.065(41)\\
\hline
\end{tabular}
\end{center}
\end{table}

In Fig.~\ref{fig:vdepnt12}, we plot the results of 
$\Delta (\epsilon -3p)/T^4$ (red) and $\Delta (\epsilon+p)/T^4$ (blue) at $t=0$ measured on $N_s^3 \times 12$ lattices with $N_s=48,$ $64$, $72$, and $96$ 
as functions of the inverse spatial volume in lattice units $1/N_s^3$.
The lattice spacing $a = 1/(N_t T_c)$ is the same for all data shown in~Fig.~\ref{fig:vdepnt12}. 
We find that $\Delta (\epsilon -3p)/T^4$ and $\Delta (\epsilon +p)/T^4$ decrease as the spatial volume increases.
Therefore, we have to take the finite-volume effect into account.

We now perform the continuum extrapolation $a\to0$ using the results shown in Table~\ref{tab2}.
To study the finite-volume effect, we perform this on lattices with fixed spatial volume $V$.
As discussed in Sec.~\ref{sec:method12}, this is achieved by fixing the aspect ratio $N_s/N_t$ in our study.
In Fig.~\ref{fig:conlim8}, we plot $\Delta (\epsilon -3p) /T^4$ and $\Delta (\epsilon +p) /T^4$ at $t=0$ as functions of $1/N_t^2 = (T_c a)^2$ for $N_s /N_t =8$. 
The left and middle panels are the results adopting $\mu=\mu_0$ and $\mu_d$, respectively.
We also show the results with NLO matching coefficients adopting $\mu=\mu_d$, discussed in~Appendix~\ref{sec:leading}.
Corresponding results for $N_s /N_t =6$ are shown in~Fig.~\ref{fig:conlim6}.

We do the $a\to0$ extrapolation by fitting these results by a linear function of $(T_c a)^2=1/N_t^2$. 
We find that the lattice spacing dependence is small in these observables.
However, we also note that the slight slope toward the continuum limit changes its sign between $N_s /N_t =8$ and 6.
This may be suggesting that, when we consider various systematic errors, the lattice spacing dependence is approximately absent within errors. 
The continuum-extrapolated values by the linear fit are plotted by the filled square and filled circle at $1/N_t^2=0$ in~Figs.~\ref{fig:conlim8} and~\ref{fig:conlim6}.
The open square and open circle at $1/N_t^2\approx0$ are the results of method~2 discussed in the next subsection. 

Our final results of method~1 for the latent heat are summarized in the left columns of~Table~\ref{tab3}. 
We find that the results for $\Delta \epsilon /T^4$, $\Delta (\epsilon -3p)/T^4$ and 
$\Delta(\epsilon +p)/T^4$, obtained using the NNLO or NLO matching coefficients and with the $\mu_0$ or $\mu_d$ scale, are all very consistent with each other at each aspect ratio.
From the results of $\Delta (\epsilon -3p)/T^4$ and $\Delta(\epsilon +p)/T^4$ with the NNLO matching coefficients and the $\mu_0$ scale, we find the pressure gaps to be 
\begin{eqnarray}
\Delta p /T^4 = 0.015(17) \ \  (N_s/N_t=8), \ \ \ \ 
\Delta p /T^4 = 0.015(14) \ \  (N_s/N_t=6), 
\label{eq:dp-method1}
\end{eqnarray}
which are only about 1\% of the latent heat and are consistent with zero.


\begin{figure}[t]
\centering
\includegraphics[width=7.5cm]{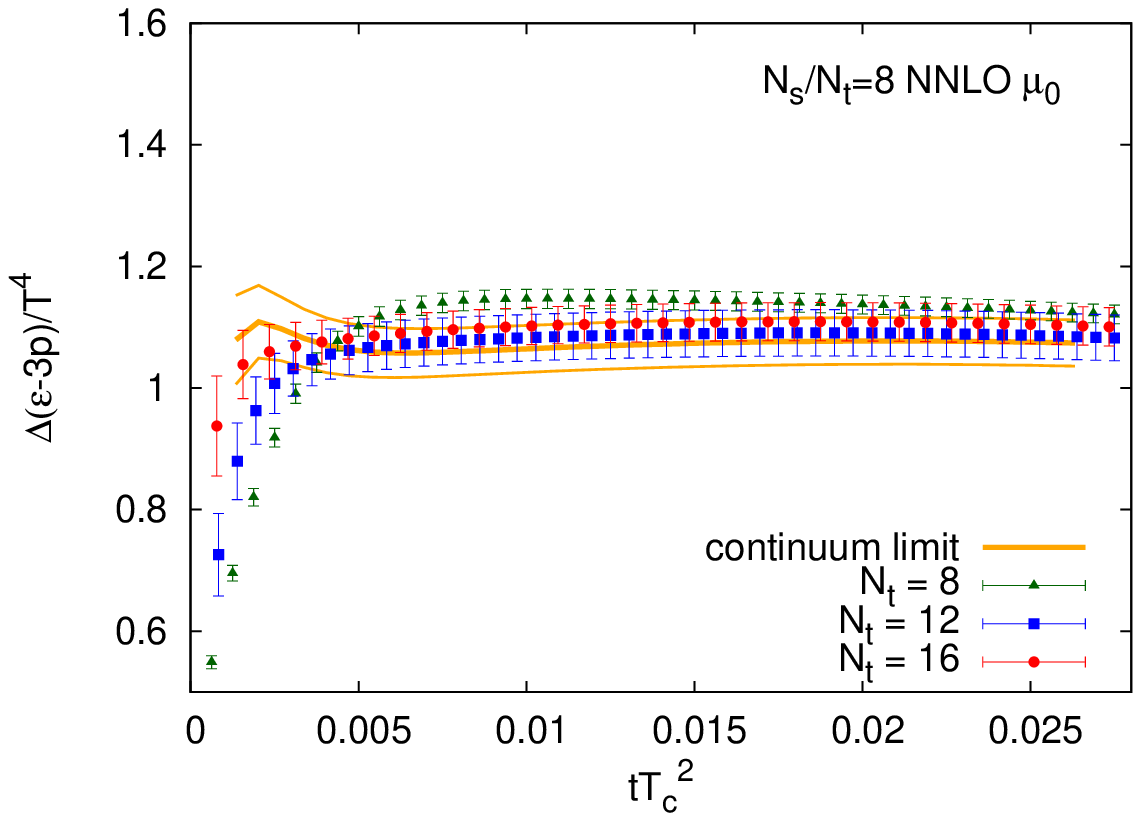}
\hspace{1mm}
\includegraphics[width=7.5cm]{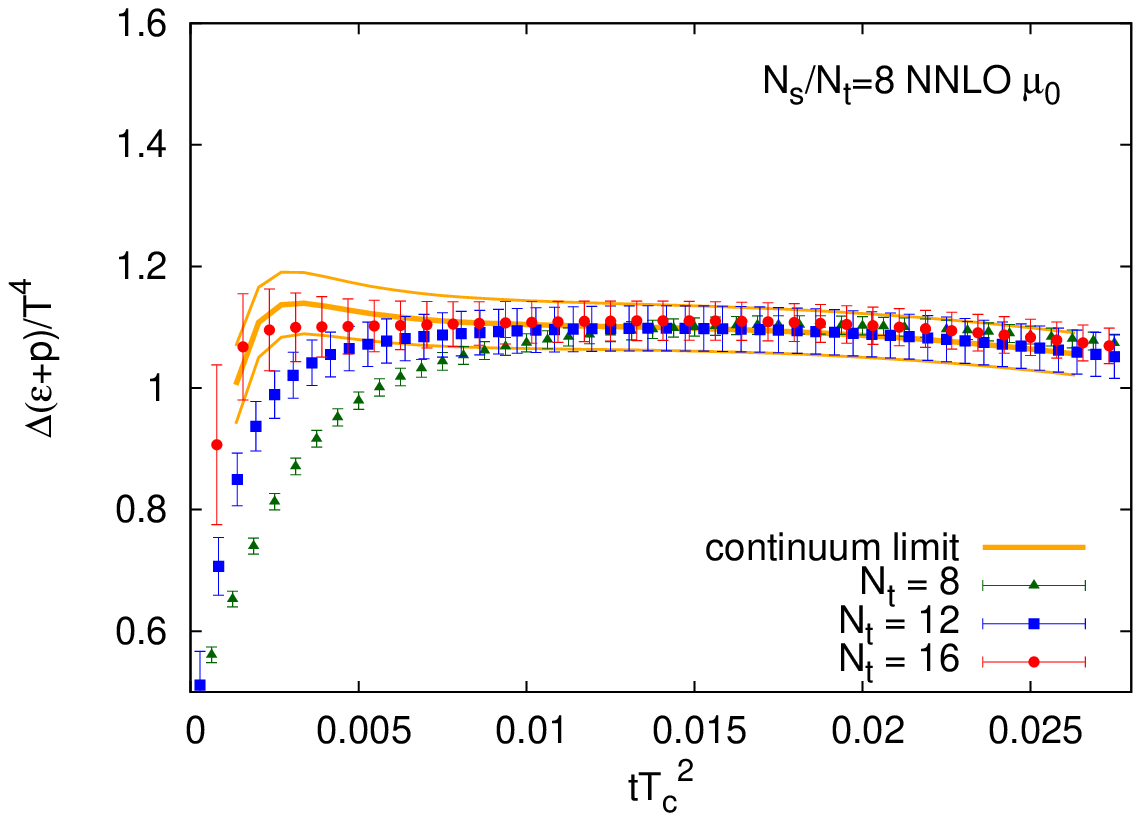}
\\
\includegraphics[width=7.5cm]{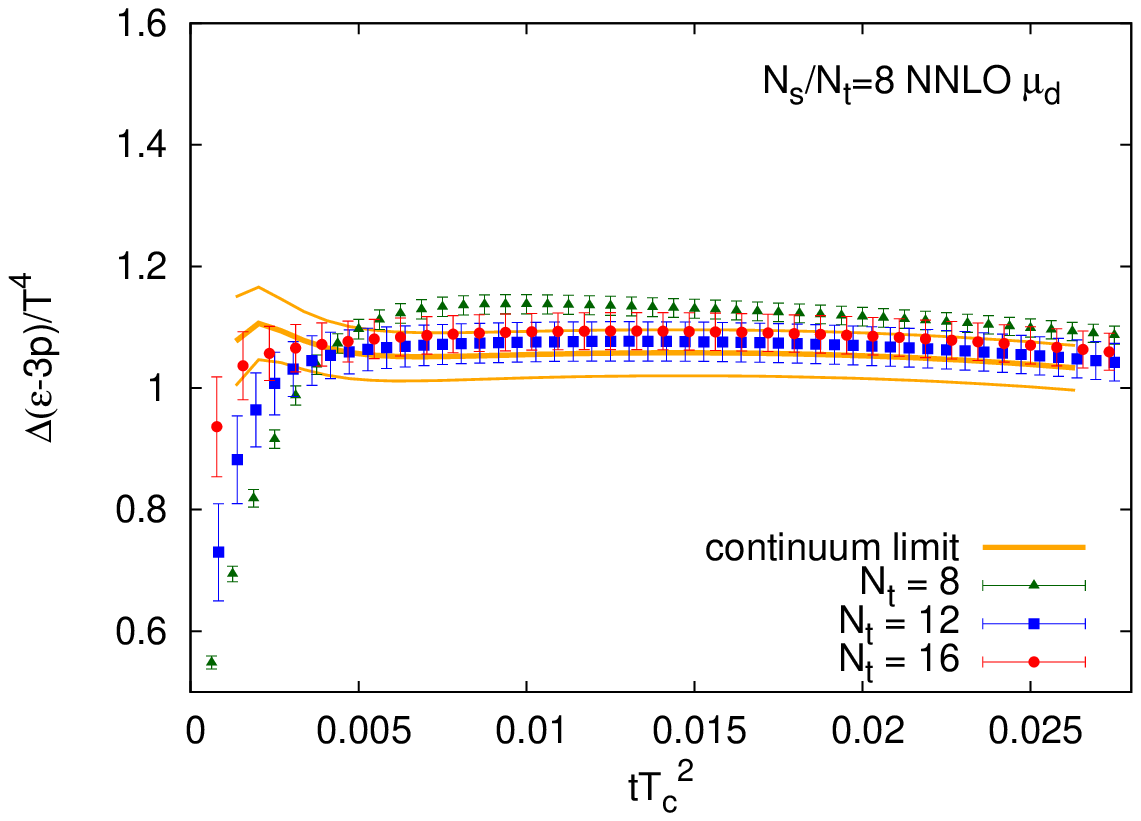}
\hspace{1mm}
\includegraphics[width=7.5cm]{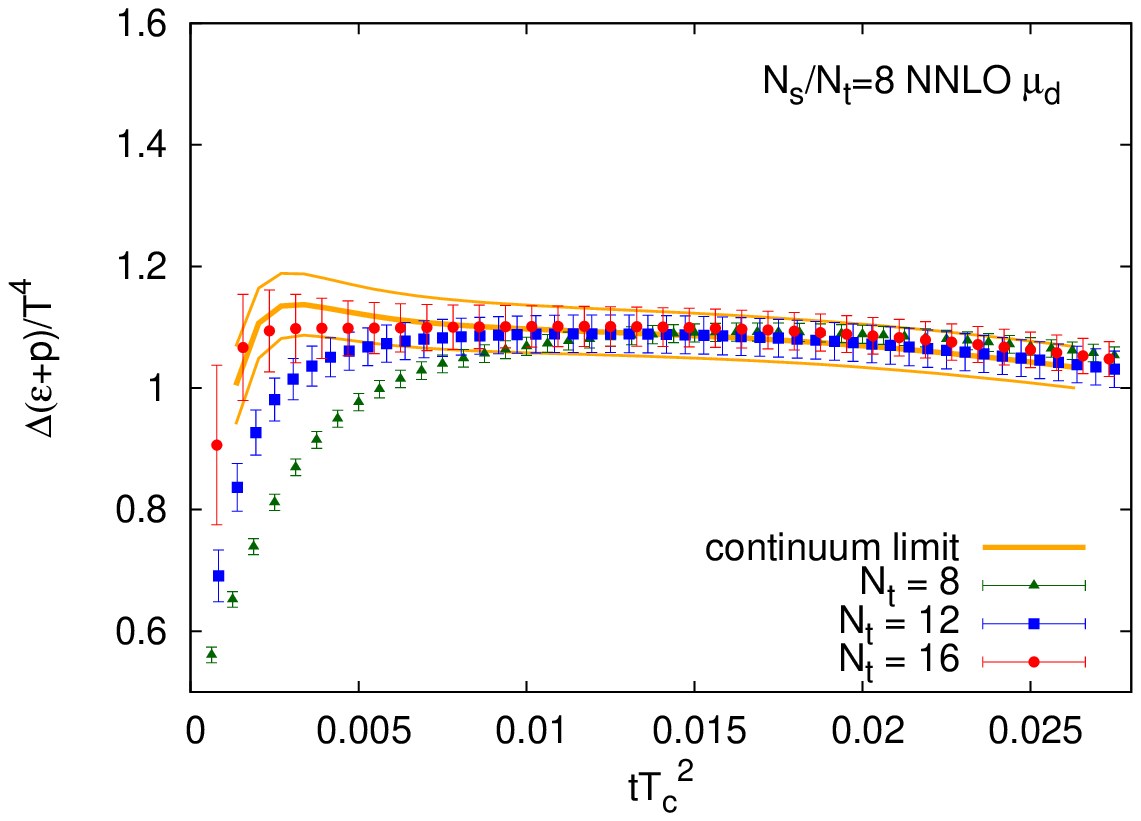}
\caption{$\Delta (\epsilon -3p)/T^4$ (left) and $\Delta (\epsilon +p)/T^4$ (right) with the NNLO matching coefficients adopting $\mu=\mu_0$ (top) and 
$\mu=\mu_d$ (bottom) on $N_s/N_t=8$ lattices.
Three orange curves represent the result in the continuum limit. 
(Thick orange curves represent the central values and thin curves represent the range of their errors.)
}
\label{fig:adep8}
\end{figure}

\begin{figure}[t]
\centering
\includegraphics[width=7.5cm]{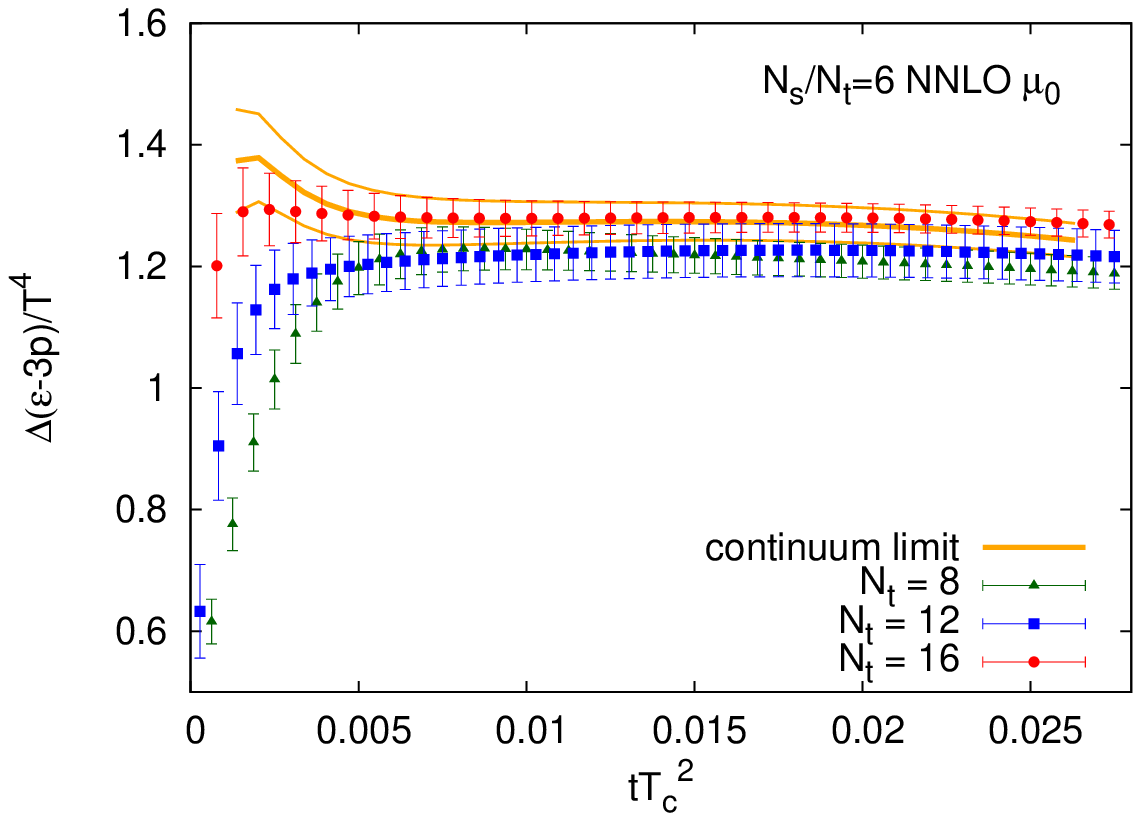}
\hspace{1mm}
\includegraphics[width=7.5cm]{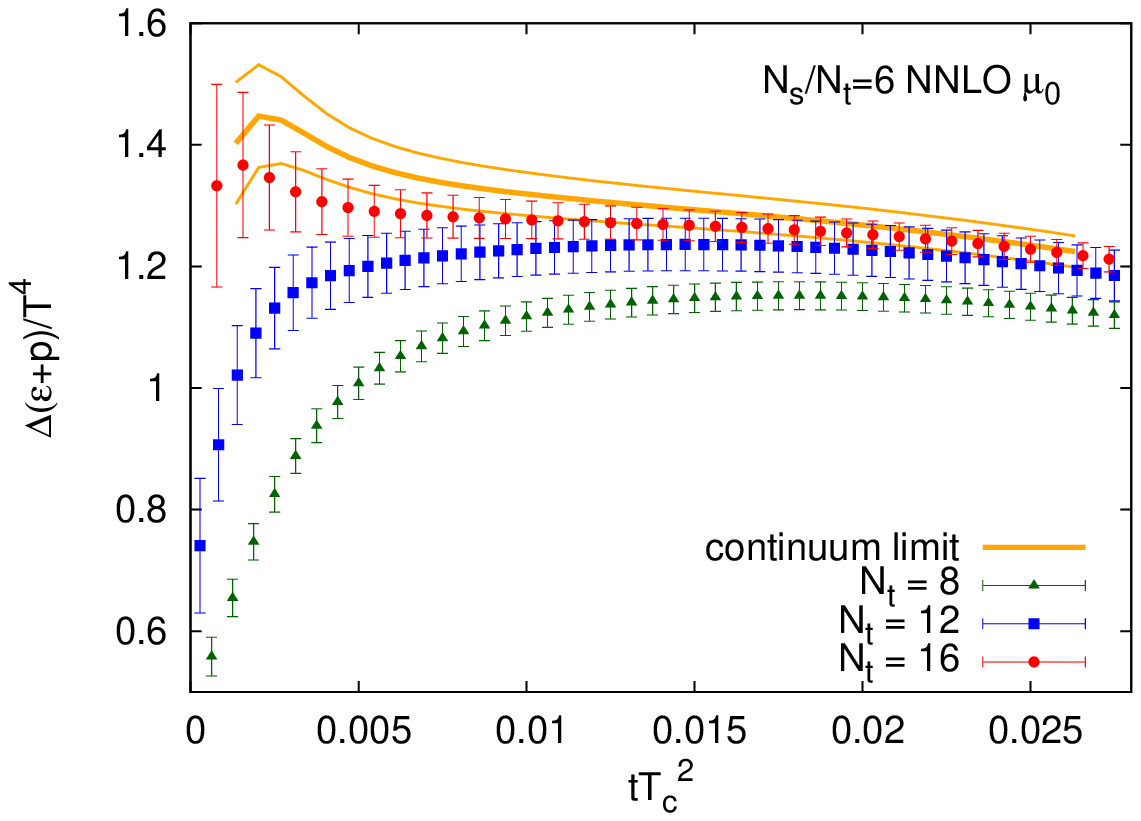}
\\
\includegraphics[width=7.5cm]{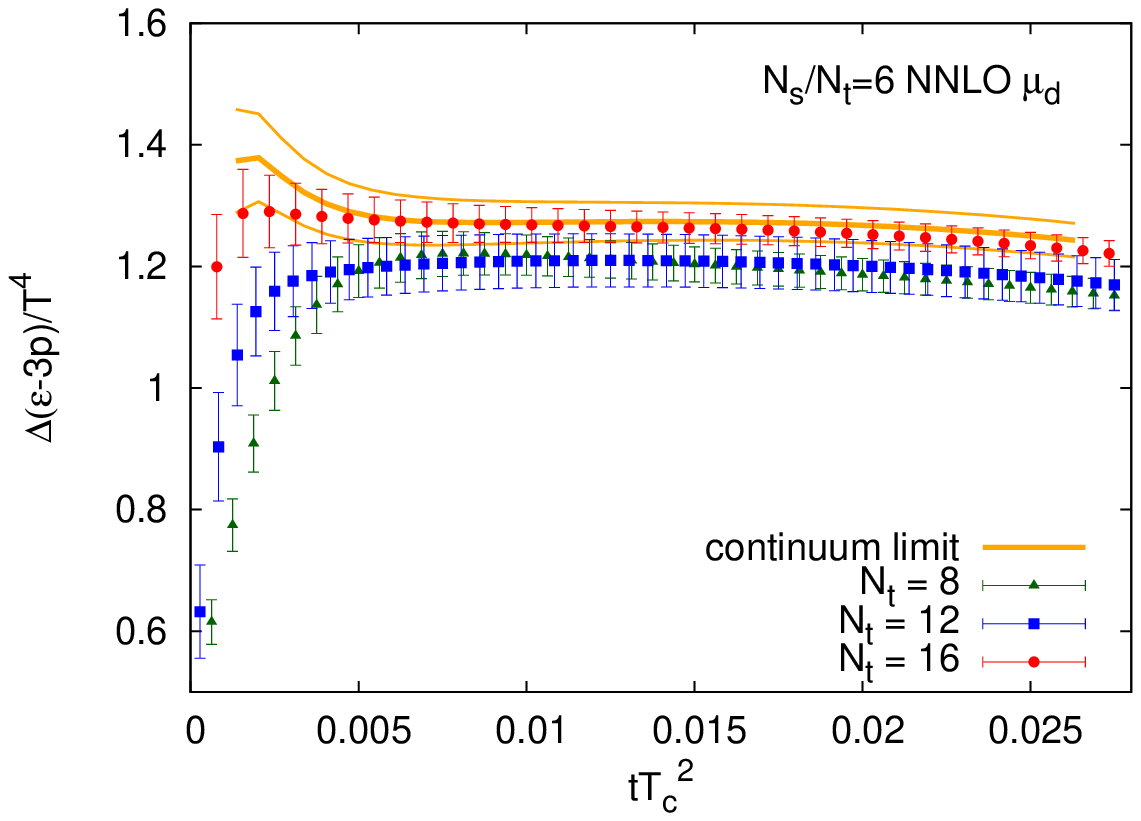}
\hspace{1mm}
\includegraphics[width=7.5cm]{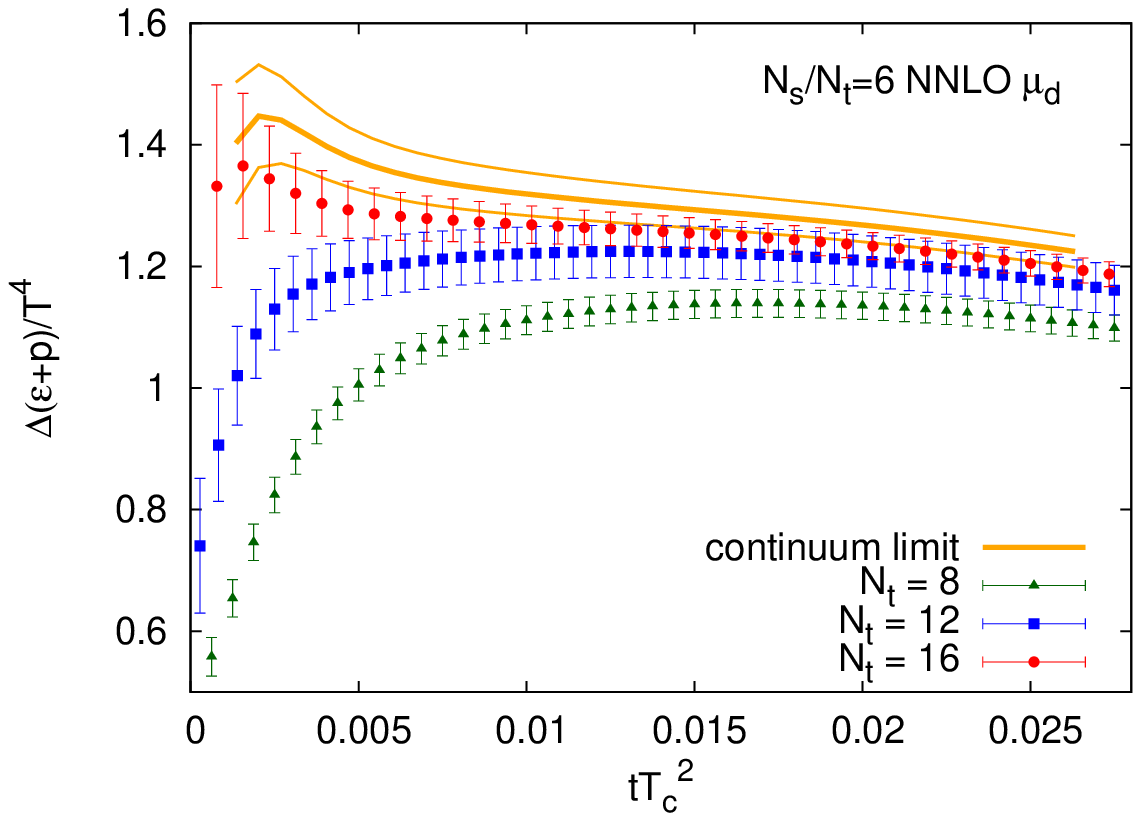}
\caption{The same as Fig.~\ref{fig:adep8}, but for $N_s/N_t=6$.
}
\label{fig:adep6}
\end{figure}

\begin{table}[t]
\caption{$\Delta \epsilon /T^4$, $\Delta (\epsilon -3p)/T^4$ and 
$\Delta(\epsilon +p)/T^4$ in the continuum limit with the NNLO and NLO matching coefficients adopting the renormalization scale $\mu_0$ or $\mu_d$.
The left and right three columns are results of method~1 ($t \to 0$ followed by $a \to 0$) and method~2 ($a \to 0$ followed by $t \to 0$), respectively.
Errors are statistical only.  Systematic errors due to variation of the fit ranges are smaller than the statistical errors.
}
\label{tab3}
\begin{center}
\begin{tabular}{cccccccc}
\hline
&&
\multicolumn{3}{c}{method~1} & 
\multicolumn{3}{c}{method~2} \\
& $\frac{N_s}{N_t}$ & $\frac{\Delta \epsilon}{T^4}$ & $\frac{\Delta (\epsilon -3p)}{T^4}$ &
$\frac{\Delta(\epsilon +p)}{T^4}$ &  $\frac{\Delta \epsilon}{T^4}$ &
$\frac{\Delta (\epsilon -3p)}{T^4}$ & $\frac{\Delta(\epsilon +p)}{T^4}$ \\
\hline
NNLO $\mu_0$ & 6 & 1.314(45)&1.269(44)&1.329(50)&1.349(38)&1.282(39)&1.372(43)\\
NNLO $\mu_d$ & 6 & 1.317(45)&1.274(44)&1.333(50)&1.355(38)&1.294(39)&1.375(43)\\
NLO $\mu_d$  & 6 & 1.293(43)&1.276(45)&1.301(48)&1.332(37)&1.290(39)&1.346(40)\\
\hline
NNLO $\mu_0$ & 8 & 1.095(40)&1.051(37)&1.113(42)&1.117(40)&1.061(39)&1.135(43)\\
NNLO $\mu_d$ & 8 & 1.099(40)&1.057(37)&1.115(42)&1.121(41)&1.072(40)&1.138(43)\\
NLO $\mu_d$  & 8 & 1.056(40)&1.030(38)&1.068(42)&1.078(40)&1.037(40)&1.091(42)\\
\hline
\end{tabular}
\end{center}
\end{table}

\begin{figure}[t]
\centering
\includegraphics[width=7.5cm]{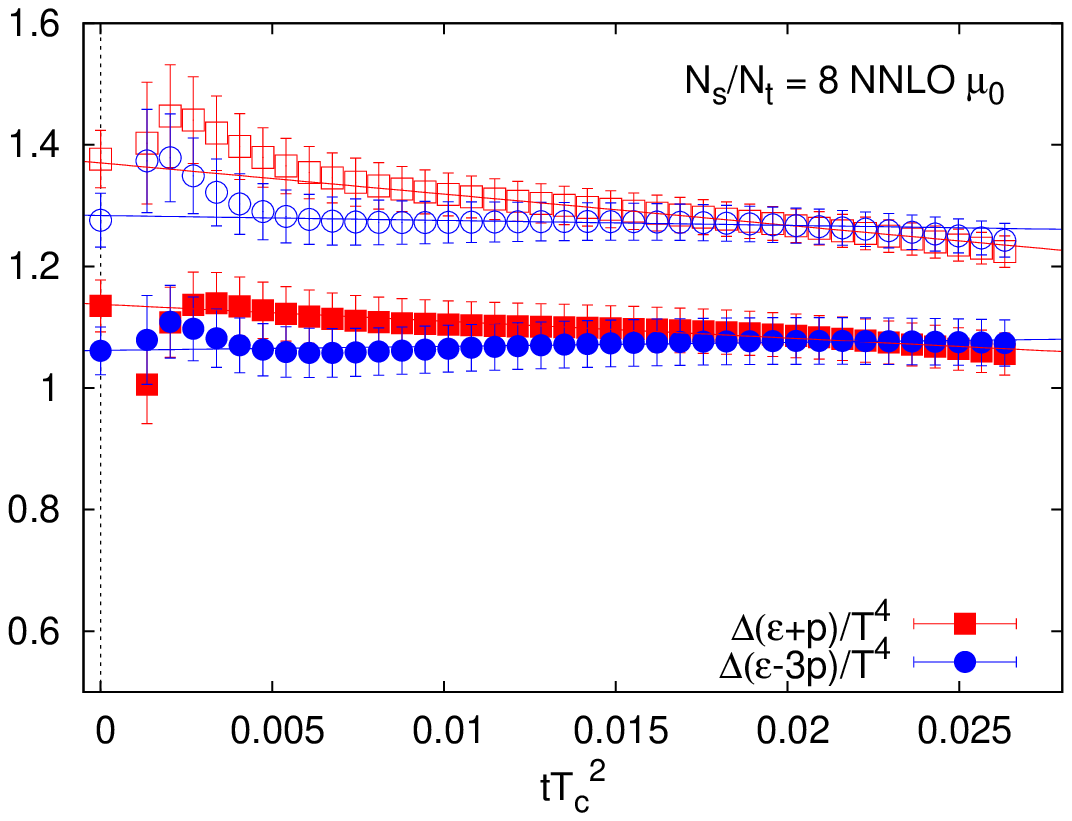}
\hspace{1mm}
\includegraphics[width=7.5cm]{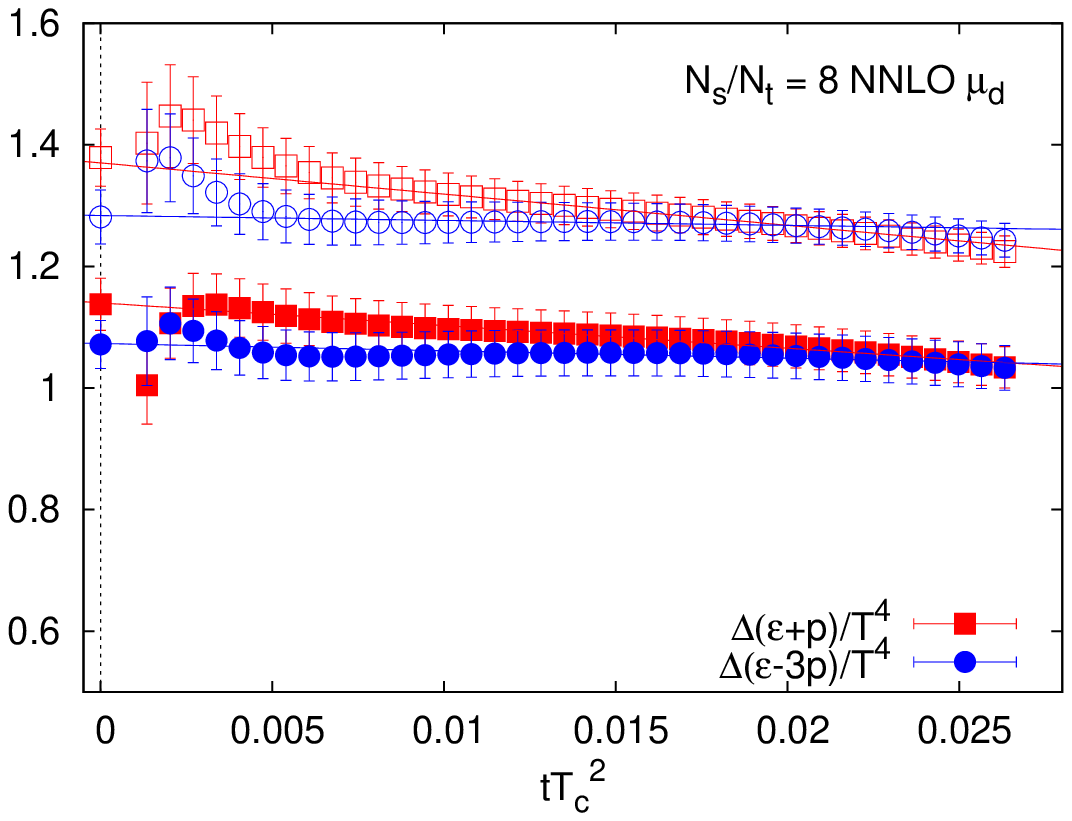}
\\
\includegraphics[width=7.5cm]{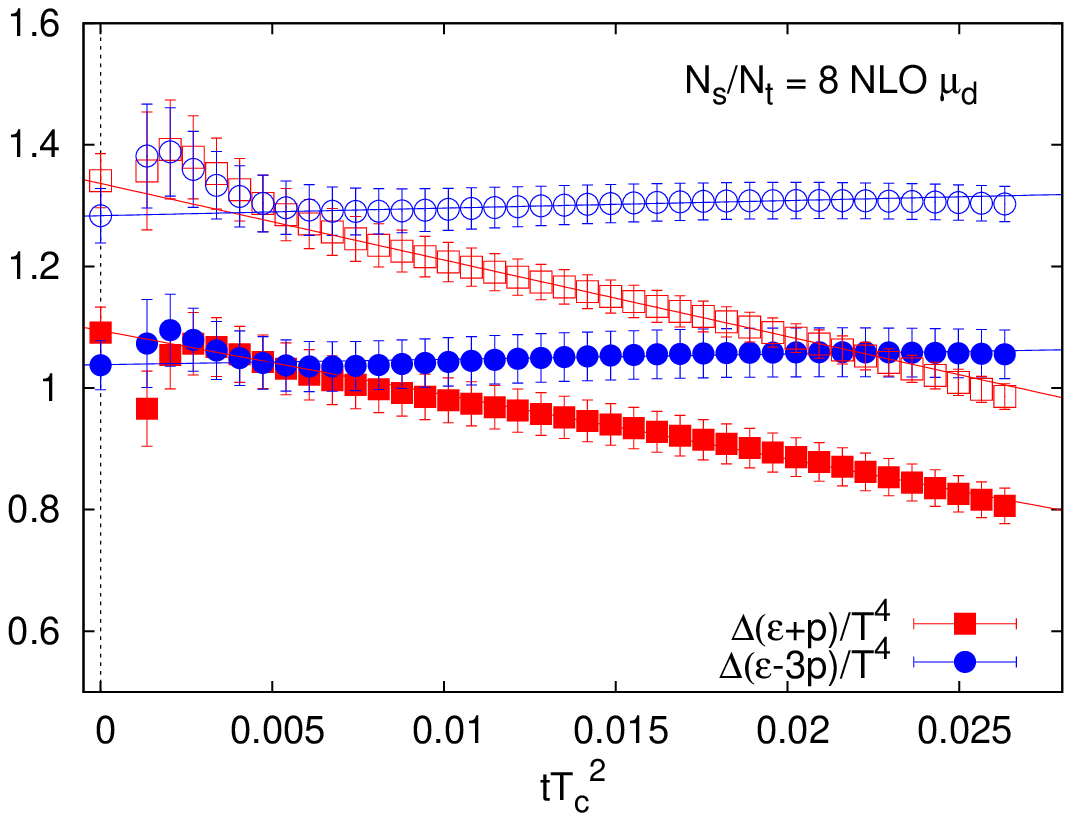}
\caption{$\Delta (\epsilon+p)/T^4$ (red square) and $\Delta (\epsilon-3p)/T^4$ 
(blue circle) in the continuum limit with the NNLO matching coefficients adopting $\mu=\mu_0$ (top left) and $\mu=\mu_d$ (top right),
and with the NLO matching coefficients adopting $\mu=\mu_d$ (bottom).
The filled symbols are the results of $N_s / N_t =8$ and the open symbols are those of $N_s / N_t =6$. 
The symbols on the vertical axes are the results in the continuum limit.
}
\label{fig:cont}
\end{figure}

\begin{figure}[t]
\centering
\includegraphics[width=7cm]{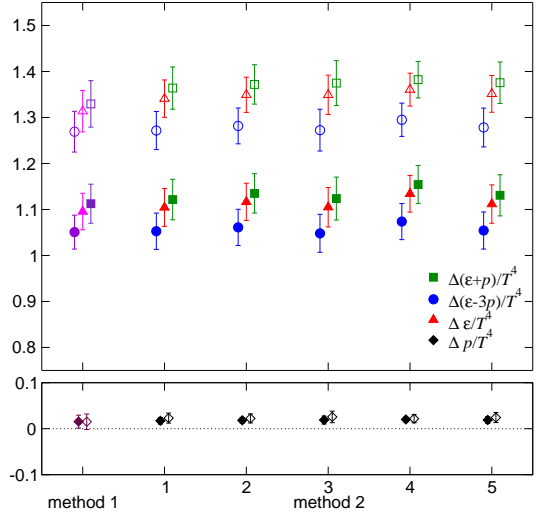}
\caption{Fit range dependence of $\Delta \epsilon/T^4$ (triangle), $\Delta (\epsilon+p)/T^4$ (square), $\Delta (\epsilon-3p)/T^4$ (circle), and $\Delta p/T^4$ (diamond), determined 
in the continuum and $t=0$ limits by method~2, with the NNLO matching coefficients adopting the $\mu_0$ scale.
The filled symbols are the results of $N_s / N_t =8$ and the open symbols are those of $N_s / N_t =6$. 
The numbers 1--5 on the horizontal axis are for the fit ranges 1--5 explained in the text. 
For comparison, results of method~1 are also shown at the left end of the plot.
}
\label{fig:fitdep}
\end{figure}

\subsection{Method 2: $a\to0$ followed by $t\to0$}
\label{continuum}

Next, we adopt method~2 to take the double limit $(t,a)\to(0,0)$, i.e., we first take the $a \to 0$ limit at each flow time $t$ in physical units, and then take the $t\to 0$ limit.
We do this by fixing the aspect ratio $N_s/N_t$ to take the finite size effect into account.
In Fig.~\ref{fig:adep8},
we plot the results of $\Delta(\epsilon -3 p)/T^4$ (left) and $\Delta(\epsilon + p)/T^4$ (right)
as functions of the flow time in a physical unit, $t T_c^2 =t/(aN_t)^2$, on lattices with $N_s/N_t=8$.
Figure~\ref{fig:adep6} shows corresponding results for $N_s/N_t=6$.
For these figures, the NNLO matching coefficients are used, and the results adopting $\mu=\mu_0$ and $\mu_d$ are given in the top and bottom panels of each figure, respectively.
The red, green and blue symbols are for the results of $N_t=8$, $12$ and $16$.
Recall that the lattice spacing $a = 1/(N_t T_c)$ is varied by $N_t$.
From these figures, we find that the lattice spacing dependence becomes rapidly smaller as the flow time increases. 

We take the $a\to0$ limit by fitting the data with a linear function of $1/N_t^2$ at each flow time $t T_c^2$. 
The orange curves in Figs.~\ref{fig:adep8} and \ref{fig:adep6} are the results in the continuum limit. The center curve represents the central value, and the upper and lower curves represent the statistical error.
For information, we also show the results obtained outside the linear window in these figures.
In~Fig.~\ref{fig:cont}, we show the results of $\Delta(\epsilon -3p)/T^4$ (circle) and  $\Delta(\epsilon + p)/T^4$ (square) in the continuum limit as functions of $tT_c^2$ for $N_s / N_t =8$ (filled symbols) and $N_s / N_t =6$ (open symbols). 
The results adopting $\mu=\mu_0$ and $\mu_d$ are given in the upper left and upper right panels.
We find that the dependence on $\mu$ is almost negligible in these quantities.
In the bottom panel of~Fig.~\ref{fig:cont}, we also show the results using data with NLO matching coefficients with the $\mu_d$ scale given in~Appendix~\ref{sec:leading}.
Comparing the upper and lower panels, we find that, by adopting the NNLO matching coefficients, the slopes in $t$ are much reduced and the pressure gap $\Delta p/T^4$ at finite $t$ is much suppressed.

We then perform $t \to 0$ extrapolation using the continuum-extrapolated results.
We avoid the data in the small-$t$ region where contamination of the $O(a^2/t)$ terms is suspected. 
To estimate a systematic error due to the $t \to 0$ extrapolation, we test the following five
fit ranges:
Range 1: $0.010 < t T_c < 0.020$,
Range 2: $0.010 < t T_c < 0.025$,
Range 3: $0.005 < t T_c < 0.020$,
Range 4: $0.015 < t T_c < 0.025$, and
Range 5: $0.005 < t T_c < 0.025$.
In~Fig.~\ref{fig:fitdep}, we show the results of these liner extrapolations in $tT_c^2$ using data with the NNLO matching coefficients and the $\mu_0$ scale. 
The triangle, square and circle symbols are for $\Delta \epsilon /T^4$, $\Delta (\epsilon +p) /T^4$ and $\Delta (\epsilon -3p)$ at $t=0$, and the filled and open symbols are the results of $N_s / N_t =8$ and $6$, respectively.
We find that the results adopting different fit ranges are very consistent with each other.
In~Fig.~\ref{fig:fitdep}, we also show the results of method~1 obtained in the previous subsection at the left end of the plot, 
which are also consistent with the results of method~2 within statistical errors.

We note that $\Delta (\epsilon +p) /T^4$ and $\Delta (\epsilon -3p) /T^4$ at $t = 0$ are very consistent with each other within the statistical errors.
Similar results are obtained also with the $\mu_d$ scale and the NLO matching coefficients. 

The results of the direct calculation of $\Delta p$ are plotted by the diamond symbols in the bottom panel of~Fig.~\ref{fig:fitdep}. 
We find that the values of $\Delta p$ are only about 1\% of the latent heat and are consistent with the results of method~1 given in Eq.~(\ref{eq:dp-method1}).
We also note that the values of $\Delta p$ are even smaller than the errors of the latent heat.
Due to correlation between $\Delta (\epsilon +p) /T^4$ and $\Delta (\epsilon -3p) /T^4$, the jackknife statistical errors for $\Delta p$ turned out to be quite small in comparison to the errors by the error propagation formula using the errors of $\Delta (\epsilon +p) /T^4$ and $\Delta (\epsilon -3p) /T^4$.
With the small errors, the mean values of $\Delta p$ deviate from zero by about 2-3 $\sigma$ statistical errors depending on the fit range.
We find that these nonvanishing values in the $t\to0$ limit originate solely from the data at $N_t=8$ used in the continuum extrapolation at each fixed flow time $t$:  
As seen from~Fig.~\ref{fig:flow8}, the values of $\Delta p$ for $N_t=8$ clearly deviate from zero, while those for $N_t=12$ and 16 are consistent with zero in the fit range of the $t\to0$ extrapolation as shown in~Figs.~\ref{fig:flow12} and \ref{fig:flow16}. 
When we remove the data on the coarsest lattice $N_t=8$, we obtain $\Delta p$ consistent with zero. 
Thus, taking account of the systematic error due to the continuum extrapolation which is larger than the statistical error for the case of $\Delta p$, we think that $\Delta p$ is consistent with zero.

Our final results of method~2 for the latent heat are summarized in the right columns of~Table~\ref{tab3}, 
for which we adopt the results with the NNLO matching coefficients and the $\mu_0$ scale using the fit range~2.
Errors are statistical only.
As shown in~Fig.~\ref{fig:fitdep}, systematic errors due to the choice of the fit range are smaller than the statistical errors.

\subsection{Results of the SF$t$X method for the latent heat}
\label{final}

Finally, we summarize our results of the SF$t$X method for the latent heat. 
From~Table~\ref{tab3} and~Fig.~\ref{fig:fitdep}, we see that the results of method~1 and method~2 agree well with each other.
We take our central values from method~2 with the NNLO matching coefficients and the $\mu_0$ scale. 
We obtain for~$N_s/N_t=8$
\begin{eqnarray}
\Delta (\epsilon -3p) /T^4 &=& 1.061(39),
\\
\Delta (\epsilon +p) /T^4 &=& 1.135(43),
\\
\Delta \epsilon /T^4 &= &1.117(40),
\end{eqnarray}
and for $N_s/N_t=6$
\begin{eqnarray}
\Delta (\epsilon -3p) /T^4 &=& 1.282(39),
\\
\Delta (\epsilon +p) /T^4 &=& 1.372(43),
\\
\Delta \epsilon /T^4 &=& 1.349(38).
\end{eqnarray}
Systematic errors estimated from the differences between method~1 and method~2 as well as among different fit ranges are smaller than the statistical errors quoted in these equations.
Performing heuristic extrapolations of these two data points to the thermodynamic limit, 
we find $\Delta\epsilon/T^4 \sim 0.95 \pm 0.07$ by a $1/V$ linear fit and $\Delta\epsilon/T^4 \sim 1.23 \pm 0.03$ by a constant fit.

\begin{figure}[t]
\centering
\includegraphics[width=7.5cm]{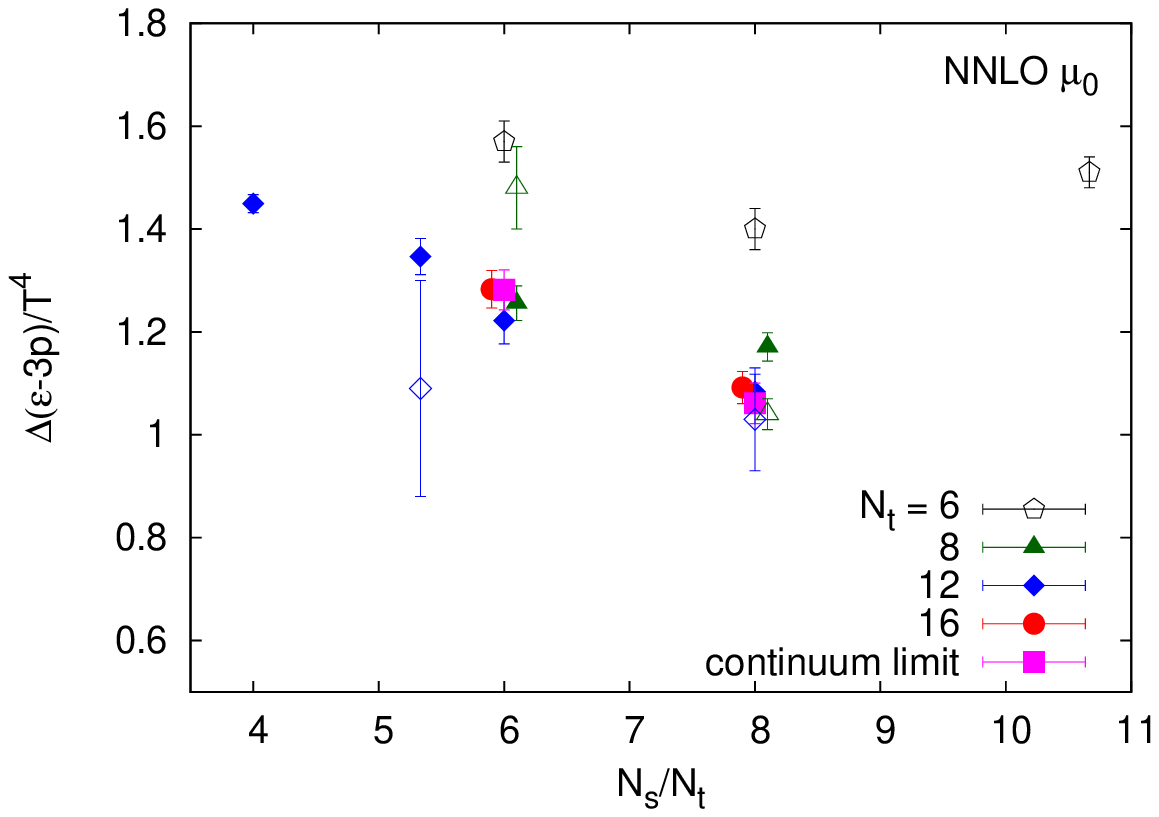}
\hspace{-2mm}
\includegraphics[width=7.5cm]{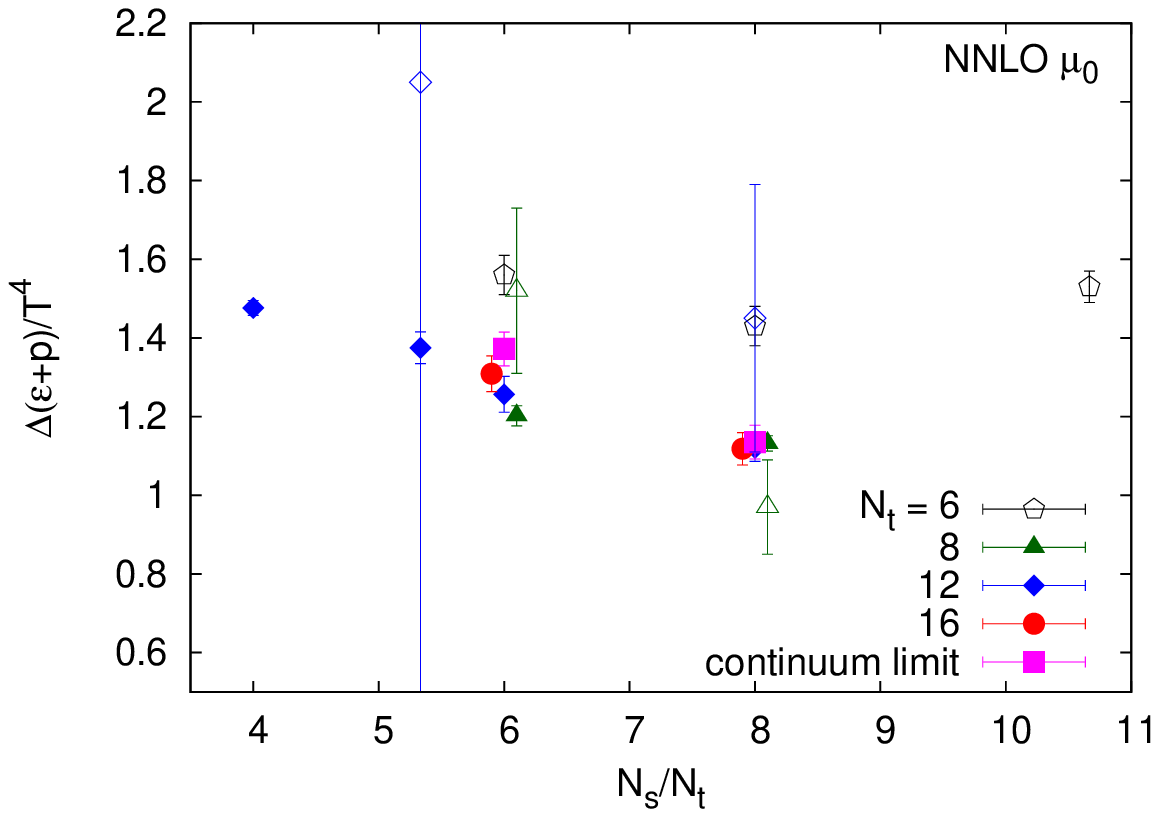}
\caption{Comparison of $\Delta (\epsilon-3p)/T^4$ (left) and $\Delta (\epsilon+p)/T^4$ (right) between the derivative method and SF$t$X method as functions of $N_s /N_t =\sqrt[3]{V} T_c$ computed by the NNLO calculation with $\mu=\mu_0$.
The open symbols are the result of the derivative method \cite{shirogane16}.
We shifted the value of $N_s/N_t$ to avoid overlapping symbols.
The original values of $N_s/N_t$ are 4, $16/3,$ 6, 8, and $32/3.$
}
\label{fig:vcompmd}
\end{figure}

\begin{figure}[t]
\centering
\includegraphics[width=7.5cm]{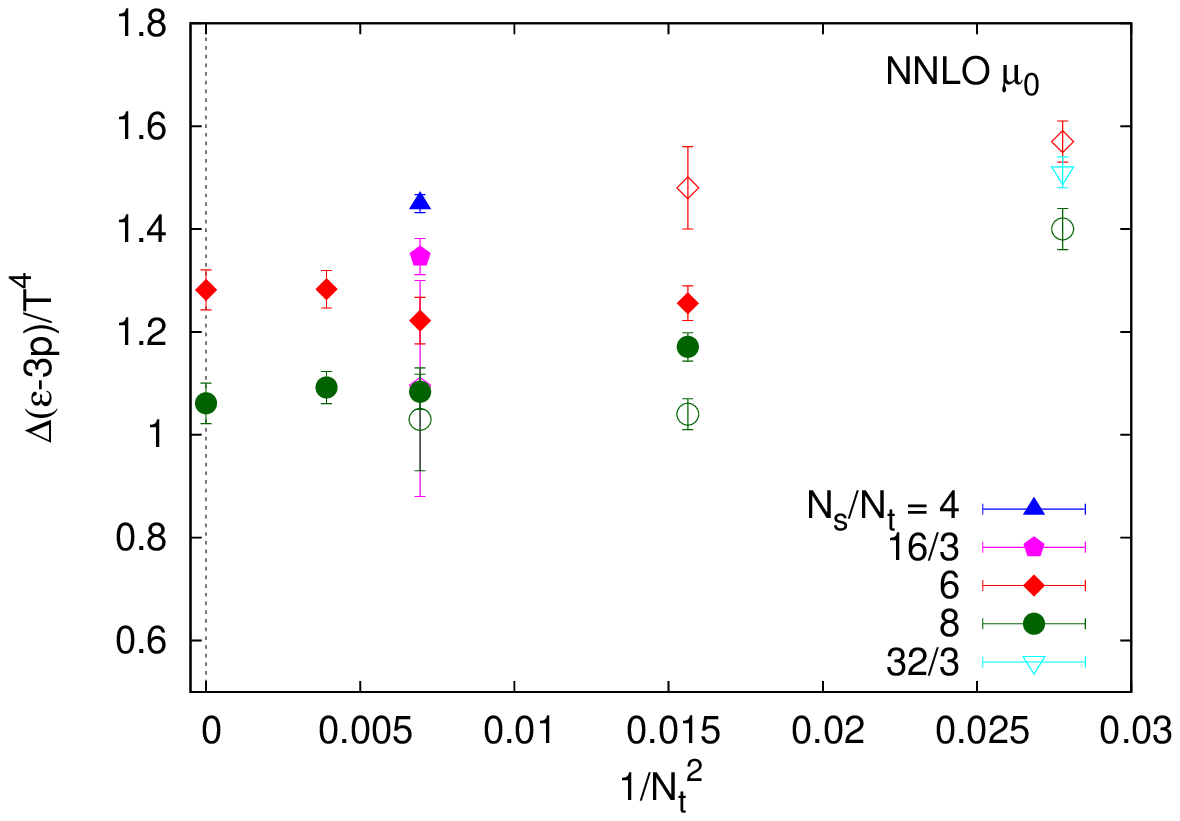}
\hspace{-2mm}
\includegraphics[width=7.5cm]{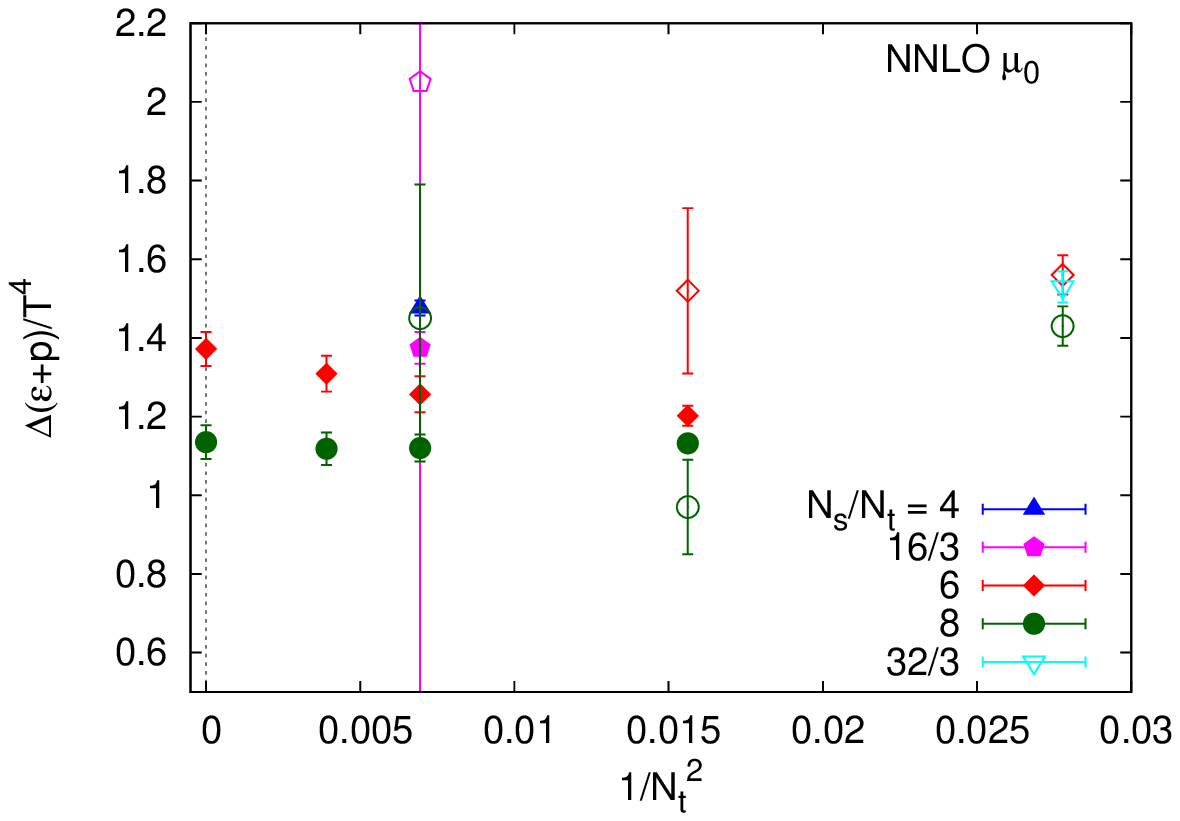}
\caption{Comparison of $\Delta (\epsilon-3p)/T^4$ (left) and $\Delta (\epsilon+p)/T^4$ (right) between the SF$t$X method and the derivative method as functions of $1/N_t^2 =a^2 T_c^2$.
Filled symbols are the results of the SF$t$X method with the NNLO matching coefficients and the $\mu_0$ scale, and open symbols are the results of the derivative method given in~Ref.~\cite{shirogane16}.
}
\label{fig:acompmd}
\end{figure}

\subsection{Comparison with the derivative method}
\label{derivative}

Finally, we compare our results of the SF$t$X method with those obtained by the derivative method~\cite{shirogane16}.
The computational cost for the SF$t$X method is much smaller than that for the derivative method, since the nonperturbative calculation of the Karsch coefficients needed in the derivative method requires large systematic study with high statistics. 
We note that the statistical errors are reduced with the SF$t$X method thanks the smearing nature of the gradient flow and also the avoidance of the nonperturbative Karsch coefficients.

In~Ref.~\cite{shirogane16}, using the derivative method, 
a rather low value of $\Delta \epsilon /T^4=0.75 \pm 0.17$ was obtained for the latent heat in the continuum limit using data at $N_t =6$, 8 and 12.
In~Ref.~\cite{shirogane16}, because the spatial volume dependence in the latent heat was found to be small in comparison with the statistical errors, which are much larger than the errors in the current study mainly due to the error from the nonperturbative Karsch coefficients, spatial volume dependence was assumed to be absent. 

To avoid uncertainties due to the continuum and infinite spatial volume extrapolations, we directly compare the results of $\Delta(\epsilon - 3p)/T^4$ and $\Delta(\epsilon + p)/T^4$ at finite lattice spacings and spatial volumes in~Fig.~\ref{fig:vcompmd}.
The horizontal axis is $N_s /N_t = \sqrt[3]{V} T_c$.
The filled symbols are the results of the SF$t$X method in the $t\to0$ limit with the NNLO matching coefficients and the $\mu_0$ scale, obtained on the lattices with $N_t =8$ (red diamond), 12 (green circle) and 16 (blue triangle), respectively (Table~\ref{tab2}).
The magenta square symbols are for the results in the continuum limit at $N_s /N_t =6$ and 8 by method 2. 
The open symbols are the results of the derivative method obtained on the lattices with $N_t =6$ (black pentagon), 8 (red diamond) and 12 (green circle), respectively~\cite{shirogane16}.
The large statistical errors for $\Delta(\epsilon + p)/T^4$ with the derivative method are caused by the large errors of nonperturbative Karsch coefficients which increase rapidly with $N_t$.
From these plots, we find that the SF$t$X method and the derivative method lead to quite similar results at each~$N_s /N_t$.

As discussed in Sec.~\ref{continuum}, the lattice spacing dependence of the latent heat by the SF$t$X method is small for $N_t=8$, 12, 16. 
The lattice spacing dependence in the results by the derivative method is also small for $N_t = 8$ and 12.
The left and right panels of~Fig.~\ref{fig:acompmd} are the results of $\Delta(\epsilon - 3p)/T^4$ and $\Delta(\epsilon + p)/T^4$ as functions of $1/N_t^2 =a^2 T_c^2$, i.e., the lattice spacing squared normalized by $T_c$.
The filled and open symbols are the results by the SF$t$X method and the derivative method, respectively.
The red diamonds and green circles are for $N_s /N_t =6$ and 8, respectively. 
We note that the data at $N_t=6$ with the derivative method, shown at the right end of these plots, show deviation from the general tendency of data on finer lattices, suggesting a large discretization error at $N_t=6$. 
As suggested from Fig.~9 of Ref.~\cite{shirogane16}, by removing the data at $N_t=6$, the derivative method may also lead to a larger latent heat in the continuum, though a reliable extrapolation is not possible with only two data points.

\begin{figure}[th]
\centering
\includegraphics[width=8.0cm]{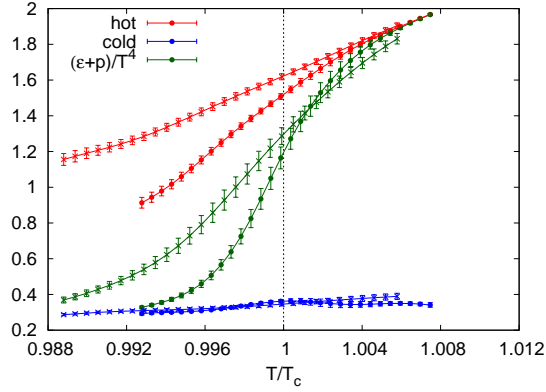}
\caption{$(\epsilon+p)/T^4$ by the SF$t$X method calculated using configurations in the hot phase (red), the cold phase (blue) and all configurations (green) with the NNLO matching coefficients and the $\mu_0$ scale, obtained on the $48^3 \times 8$ (cross) and $64^3 \times 8$ (circle) lattices. 
}
\label{fig:hisflow}
\end{figure}

\begin{figure}[t]
\centering
\includegraphics[width=8.0cm]{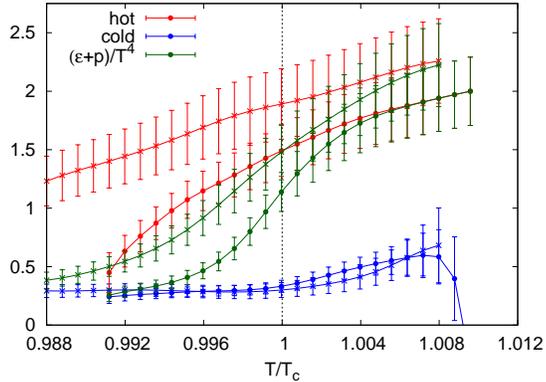}
\caption{The same as Fig.~\ref{fig:hisflow}, but by the derivative method. 
}
\label{fig:hisder}
\end{figure}

\section{Hysteresis of the energy density near $T_c$}
\label{hysteresis}

In Sec.~\ref{results}, we found that the latent heat is clearly dependent on the spatial volume of the system. 
At a first-order phase transition point, however, because the correlation length does not diverge, the spatial volume dependence should be mild when the volume is sufficiently large.
To find the origin of the spatial volume dependence in our latent heat, we study $(\epsilon + p) /T^4$ separately in the hot and cold phases at temperatures near $T_c$ using the multipoint reweighting method. 

The results obtained on $48^3 \times 8$ and $64^3 \times 8$ lattices are plotted by the cross and circle symbols in Fig.~\ref{fig:hisflow}. 
The flow time is $t/a^2=1.4$ --- as shown in Sec.~\ref{tto0}, the result is about the same as that in the $t\to0$ limit at this value of $t$. 
The NNLO matching coefficients with the $\mu_0$ scale are used.
The red symbols are for the results obtained in the hot phase and the blue symbols are for the cold phase, while the green symbols show the results without the phase separation.
The horizontal axis is the temperature $T=1/(N_t a)$ normalized by the critical temperature $T_c$, 
where the relation between the lattice spacing $a$ and $\beta$ is determined by the critical point $\beta_c$ as function of $N_t$~\cite{shirogane16}. 
The results for the hot phase below $T_c$ and for the cold phase above $T_c$ are those obtained in corresponding metastable states.
We find that the spatial volume dependence appears only in the (metastable) hot phase around $T_c$, while no apparent volume dependence is visible in the (metastable) cold phase.
We thus conclude that the spatial volume dependence of the latent heat is due to that in the contribution of the hot phase.
From this figure, we also note that latent heat is sensitive to the value of the critical point $\beta_c$.
A careful determination is required for $\beta_c$.

In Fig.~\ref{fig:hisder}, we show the corresponding plot by the derivative method, obtained on the $48^3 \times 8$ (cross) and $64^3 \times 8$ (circle) lattices~\cite{ejiri98,shirogane16}. 
In this calculation, because our nonperturbative determination of the Karsch coefficients is possible only at $\beta_c$~\cite{ejiri98,shirogane16},
we use the Karsch coefficients obtained at $\beta_c$ at other values of $\beta$ around $\beta_c$. 
We see similar spatial volume dependence of the results in the (metastable) hot phase, 
though the statistical errors are large.

To draw a definite statement about the nature of the metastable state, we need more statistics, because the effective number of configurations for the metastable states after the phase separation is not large in the reweighting calculation when $T$ is not sufficiently close to $T_c$. 
Further study with higher statistics is needed to precisely determine the volume dependence of the hot phase, and thus to carry out a reliable extrapolation of the latent heat to the thermodynamic limit.

\section{Conclusions}
\label{conclusion}

To study the latent heat and the pressure gap at the first-order deconfining phase transition temperature in the SU(3) Yang-Mills theory,
we performed simulations on lattices with various spatial volumes and lattice spacings around the phase transition temperature $T_c$.
Separating generated configurations into hot and cold phases using the value of the Polyakov loop 
and adjusting the temperature to $T_c$ by the multipoint reweighting method, 
we calculated the energy density and the pressure gap in each phase at $T_c$, using the small flow-time expansion (SF$t$X) method based on the gradient flow. 

We confirmed that the pressure gap between the hot and cold phases is consistent with zero, as expected from the dynamical balance of two phases at~$T_c$. 
Our results for the latent heat in the continuum limit are summarized in~Sec.~\ref{final}.
We obtained $\Delta \epsilon /T^4 = 1.117 \pm 0.040$ for the spatial volume corresponding to the aspect ratio $N_s/N_t=8$, 
and $1.349 \pm 0.038$ for $N_s/N_t=6$.
From a study of hysteresis curves for $(\epsilon + p) /T^4$ around $T_c$, we note that $(\epsilon + p) /T^4$ in the (metastable) deconfined phase is sensitive to the spatial volume, 
while that in the confined phase is insensitive.
The value of $(\epsilon + p) /T^4$ decreases as the volume increases in the (metastable) deconfined phase. 
Study with higher statistics is needed to precisely determine the latent heat in the thermodynamic limit.

Furthermore, using the systematic data at various lattice spacings, we examined the effect of alternative procedures in the SF$t$X method. 
We compared two orders of the double extrapolation $(a,t)\to(0,0)$ --- method~1 (first $t\to0$ and then $a\to0$) and method~2 (first $a\to0$ and then $t\to0$). 
We also studied the effects of the renormalization scale and the truncation of higher-order terms in the matching coefficients.
We confirmed that the final results with alternative procedures are all consistent with each other. 
We also found that the use of NNLO matching coefficients improves the signal of the latent heat with the SF$t$X method.

\section*{Acknowledgements}

We thank other members of the WHOT-QCD Collaboration for discussions and comments.
This work was in part supported by JSPS KAKENHI Grant Numbers JP20H01903,
JP19H05146, JP19H05598, JP19K03819, JP18K03607, JP17K05442 and JP16H03982, 
and the Uchida Energy Science Promotion Foundation.
This research used computational resources of COMA, Oakforest-PACS, and Cygnus provided by the Interdisciplinary Computational Science Program of Center for Computational Sciences, University of Tsukuba,
K and other computers of JHPCN through the HPCI System Research Projects (Project ID:hp17208, hp190028, hp190036, hp200089) and JHPCN projects (jh190003, jh190063, jh200049), OCTOPUS at Cybermedia Center, Osaka University, ITO at Research Institute for Information Technology, Kyushu University, and Grand Chariot at Information Initiative Center, Hokkaido University, 
SR16000 and BG/Q by the Large Scale Simulation Program of High Energy Accelerator Research Organization (KEK) (Nos.\ 14/15-23, 15/16-T06, 15/16-T-07, 15/16-25, 16/17-05). 
The authors also thank the Yukawa Institute for Theoretical Physics at Kyoto University for the workshop YITP-W-19-09.


\appendix

\section{Energy-momentum tensor with NLO matching coefficients}
\label{sec:leading}

\begin{figure}[t]
\centering
\includegraphics[width=7.5cm]{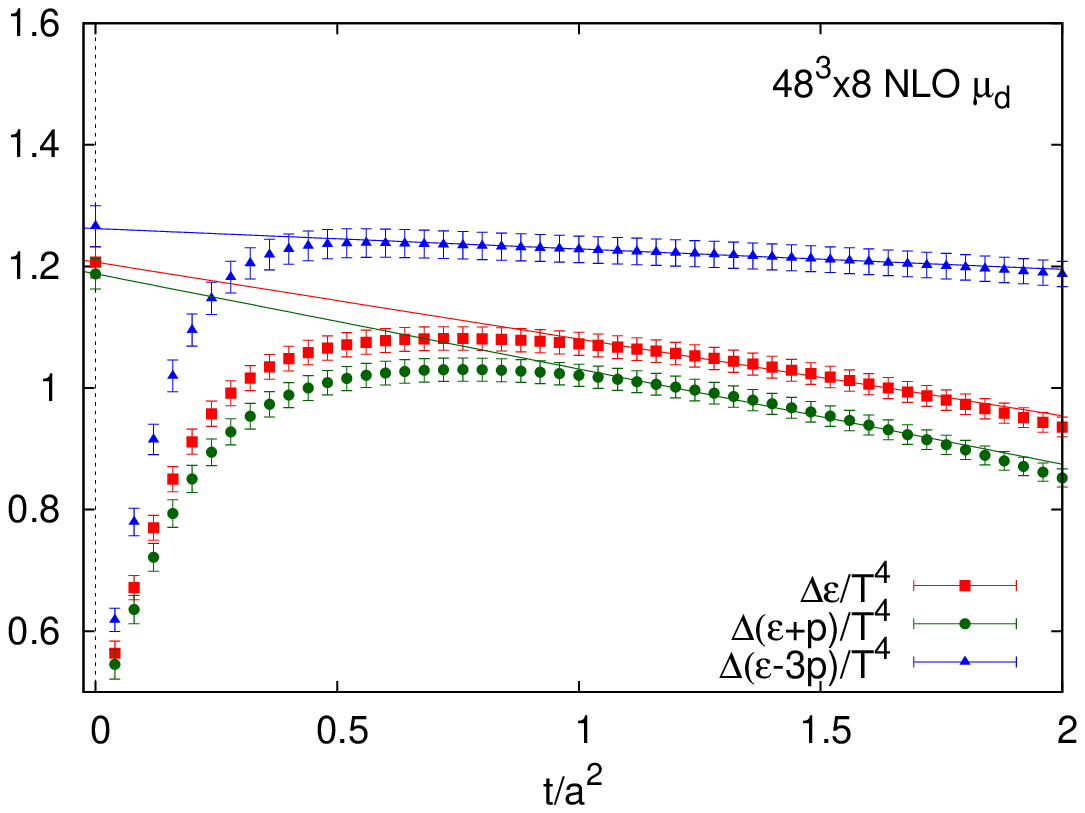}
\hspace{1mm}
\includegraphics[width=7.5cm]{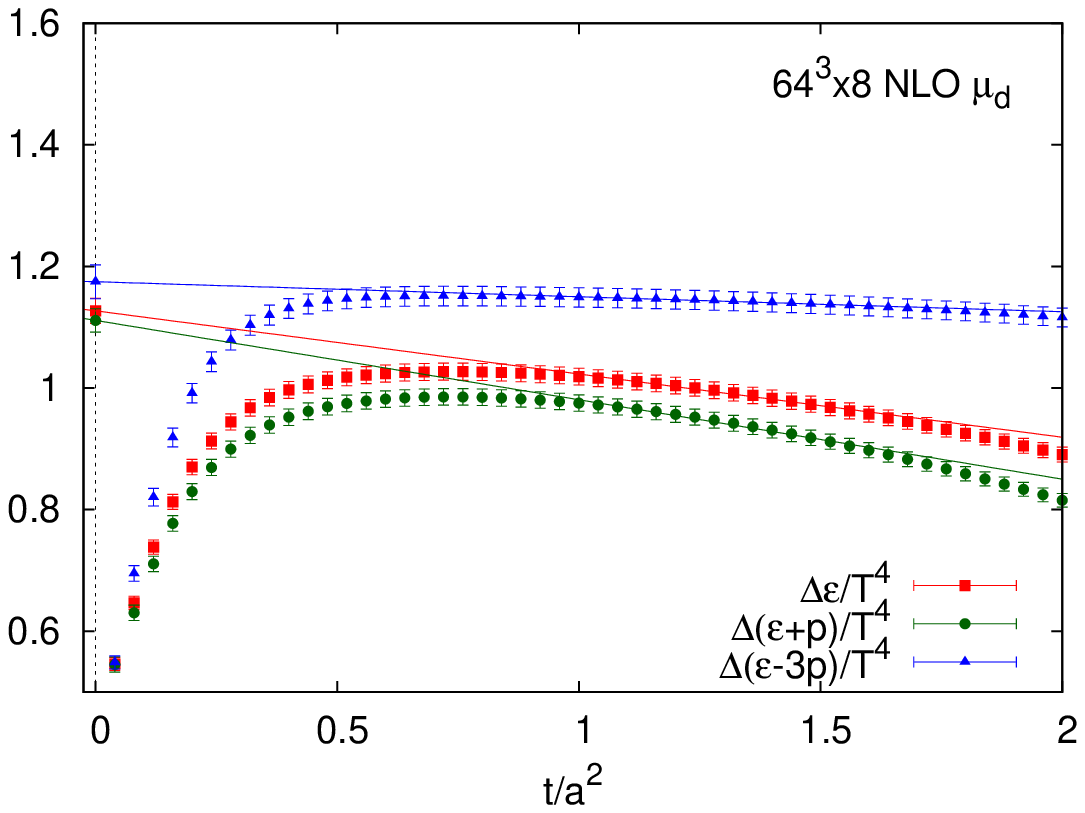}
\caption{
$\Delta (\epsilon -3p)/T^4$ (blue), $\Delta (\epsilon +p)/T^4$ (green) and $\Delta \epsilon /T^4$ (red) calculated on $48^3 \times 8$ (left) and $64^3 \times 8$ (right) lattices with the NLO matching coefficients. 
The horizontal axis is flow time in a lattice unit $t/a^2$.
The straight lines are the results of $t \to 0$ linear extrapolation.
}
\label{fig:flow8lo}
\end{figure}

\begin{figure}[t]
\centering
\includegraphics[width=7.5cm]{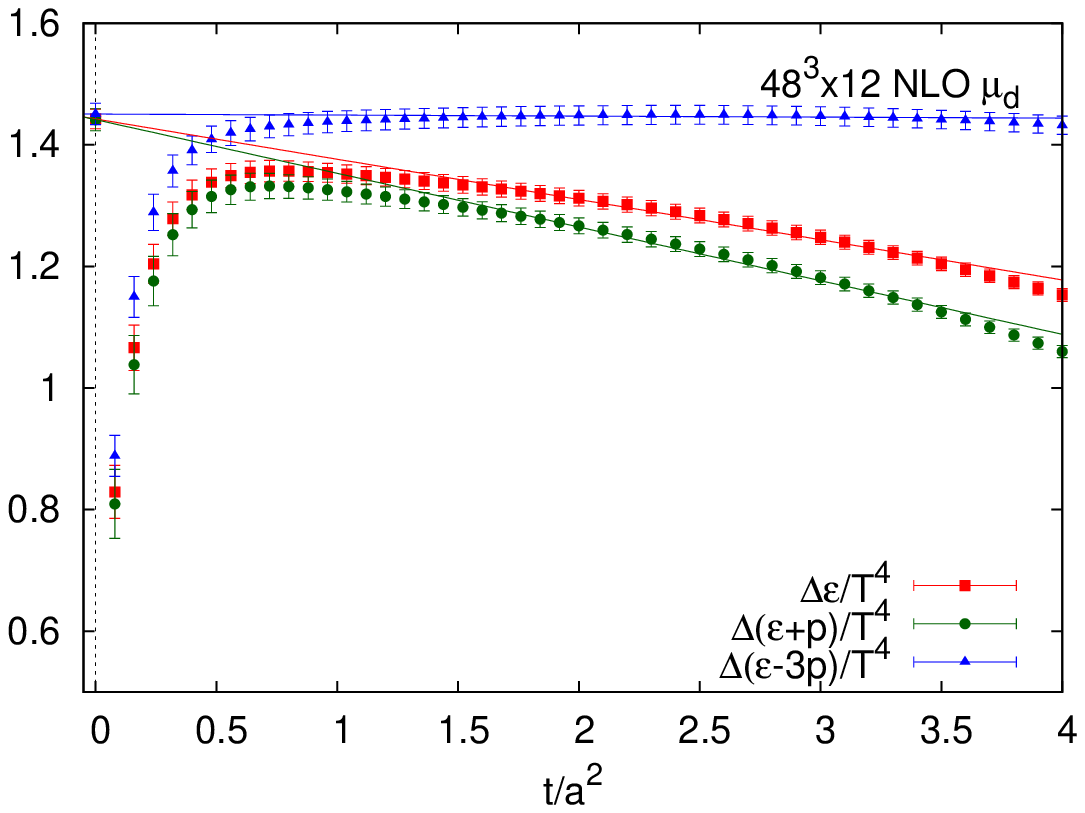}
\hspace{1mm}
\includegraphics[width=7.5cm]{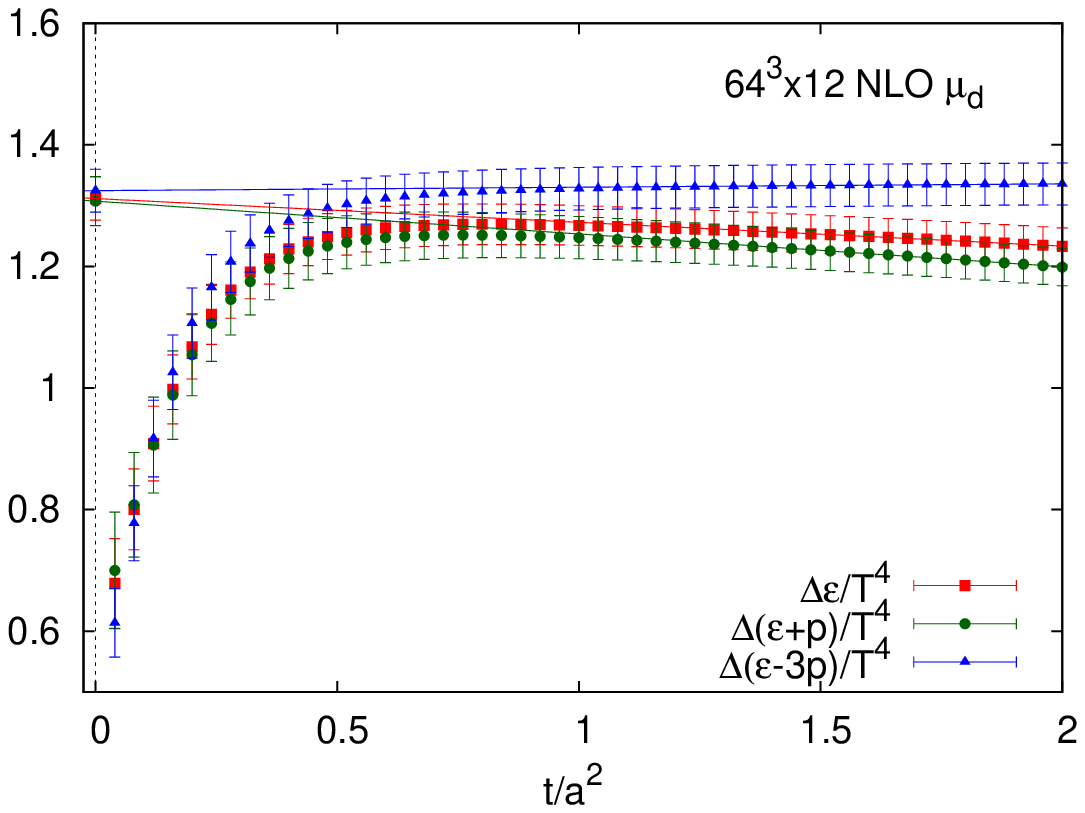}
\\
\includegraphics[width=7.5cm]{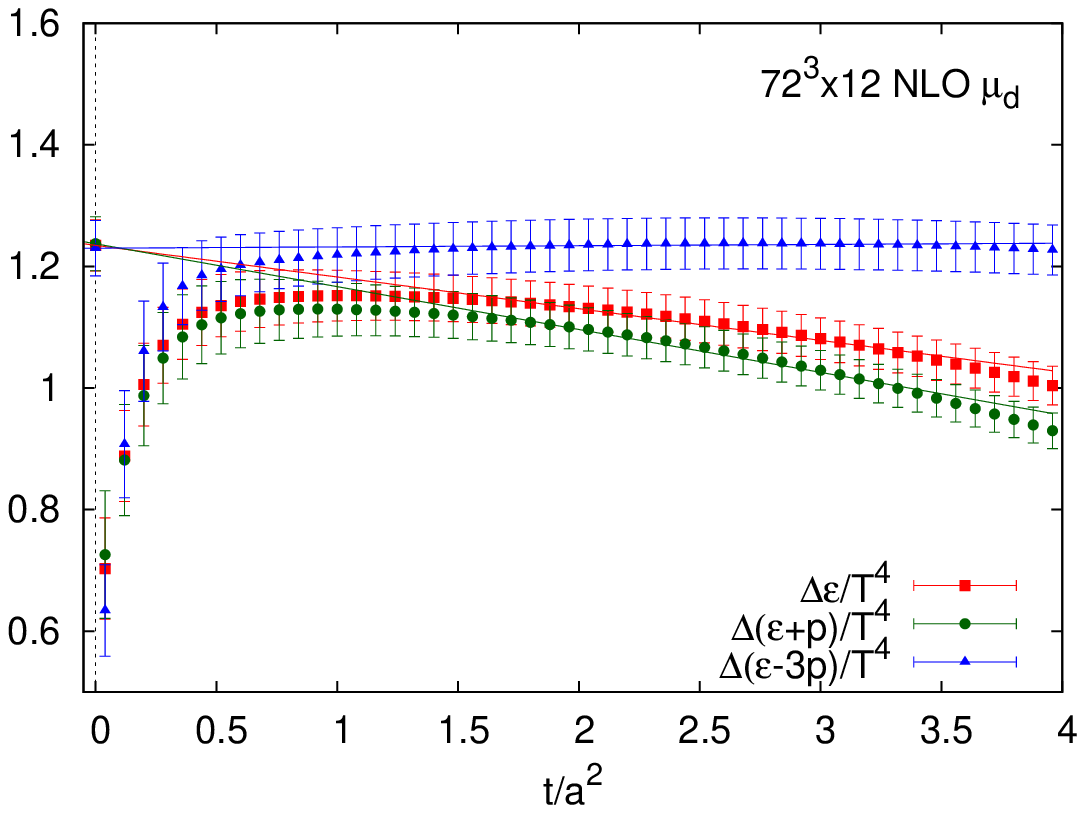}
\hspace{1mm}
\includegraphics[width=7.5cm]{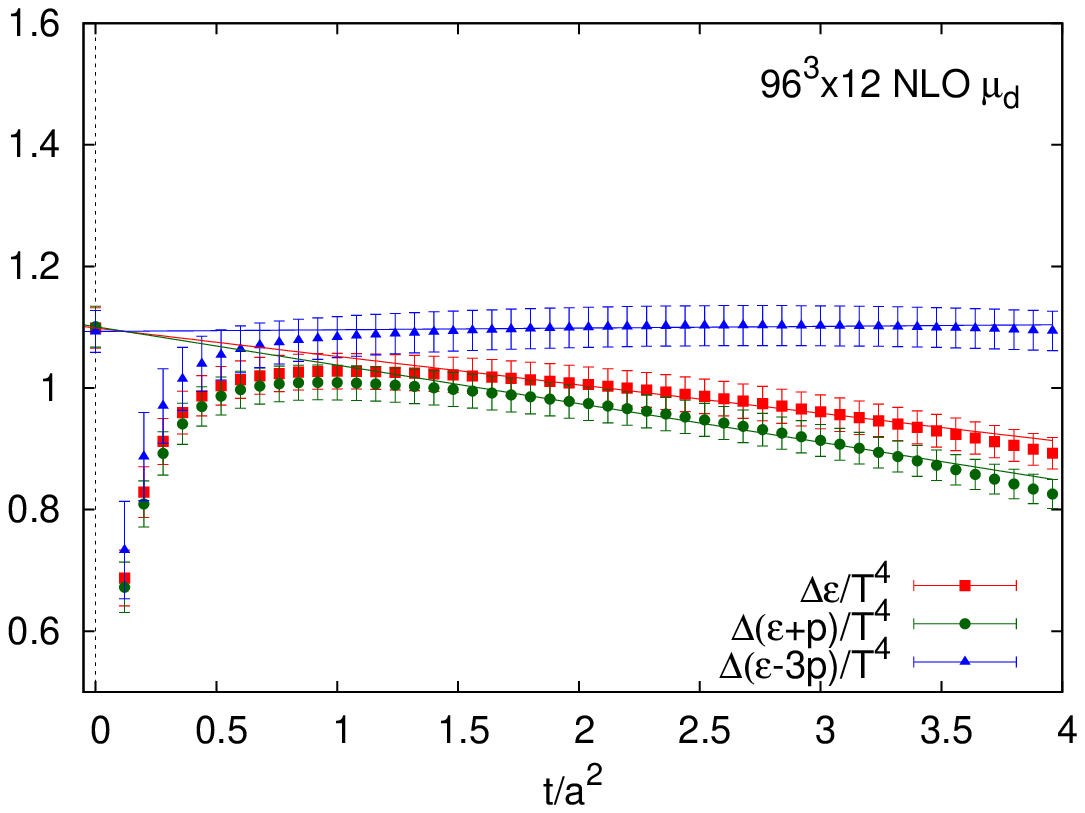}
\caption{The same as Fig.~\ref{fig:flow8lo}, but on the $48^3 \times 12$ (top left), $64^3 \times 12$ (top right), $72^3 \times 12$ (bottom left) and $96^3 \times 12$ (bottom right) lattices.
}
\label{fig:flow12lo}
\end{figure}

\begin{figure}[t]
\centering
\includegraphics[width=7.5cm]{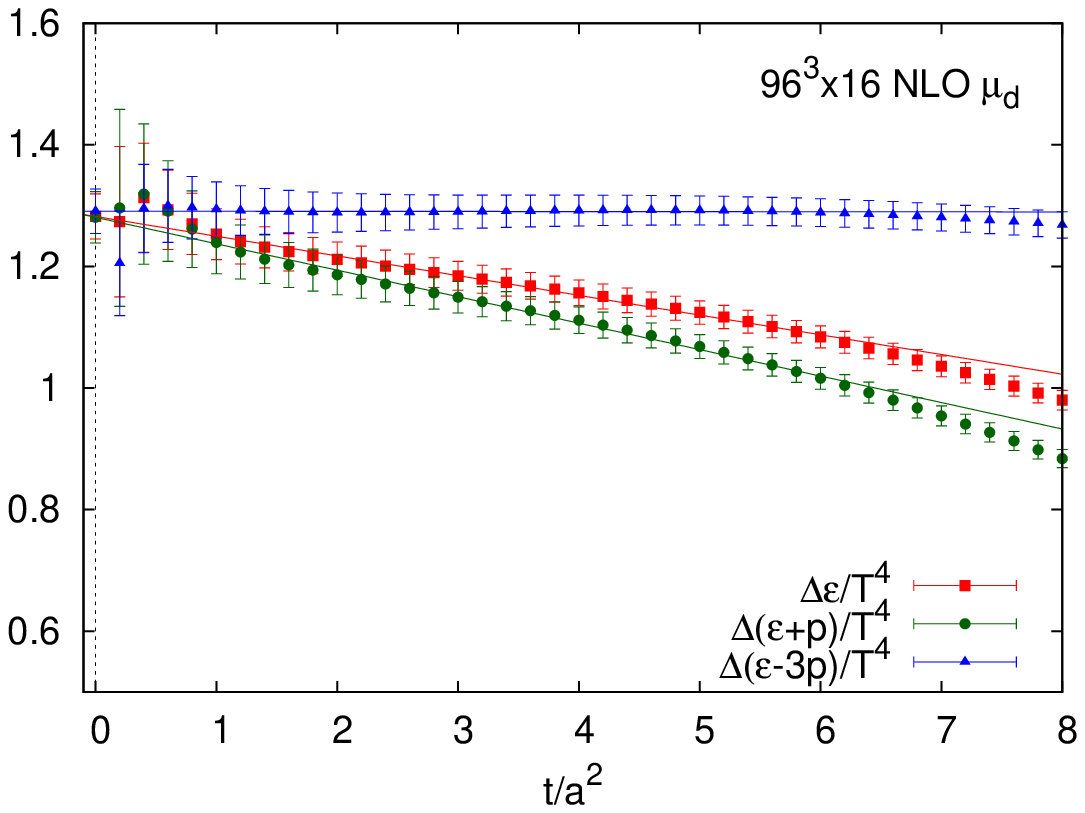}
\hspace{1mm}
\includegraphics[width=7.5cm]{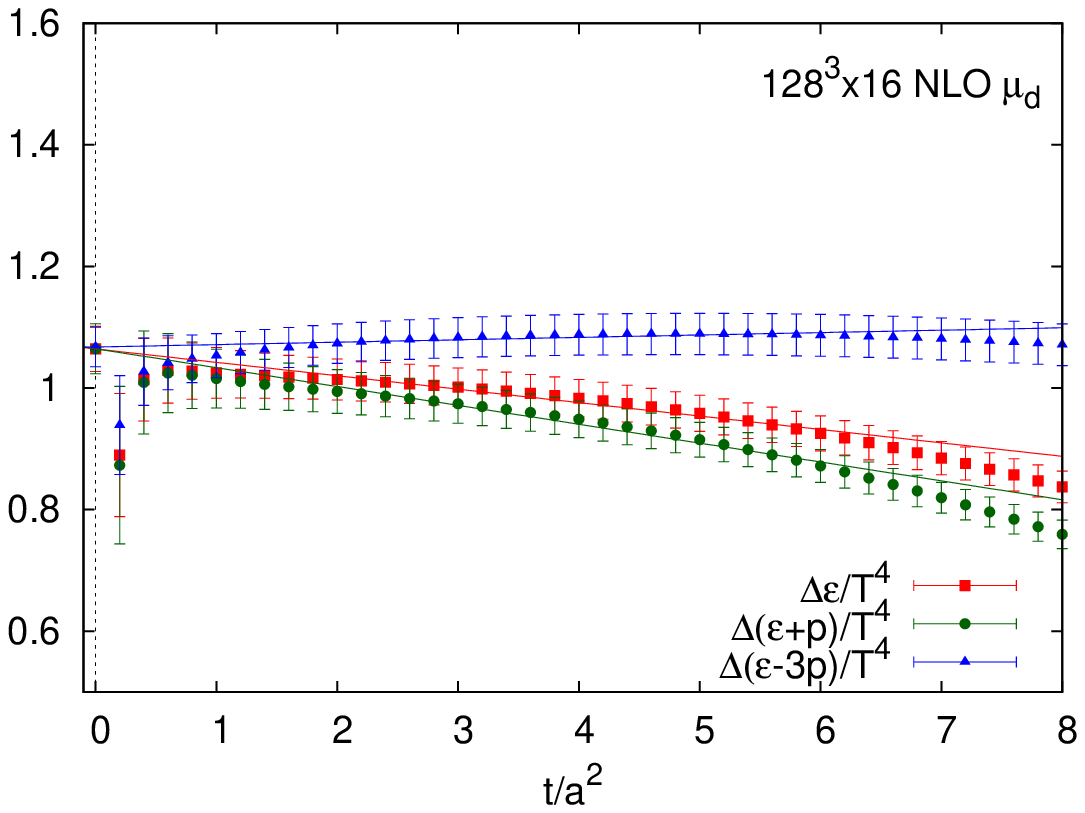}
\caption{The same as Fig.~\ref{fig:flow8lo}, but on the $96^3 \times 16$ (left) and $128^3 \times 16$ (right) lattices.
}
\label{fig:flow16lo}
\end{figure}

In this appendix, we calculate the energy-momentum tensor using the NLO matching coefficients, following the original paper of the SF$t$X method \cite{Suzuki:2013gza}:
\begin{eqnarray}
T_{\mu\nu}^R(x)
=\lim_{t\to0}\left\{\frac{1}{\alpha_U(t)}U_{\mu\nu}(t,x)
   +\frac{\delta_{\mu\nu}}{4\alpha_E(t)}
   \left[E(t,x)-\left\langle E(t,x)\right\rangle_0 \right]\right\},
\label{eq:(4)}
\end{eqnarray}
where
\begin{eqnarray}
  \alpha_U(t) &=& \frac{1}{c_1(t)} \;=\; 
g^2
\left[1+2 \frac{\beta_0}{(4\pi)^2} \bar{s}_1 g^2 +O(g^4)\right] 
\\
\alpha_E(t) &=& \frac{1}{c_2(t)} \;=\; 
\frac{(4\pi)^2}{2 \beta_0}\left[1+
2\frac{\beta_0}{(4\pi)^2} \bar{s}_2 g^2 +O(g^4)\right] .
\end{eqnarray}
For the NLO calculation in this appendix, following Ref.~\cite{Suzuki:2013gza}, we keep terms up to the next-to-leading order only in $\alpha_U(t)$ and $\alpha_E(t)$
and adopt the $\mu_d$ scale. 
We then have
\begin{eqnarray}
\bar{s}_1 &=& \frac{7}{22}+\frac{1}{2}\gamma_E-\ln2\simeq -0.08635752993, \\
\bar{s}_2 &=& \frac{21}{44}-\frac{\beta_1}{2\beta_0^2}
=\frac{27}{484} \simeq 0.05578512397.
\end{eqnarray}

\begin{figure}[t]
\centering
\includegraphics[width=7.5cm]{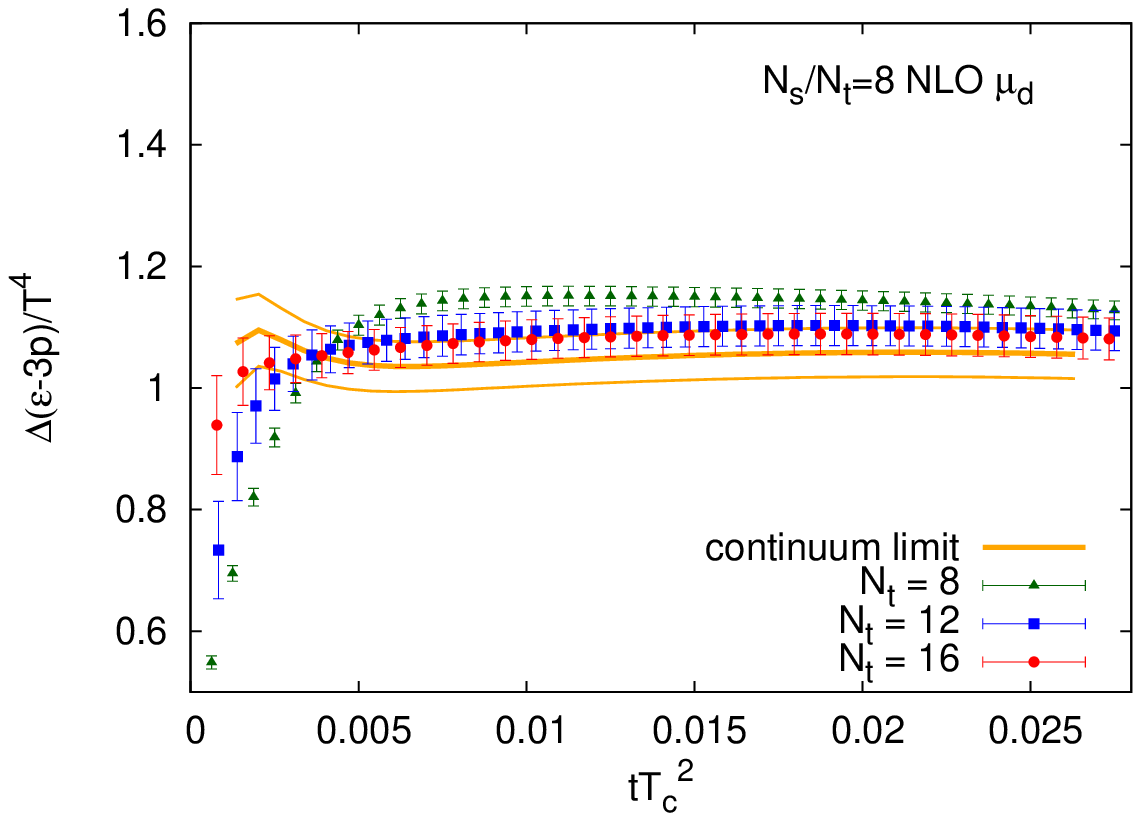}
\hspace{1mm}
\includegraphics[width=7.5cm]{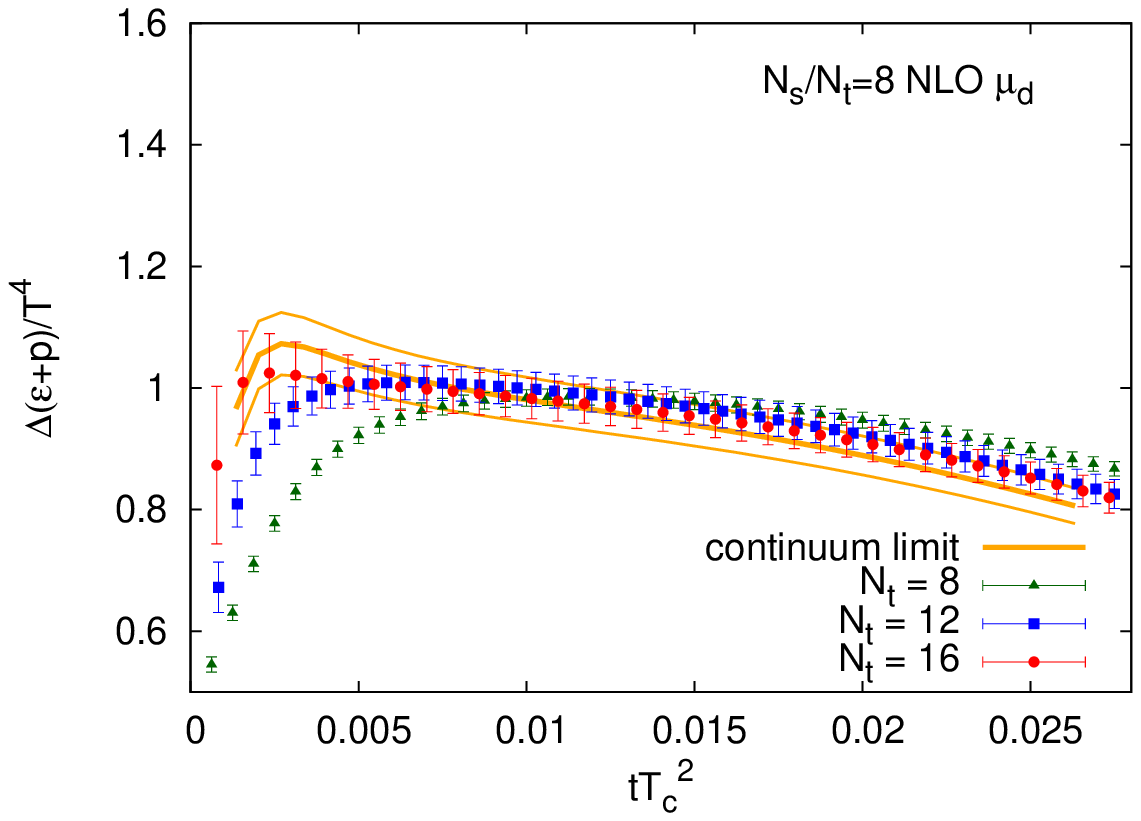}
\caption{$\Delta (\epsilon -3p)/T^4$ (left) and $\Delta (\epsilon +p)/T^4$ (right) with the NLO matching coefficients, obtained on the $N_s/N_t=8$ lattices.
Three orange curves represent the result in the continuum limit. 
(Thick orange curves represent the central values and thin curves represent the range of their errors.)
}
\label{fig:adep8lo}
\end{figure}

\begin{figure}[t]
\centering
\includegraphics[width=7.5cm]{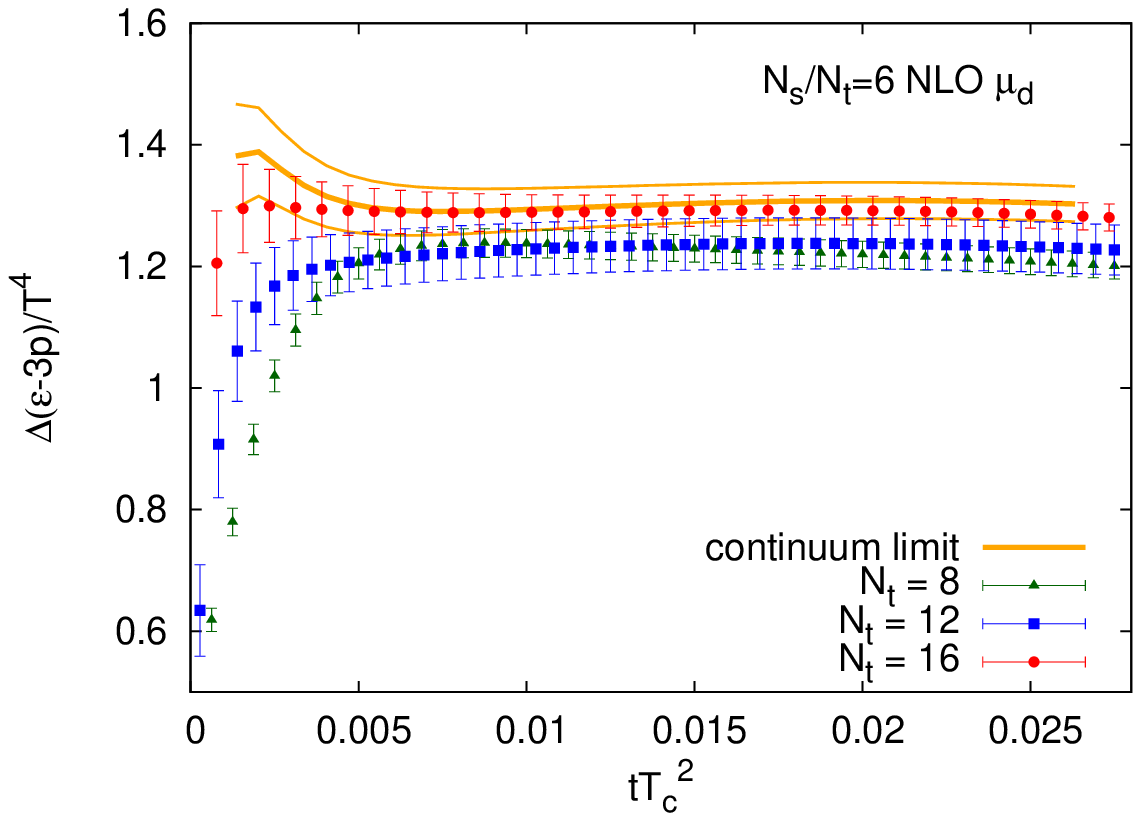}
\hspace{1mm}
\includegraphics[width=7.5cm]{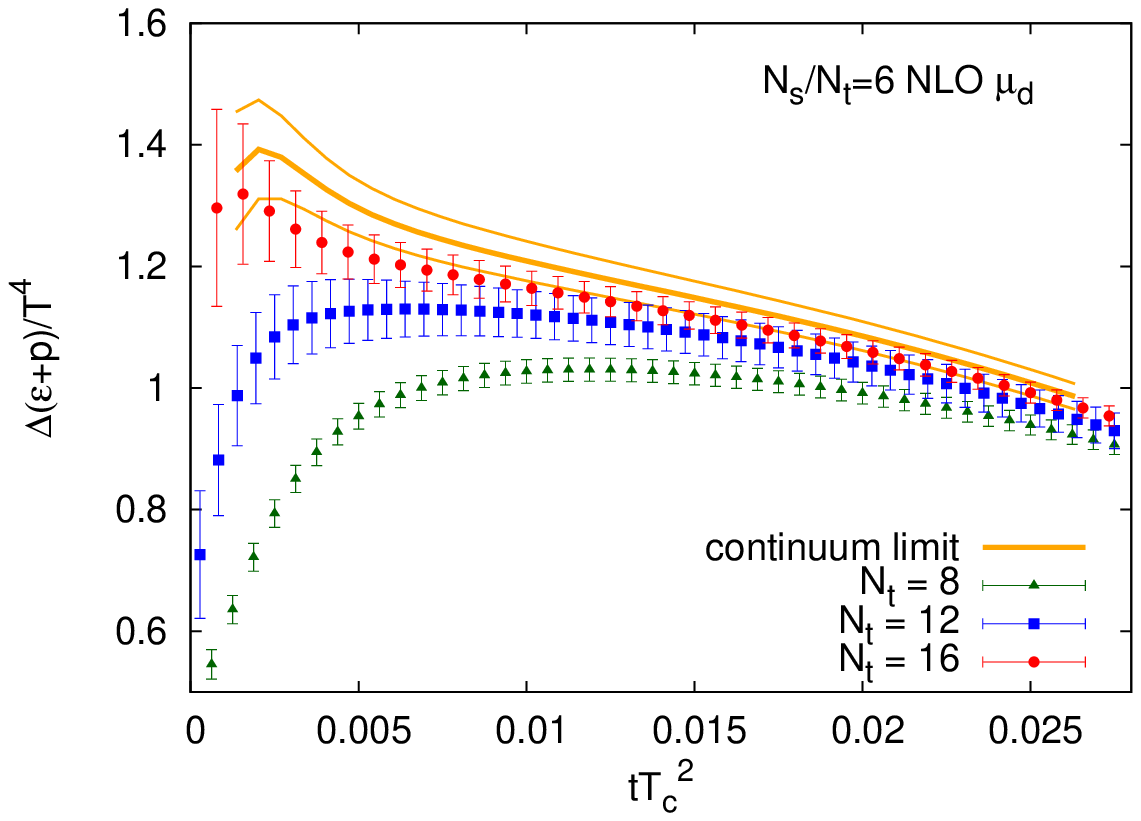}
\caption{The same as Fig.~\ref{fig:adep8lo}, but on the $N_s/N_t=6$ lattices.
}
\label{fig:adep6lo}
\end{figure}

The details of the simulations are the same as the NNLO calculations given in Sec.~\ref{parameter}.
The results of $\Delta(\epsilon -3 p)/T^4$, $\Delta(\epsilon + p)/T^4$ and $\Delta \epsilon/T^4$ are plotted as functions of flow time $t/a^2$ 
in~Figs.~\ref{fig:flow8lo}, \ref{fig:flow12lo}, and~\ref{fig:flow16lo}.
The blue, green, and red symbols are the results of $\Delta(\epsilon -3 p)/T^4$, $\Delta(\epsilon + p)/T^4$, and $\Delta \epsilon/T^4$, respectively.

\underline{\textit{Method 1: $t\to0$ followed by $a\to0$}} \hspace{2mm}
The results of the $t \to 0$ extrapolation on each finite lattice are shown in~Figs.~\ref{fig:flow8lo}, \ref{fig:flow12lo}, and~\ref{fig:flow16lo} by the lines, and the results in the $t\to0$ limit are summarized in~Table~\ref{tab2}.
Continuum extrapolation of the results given in~Table~\ref{tab2} is discussed in~Sec.~\ref{tto0}. 

\underline{\textit{Method 2: $a\to0$ followed by $t\to0$}} \hspace{2mm}
To carry out the continuum extrapolation at each $t$ in physical units, we replot the data shown in~Figs.~\ref{fig:flow8lo}, \ref{fig:flow12lo}, and~\ref{fig:flow16lo} as a function of $tT_c^2$. 
For the aspect ratio $N_s/N_t=8$ and 6, we obtain Figs.~\ref{fig:adep8lo} and \ref{fig:adep6lo}. 
The red, green and blue symbols are the results of $N_t=8$, $12$ and $16$.
These figures correspond to~Figs.~\ref{fig:adep8} and \ref{fig:adep6} with the NNLO matching coefficients. 
We find that the difference between the results with different lattice spacing becomes smaller as the flow time $t$ increases.
Repeating the analyses of~Sec.~\ref{continuum}, we obtain the continuum limit shown by the orange curves in Figs.~\ref{fig:adep8lo} and \ref{fig:adep6lo}. 
These results of $\Delta(\epsilon + p)/T^4$ and  $\Delta(\epsilon -3 p)/T^4$ in the continuum limit are summarized in Fig.~\ref{fig:cont} as functions of $t$, 
and the results of the $t \to 0$ extrapolation are given in Table \ref{tab3}.

\section{Choice of lattice field strength operator}
\label{plaquette}

\begin{figure}[t]
\centering
\includegraphics[width=7.5cm]{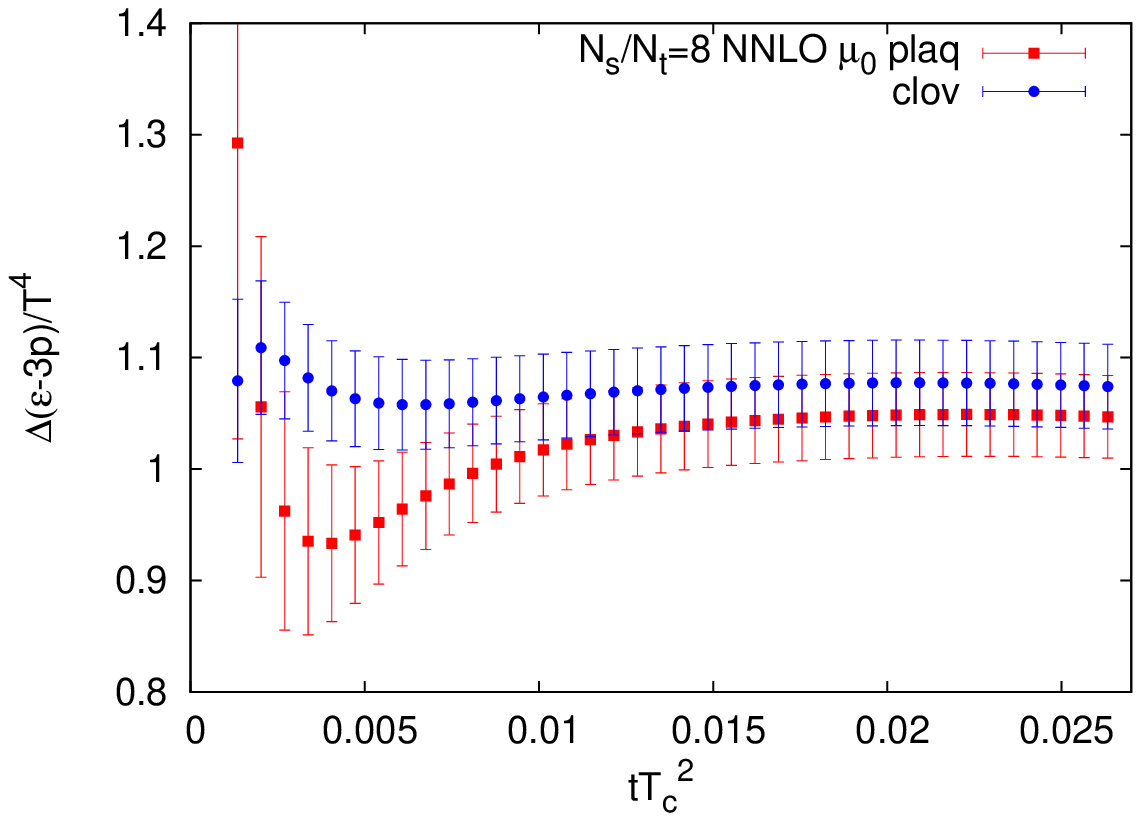}
\hspace{1mm}
\includegraphics[width=7.5cm]{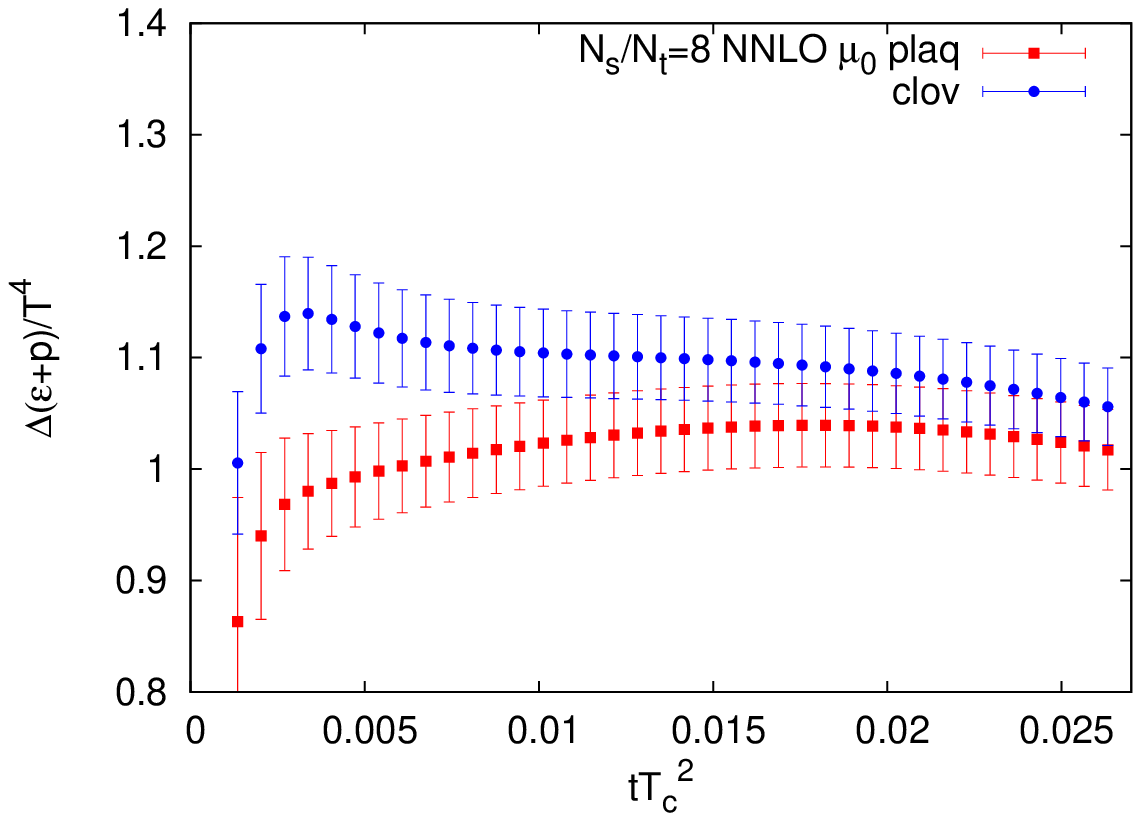}
\\
\includegraphics[width=7.5cm]{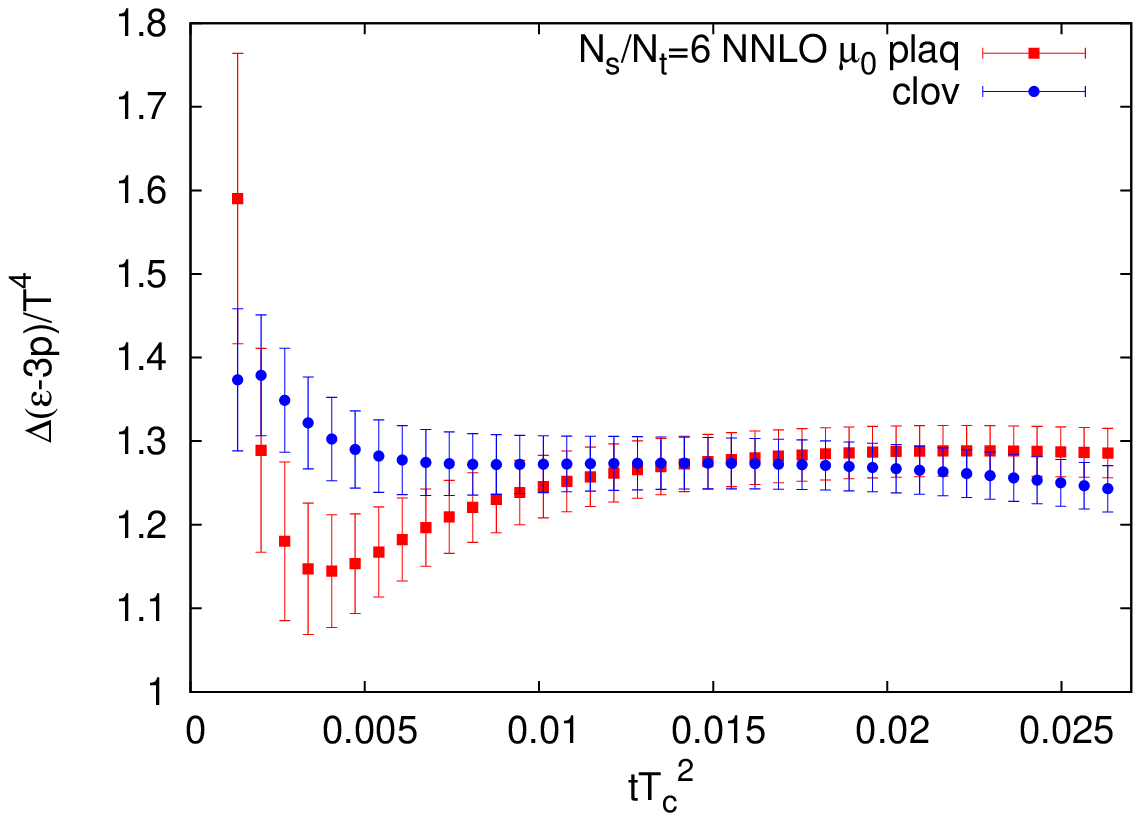}
\hspace{1mm}
\includegraphics[width=7.5cm]{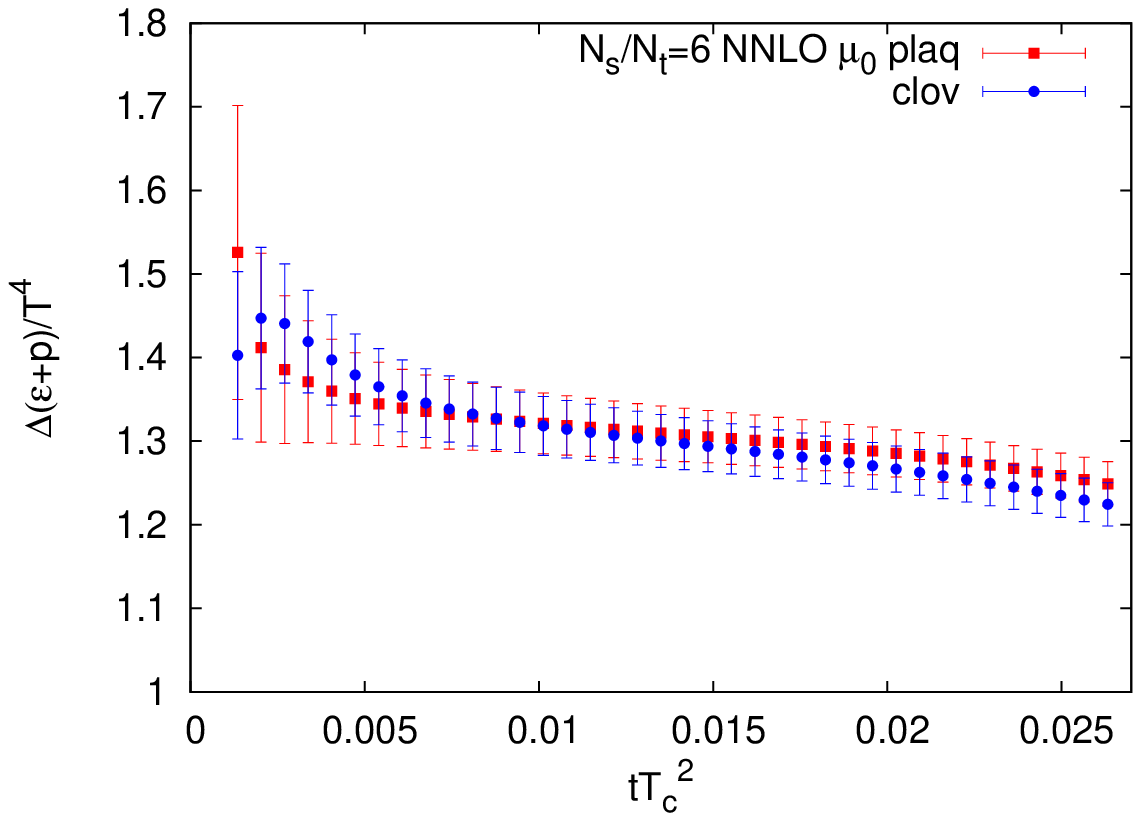}
\caption{
$\Delta (\epsilon-3p)/T^4$ using clover-shaped operator (red) and plaquette (blue) in the continuum limit for $N_s/N_t=8$ (upper) and 6 (lower)
by the NNLO calculation with $\mu=\mu_0$, respectively.
}
\label{fig:plaqclov}
\end{figure}

In this appendix, we test the influence of lattice operators for the field strength $G_{\mu \nu}(t,x)$ to define the operators $U_{\mu\nu}(t,x)$ and $E(t,x)$ in~Eqs.~(\ref{eq:UE1}) and (\ref{eq:UE}).
Two major choices for $G_{\mu \nu}(t,x)$ are the plaquette and clover-shaped lattice operators.

In the left and right panels of~Fig.~\ref{fig:plaqclov}, we compare the results of 
$\Delta(\epsilon-3p)/T^4$ and $\Delta(\epsilon+p)/T^4$ in the continuum limit obtained by the clover-shaped and plaquette operators adopting method~2 discussed in Sec.~\ref{sec:method12}.
The upper and lower panels are the results for $N_s/N_t=8$ and 6, respectively. 
The blue and red symbols are for results with the clover-shaped and plaquette operators, respectively. 
The two results agree well with each other at sufficiently large $t$. 
The disagreement at $tT_c^2 \simle 0.01$ is caused by the lattice discretization error, and the data there should be removed in the $t\to 0$ extrapolation needed in the SF$t$X method.
Because we have a wider range of $t$ for the linear $t\to0$ extrapolation with the clover-shaped operator, we use that for the calculation of $U_{\mu\nu}(t,x)$ and $E(t,x)$ in this study.
With the clover-shaped operator, we may use data down to $tT_c^2 \sim 0.005$.


\end{document}